\documentclass[aps,prd,preprintnumbers,nofootinbib,
superscriptaddress,fleqn,floatfix,tightenlines,10pt,
byrevtex]{revtex4-2}
\usepackage{amsmath,amssymb}
\usepackage{graphicx}
\usepackage{bm}
\usepackage{slashed}
\usepackage[hypertexnames=false,colorlinks=true,citecolor=blue,urlcolor=blue,
linkcolor=black]{hyperref}
\usepackage{tikz}
\makeatletter
\let\revtex@makecaption\@makecaption
\makeatother
\usepackage{subcaption}
\makeatletter
\AtBeginDocument{\let\@makecaption\revtex@makecaption}
\makeatother
\usepackage{placeins}

\newcommand{\signedcutgraphic}[2][]{%
  \begin{tikzpicture}
    \node[inner sep=0] (signedcutimage) {\includegraphics[#1]{#2}};
    \node[
      anchor=south,
      fill=white,
      minimum width=28mm,
      minimum height=5mm,
      inner sep=0pt
    ] at ([xshift=4mm]signedcutimage.south)
      {\scriptsize $b_y\,[\mathrm{fm}]$};
  \end{tikzpicture}%
}

\begin{document}
\preprint{INHA-NTG-11/2026}
\title{Transverse distributions of the energy-momentum tensor for a 
spin-\texorpdfstring{$3/2$}{3/2} baryon}

\author{Hui-Jae Lee}
\email{hjlee6674@inha.edu}
\affiliation{Department of Physics, Inha University, Incheon 22212,
Republic of Korea}

\author{Ki-Hoon Hong}
\email{kihoon.hong@inha.ac.kr}
\affiliation{Institute of Quantum Science, Inha University, Incheon 22212,
Republic of Korea}

\author{June-Young Kim}
\email{jun-young.kim@inha.ac.kr}
\affiliation{Department of Physics, Inha University, Incheon 22212,
Republic of Korea}
\affiliation{Institute of Quantum Science, Inha University, Incheon 22212,
Republic of Korea}

\author{Hyun-Chul Kim}
\email{hchkim@inha.ac.kr}
\affiliation{Department of Physics, Inha University, Incheon 22212,
Republic of Korea}
\affiliation{Institute of Quantum Science, Inha University, Incheon 22212,
Republic of Korea}
\affiliation{School of Physics, Korea Institute for Advanced Study
(KIAS), Seoul 02455, Republic of Korea}

\begin{abstract}
We develop a multipole description of transverse
distributions of the energy-momentum tensor for a spin-$3/2$
baryon in frames connected by a longitudinal boost.  In the
transverse Breit frame, the $T^{00}$, $T^{03}$, and $T^{33}$ matrix
elements are expressed through seven multipole form factors for
energy, angular momentum, and stress.  At finite longitudinal
momentum, we factorize the Lorentz mixing of these three components
from the spin-$3/2$ Wigner rotations of the external states.
The resulting elastic-frame matrix elements contain six transverse
multipoles, whose Fourier transforms define the distributions of
energy, longitudinal momentum, and longitudinal momentum flux.  We
also calculate $T^{++}$, $T^{+-}$, and $T^{--}$ directly with
light-front Rarita-Schwinger spinors.  The elastic-frame construction
provides a continuous interpolation from the transverse Breit frame
to the infinite-momentum frame.  In this limit, the Wigner rotation becomes
the Melosh rotation, and the leading elastic-frame matrix elements
reproduce the corresponding light-front results.
Using the $\Delta$-baryon gravitational form factors obtained in the
Skyrme model as a representative numerical input, we find that the
energy distribution is dominated by the energy monopole defined in the
transverse Breit frame and changes only weakly under longitudinal boosts.
Through boost mixing, this monopole provides the dominant contribution
to the longitudinal momentum distribution and its flux at finite $P_z$.
For a longitudinally polarized spin-$3/2$ target, the distributions contain
only the monopole contributions, whereas those of a transversely polarized
target exhibit spin-dependent quadrupole and octupole deformations and a
dipole that shifts their maxima.
\end{abstract}

\date{\today}
\maketitle

\section{Introduction}
\label{sec:introduction}

The energy-momentum tensor~(EMT) of quantum chromodynamics~(QCD) is
the conserved current associated with space-time translations.  Its
hadronic matrix elements are parametrized by gravitational form
factors~(GFFs), which encode the distributions of energy and momentum
as well as the mass and angular-momentum sum
rules~\cite{Pagels:1966zza,Ji:1995sv, 
Ji:1996ek,Leader:2013jra,Burkert:2023wzr}.  Mellin moments of
generalized parton distributions~(GPDs) provide access to the same
form factors in hard exclusive reactions~\cite{Goeke:2001tz,
Diehl:2003ny,Belitsky:2005qn}.  The stress components of the EMT
contain further information on pressure and shear forces.  Their
spatial integrals obey the von Laue condition for a stable isolated
system, while their spatial dependence is governed by the hadronic
$D$-term and, for higher-spin targets, by the associated
mechanical form factors~\cite{vonLaue:1911,Polyakov:2002yz,Polyakov:2018zvc,
Lorce:2018egm}.

Turning the GFFs into spatial distributions of energy, momentum, and
internal forces requires a definition of a density for a relativistic
bound system.  Spatial distributions have traditionally been defined
as three-dimensional Fourier transforms of the form factors in the
Breit frame.  This definition was applied to the EMT in
Refs.~\cite{Polyakov:2002yz,Goeke:2007fp,Polyakov:2018zvc}.  For a
relativistic system, however, the Breit-frame distributions
 cannot be
interpreted as probability densities: the initial and final states
carry different momenta, so the matrix element is not an expectation
value in a single localized state.  When the localization scale
becomes comparable to the Compton wavelength, relativistic
wave-packet and recoil effects can no longer be neglected.
These effects complicate the interpretation of the resulting
three-dimensional distributions as intrinsic densities of the
hadron~\cite{Jaffe:2020ebz,Epelbaum:2022fjc,Lorce:2025oot}.  A
relativistically consistent framework is therefore needed to relate
the GFFs to spatial distributions of energy, pressure, and shear
forces inside a hadron.

Over the past two decades, such a framework has been developed
in terms of transverse distributions based on light-front~(LF)
kinematics~\cite{Dirac:1949cp,Kogut:1969xa,Soper:1971wn,
Brodsky:1997de}. The Fourier transform of GPDs at zero skewness with
respect to the transverse momentum transfer yields the spatial
distributions of partons in the transverse
plane~\cite{Burkardt:2000za, Burkardt:2002hr}. Since transverse boosts   
form a Galilean subgroup of the LF
kinematics, this construction avoids the recoil ambiguity of
three-dimensional localization, and the resulting,
zero-skewness impact-parameter parton distributions
 admit
a probabilistic interpretation.  The 
framework was subsequently extended to transverse spin
densities~\cite{Diehl:2005jf} and to transverse charge distributions
extracted from empirical electromagnetic form factors of the nucleon,
the deuteron, and the $N\to\Delta$
transition~\cite{Miller:2007uy,Carlson:2007xd,Carlson:2009ovh,
Miller:2010nz}. More recently, the transverse-density framework
 has
been extended to the EMT.  LF distributions of momentum and
internal forces were formulated for spin-$1/2$ and spin-$1$
hadrons~\cite{Freese:2021czn,Freese:2022yur}, and Abel-tomography
relations were established,
between the three-dimensional Breit-frame distributions and
their two-dimensional LF counterparts
~\cite{Panteleeva:2021iip,Kim:2021jjf}.  The frame
dependence of the nucleon EMT distributions has been traced from the
ultrarelativistic to the nonrelativistic
regime~\cite{Lorce:2022cle} and analyzed for a polarized
nucleon~\cite{Won:2025dgc, Won:2026ljg}. Taken together, these developments provide
well-defined spatial representations of the energy density, the
pressure, and the shear forces, related, through the GFFs,
to GPDs extracted from hard exclusive
reactions~\cite{Diehl:2003ny,Belitsky:2005qn,Burkert:2023wzr}. 

All these distributions are defined in the same kinematics:
the momentum transfer is purely transverse, and the longitudinal
momentum of the hadron remains unchanged~\cite{Drell:1969km,West:1970av}.  The
frames with this property constitute the elastic-frame~(EF) family, parametrized
by the longitudinal momentum of the
hadron~\cite{Lorce:2020onh,Panteleeva:2021iip,Kim:2021jjf,
  Lorce:2022cle,Won:2025dgc}. This family interpolates continuously between the
 transverse Breit frame and the infinite-momentum frame~(IMF).
At each longitudinal momentum, a two-dimensional Fourier transform
with respect to the momentum transfer converts the matrix elements
into spatial distributions in the transverse plane.  The transverse
distributions of a spin-$3/2$ baryon contain dipole, quadrupole, and
octupole structures in addition to the
monopole~\cite{Cotogno:2019vjb,Kim:2020lrs}.  The 
corresponding energy and mechanical multipoles have been analyzed for
spin-$1$ and spin-$3/2$ hadrons~\cite{Cosyn:2019aio,
Polyakov:2019lbq,Panteleeva:2020ejw,Kim:2022wkc}.  The longitudinal
boost acts on these structures in two ways.  First, it mixes the
components of the EMT.  Second, since the boost is not collinear with
the initial and final momenta, the spins undergo Wigner
rotations~\cite{Wigner:1939cj,Keister:1991sb}.  In the IMF limit, the
Wigner rotation becomes the Melosh rotation between canonical spin
and LF helicity~\cite{Soper:1971wn,Melosh:1974cu,
Carlson:2003je}.  The dependence on the longitudinal momentum
therefore involves the Lorentz mixing and the spin rotation together,
and both effects must be included to separate the
multipole structure at $P_z=0$ from effects generated by the boost.
This combined treatment has so far been restricted to lower spins.
For a polarized spin-$1/2$ nucleon,
Ref.~\cite{Won:2025dgc} showed that polarization effects are essential
for the longitudinal boost of the Breit-frame EMT distributions and
that the IMF recovers both the good and bad LF EMT
components.  For spin-$1$ hadrons, the multipoles 
generated by a longitudinal boost were studied~\cite{Kim:2022wkc}.  A
spin-$3/2$ baryon goes beyond both cases: the rank-three octupoles 
survive, and six transverse multipole structures appear in an EF.  To
our knowledge, a corresponding analysis, in which the complete
spin-$3/2$ multipole content is followed from the transverse Breit
frame to the IMF, has not yet been carried out. 

In this work, we perform this analysis for a spin-$3/2$ baryon.  We
employ the covariant spin-$3/2$ EMT parametrization of
Refs.~\cite{Cotogno:2019vjb,Kim:2020lrs} and obtain the multipole
form factors at $P_z=0$ by restricting the Breit-frame construction of
Ref.~\cite{Kim:2020lrs} to transverse momentum transfer.  Starting
from these form factors, we show that the longitudinal Lorentz mixing
separates from the spin-$3/2$ Wigner rotations, and we expand the
boosted matrix elements in six transverse multipole structures. Their
Fourier transforms define the transverse spatial distributions of
energy, longitudinal momentum, and longitudinal momentum flux.
We then evaluate the same EMT components directly with LF
Rarita-Schwinger spinors, using the form-factor basis defined in the
Breit frame. This comparison demonstrates, at the level of the
complete matrix elements, how the leading EF and LF terms agree in the
IMF. 

This paper is organized as follows. Section~\ref{sec:formalism32}
defines the covariant spin-$3/2$ EMT parametrization and the
irreducible spin and transverse multipole
basis. Sections~\ref{sec:EFspin32} and \ref{sec:spatial32} derive the
finite-$P_z$ EF matrix elements by factorizing the Lorentz component
mixing from the spin-$3/2$ Wigner rotations and perform their Fourier
transforms to obtain transverse distributions of energy, longitudinal
momentum, and longitudinal momentum flux. Section~\ref{sec:LFspin32}
calculates the LF EMT matrix elements directly with LF
Rarita-Schwinger spinors and matches them to the leading terms of
their EF counterparts in the $P_z\to\infty$
limit. Section~\ref{sec:results32} uses the $\Delta$-baryon EMT form
factors obtained in the Skyrme model, determines a multipole
parametrization of their dependence on momentum transfer, and studies
the $P_z$ dependence of the transverse EMT
distributions. Section~\ref{sec:summary32} summarizes the results.

\section{Covariant EMT and spin-\texorpdfstring{$3/2$}{3/2}
  tensor  algebra} 
\label{sec:formalism32}
In this section, we construct the formalism for the boost analysis. We
first present the Belinfante EMT operator and parametrize its
spin-$3/2$ matrix element in terms of ten covariant GFFs. We also
specify the spinor conventions and state normalization used throughout.
We then construct the irreducible spin and transverse tensors that
organize the matrix elements into multipoles.
\subsection{Covariant matrix element and canonical spin states}
The symmetric and gauge-invariant Belinfante
EMT
in QCD consists of quark~($q$) and gluon~($g$) parts,
which are given respectively by~\cite{Belinfante:1939,Leader:2013jra}
\begin{align}
 \hat T_q^{\mu\nu}
 &=\frac{1}{4}\bar q\left(
 \gamma^\mu i\overleftrightarrow D^{\nu}
 +\gamma^\nu i\overleftrightarrow D^{\mu}\right)q,
 \qquad \hat T_g^{\mu\nu}
 =-F_A^{\mu\lambda}F^{A\nu}{}_{\lambda}
 +\frac{1}{4}g^{\mu\nu}F_A^{\rho\lambda}F^A_{\rho\lambda}, 
 \label{eq:BelinfanteEMT32}
\end{align}
where $A$ denotes the color index and the covariant derivatives take
the form
$\overleftrightarrow D^{\mu}=\overrightarrow D^{\mu}
-\overleftarrow D^{\mu}$, with
$\overrightarrow D^\mu=\overrightarrow \partial^\mu-igA_A^\mu t^A$.
The total EMT is given by the sum of the quark and gluon contributions,
\begin{align}
\hat T^{\mu\nu}
=\sum_q \hat T_q^{\mu\nu}+\hat T_g^{\mu\nu},
\end{align}
where the sum runs over quark flavors. Although the quark and gluon
contributions are not conserved separately, the total EMT obeys
$\partial_\mu\hat T^{\mu\nu}=0$. 

With $a=q,g$ denoting the quark and gluon parts, respectively, the
corresponding matrix element between spin-$3/2$ baryon states is
parametrized in terms of ten covariant EMT form factors,
$F^a_{i,j}(t)$, as~\cite{Cotogno:2019vjb, Kim:2020lrs}
\begin{align}
&\langle p',\sigma'|\hat T_a^{\mu\nu}(0)|p,\sigma\rangle
\nonumber\\
=&-\bar u^{\alpha'}(p',\sigma')\Bigg[
\frac{P^\mu P^\nu}{m}
\left(g_{\alpha\alpha'}F^a_{1,0}(t)
-\frac{\Delta_{\alpha'}\Delta_\alpha}{2m^2}F^a_{1,1}(t)\right)
\nonumber\\
&+\frac{\Delta^\mu\Delta^\nu-g^{\mu\nu}\Delta^2}{4m}
\left(g_{\alpha'\alpha}F^a_{2,0}(t)
-\frac{\Delta_{\alpha'}\Delta_\alpha}{2m^2}F^a_{2,1}(t)\right)
+m g^{\mu\nu}
\left(g_{\alpha'\alpha}F^a_{3,0}(t)
-\frac{\Delta_{\alpha'}\Delta_\alpha}{2m^2}F^a_{3,1}(t)\right)
\nonumber\\
&+\frac{i(P^\mu\sigma^{\nu\rho}+P^\nu\sigma^{\mu\rho})\Delta_\rho}{2m}
\left(g_{\alpha'\alpha}F^a_{4,0}(t)
-\frac{\Delta_{\alpha'}\Delta_\alpha}{2m^2}F^a_{4,1}(t)\right)
\nonumber\\
&-\frac{1}{m}\Big(
\Delta^\mu g^\nu_{\alpha'}\Delta_\alpha
+\Delta^\nu g^\mu_{\alpha'}\Delta_\alpha
+\Delta^\mu g^\nu_\alpha\Delta_{\alpha'}
+\Delta^\nu g^\mu_\alpha\Delta_{\alpha'}
-2g^{\mu\nu}\Delta_{\alpha'}\Delta_\alpha
\nonumber\\
&\hspace{24mm}
-g^\mu_{\alpha'}g_\alpha^\nu\Delta^2
-g^\nu_{\alpha'}g^\mu_\alpha\Delta^2\Big)F^a_{5,0}(t)
+m\left(g_{\alpha'}^\mu g_\alpha^\nu
+g_{\alpha'}^\nu g_\alpha^\mu\right)F^a_{6,0}(t)
\Bigg]u^\alpha(p,\sigma),
\label{eq:covariantEMTspin32}
\end{align}
where $\sigma^{\mu\nu}=i[\gamma^\mu,\gamma^\nu]/2$. The ket
$|p,\sigma\rangle$ and the bra $\langle p',\sigma'|$ represent the
initial and final spin-$3/2$ baryons, respectively.
Their four-momenta are denoted by $p$ and $p'$, and their
canonical spin projections by $\sigma$ and $\sigma'$, respectively.
Both momenta satisfy the on-shell
conditions $p^{2}=p'^{2}=m^{2}$, where $m$ denotes the mass of the
spin-$3/2$ baryon. The one-particle states are
covariantly normalized by
\begin{align}
 \langle p',\sigma'|p,\sigma\rangle
 =2p^0(2\pi)^3\delta^{(3)}(\bm p'-\bm p)\delta_{\sigma'\sigma}.
 \label{eq:statenormalization32}
\end{align}
We introduce the average momentum $P=(p'+p)/2$, the momentum transfer
$\Delta=p'-p$, 
and the invariant momentum transfer $t=\Delta^{2}$.
The on-shell conditions then imply
\begin{align}
 P\cdot\Delta=0,
 \qquad
 P^2=m^2-\frac{t}{4}.
 \label{eq:covariantConstraints32}
\end{align}

In Eq.~\eqref{eq:covariantEMTspin32}, $u^\alpha(p,\sigma)$ is the
Rarita--Schwinger spinor of the initial baryon, and
$\bar u^{\alpha'}(p',\sigma')$ the adjoint spinor of the final one.
The on-shell spinor satisfies~\cite{Rarita:1941mf}
\begin{align}
 (\slashed p-m)u^\alpha(p,\sigma)=0,
 \qquad
 p_\alpha u^\alpha(p,\sigma)=0,
 \qquad
 \gamma_\alpha u^\alpha(p,\sigma)=0,
 \label{eq:RSconstraints32}
\end{align}
where $\slashed p=\gamma^\mu p_\mu$. Explicit expressions for the
canonical and LF Rarita-Schwinger spinors are given in
Appendix~\ref{app:canonicalRS32}.

The quark and gluon EMT form factors in
Eq.~\eqref{eq:covariantEMTspin32} depend on the renormalization scale.
For brevity, we omit this dependence from the notation.
Conservation of the total EMT requires $F^a_{3,0}$, $F^a_{3,1}$, and
$F^a_{6,0}$ to sum to zero over all quark and gluon contributions. For
an individual contribution $a$, however, these form factors need not
vanish because the corresponding quark or gluon EMT is not separately
conserved. At $t=0$, Poincar\'e symmetry also yields the momentum and
angular-momentum sum rules~\cite{Cotogno:2019vjb,Kim:2020lrs}, so that
the form factors satisfy
\begin{align}
 \sum_a F^a_{3,0}(t)=\sum_a F^a_{3,1}(t)
 =\sum_a F^a_{6,0}(t)=0,
 \qquad
 \sum_a F^a_{1,0}(0)=1,
 \qquad
 \sum_a F^a_{4,0}(0)=\frac32.
 \label{eq:totalconstraints32}
\end{align}
The constraints on the mass and spin form factors at zero momentum
transfer are universal consequences of Poincar\'e
covariance~\cite{Cotogno:2019xcl,Lorce:2019sbq}.

\subsection{Irreducible spin and transverse tensors}
\label{sec:ME}
To study the frame dependence of the EMT matrix elements in
Eq.~\eqref{eq:covariantEMTspin32} under longitudinal boosts, we expand
them in structures formed by contracting spin multipole operators with
irreducible transverse tensors. We first specify the spin part of this
basis. For matrix elements between spin-$S$ states, the spin multipole
rank cannot exceed $2S$~\cite{Varshalovich:1988ifq,
Cotogno:2019vjb}. The spin-$3/2$ basis therefore terminates at the
octupole and consists of $\bm1$, $S^i$, $Q^{ij}$, and $O^{ijk}$.
The rank-two and rank-three tensors are constructed from symmetric
traceless products of the spin operators. Following
Ref.~\cite{Kim:2020lrs}, we define them by
\begin{subequations}
\label{eq:Qspin32}
\begin{align}
 Q^{ij}
 &=\frac12\left[
 S^iS^j+S^jS^i-\frac{2}{3}S(S+1)\delta^{ij}\bm1
 \right],\\
 O^{ijk}
 &=\frac16\Bigg[
 S^iS^jS^k+S^jS^iS^k+S^kS^jS^i
 +S^jS^kS^i+S^iS^kS^j+S^kS^iS^j
 \nonumber\\[-2pt]
 &\hspace{14mm}
 -\frac{6S(S+1)-2}{5}
 \left(
 \delta^{ij}S^k+\delta^{ik}S^j+\delta^{jk}S^i
 \right)
 \Bigg],
 \qquad (i,j,k=1,2,3).
\end{align}
\end{subequations}
By construction, both tensors are symmetric and traceless:
$Q^{ii}=0$ and $O^{iij}=O^{iji}=O^{jii}=0$.

Under the longitudinal boosts considered below, the momentum transfer
remains in the transverse plane. The two-dimensional multipole
expansion is therefore constructed with respect to rotations about the
$z$ axis. We first classify the spin multipoles by their $z$-axis
projection $m_S$ as
\begin{subequations}
\label{eq:spinmultipoleprojections32}
\begin{align}
 \bm1,\quad Q^{33}
 &:\qquad m_S=0,
 \\[1ex]
 S^i,\quad O^{33i}
 &:\qquad m_S=\pm1,
 \\[1ex]
 Q^{ij}
 &:\qquad m_S=\pm2,
 \\[1ex]
 O^{ijk}
 &:\qquad m_S=\pm3.
\end{align}
\end{subequations}
Here $i,j,k=1,2$ denote transverse indices.

We next introduce the rank-$n$ transverse tensors $X_n$ that describe
the dependence on the direction of $\bm\Delta_\perp$:
\begin{subequations}
\label{eq:Xspin32}
\begin{alignat}{2}
 X_0(\theta_\Delta)
 &=1,
 &\qquad m_L&=0,
 \\[1.5ex]
 X_1^i(\theta_\Delta)
 &=
 \hat\Delta_\perp^i,
 &\qquad m_L&=\pm1,
 \\
 X_2^{ij}(\theta_\Delta)
 &=
 \hat\Delta_\perp^i\hat\Delta_\perp^j
   -\frac12\delta^{ij},
 &\qquad m_L&=\pm2,
 \\
 X_3^{ijk}(\theta_\Delta)
 &=
 \hat\Delta_\perp^i\hat\Delta_\perp^j\hat\Delta_\perp^k
 -\frac14\left(
 \delta^{ij}\hat\Delta_\perp^k
 +\delta^{ik}\hat\Delta_\perp^j
 +\delta^{jk}\hat\Delta_\perp^i
 \right),
 &\qquad m_L&=\pm3, \\
 &\vdots \nonumber
\end{alignat}
\end{subequations}
Here $i,j,k=1,2$ are transverse indices,
$\hat{\bm\Delta}_\perp\equiv
\bm\Delta_\perp/|\bm\Delta_\perp|$ is the unit vector in the direction
of $\bm\Delta_\perp$, and $\theta_\Delta$ is its azimuthal angle. The
quantity $m_L$ denotes the $z$-axis projection of $X_n$.

\begin{figure}[t]
\centering
\includegraphics[width=0.40\textwidth]{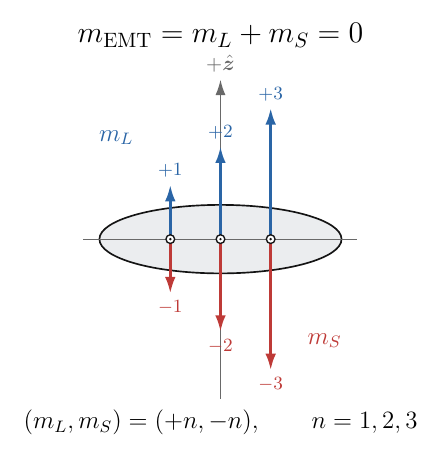}
\caption{Allowed pairs $(m_L,m_S)$ satisfying the selection rule 
$m_{\mathrm{EMT}}=m_L+m_S=0$ for the EMT components considered in
this work.}
\label{fig:1}
\end{figure}

To determine how these transverse tensors~\eqref{eq:Xspin32} combine
with the spin multipoles~\eqref{eq:spinmultipoleprojections32}, the
EMT components are likewise labeled by the projection
$m_{\mathrm{EMT}}$:
\begin{subequations}
\label{eq:EMTprojections32}
\begin{align}
 T^{00},\quad T^{03},\quad T^{33}
 &:\qquad m_{\mathrm{EMT}}=0,
 \label{eq:EMTprojections32a}\\[1ex]
 T^{0i},\quad T^{3i}
 &:\qquad m_{\mathrm{EMT}}=\pm1,
 \\[1ex]
 T^{ij}
 &:\qquad m_{\mathrm{EMT}}=0,\pm2.
\end{align}
\end{subequations}
Since the transverse distributions of energy, longitudinal momentum,
and longitudinal momentum flux follow from $T^{00}$, $T^{03}$, and
$T^{33}$, respectively, we restrict ourselves to the components in
Eq.~\eqref{eq:EMTprojections32a} with $m_{\mathrm{EMT}}=0$. For these,
rotational covariance requires
\begin{equation}
 m_{\mathrm{EMT}}=m_L+m_S=0,
 \qquad
 m_L=-m_S.
 \label{eq:projectionselection32}
\end{equation}
The allowed $(m_L,m_S)$ pairs are shown in Fig.~\ref{fig:1}. A
spin-$3/2$ baryon has $|m_S|\leq 3$, so only the tensors of rank 
$n\leq 3$ in Eq.~\eqref{eq:Xspin32} contribute. Each transverse
tensor is contracted with a spin multipole of projection
$m_S=-m_L$. These contractions form the basis for the multipole
expansion of the EMT matrix elements considered below.

\section{EF matrix elements and Wigner rotation}
\label{sec:EFspin32}

In this section, we introduce the EF kinematics for general
longitudinal momentum $P_z$. At $P_z=0$, the EF becomes the transverse
Breit frame, where we perform the multipole expansion and obtain seven
multipole form factors. We use these Breit-frame form factors, rather
than the covariant GFFs, as a basis for expressing the EF multipole
form factors at finite $P_z$. The boost produces Lorentz mixing of the
EMT components and a Wigner rotation of the canonical spins. We
separate these effects and take the $P_z\to\infty$ limit to obtain the
IMF multipole form factors.

\subsection{EF kinematics}
To study how the EMT matrix elements change under longitudinal boosts,
we work in the symmetric EF considered in
Refs.~\cite{Lorce:2020onh,Kim:2021jjf,Lorce:2022cle}. The average
momentum $P$ and the momentum transfer $\Delta$ are given as 
\begin{align}
 P^\mu&=(P^0,\bm0_\perp,P_z), \qquad
 \Delta^\mu=(0,\bm\Delta_\perp,0),
 \label{eq:EFkinematics32}
\end{align}
where
\begin{equation}
 t=\Delta^2=-\bm\Delta_\perp^2,
 \qquad
 \tau=-\frac{t}{4m^2},
 \qquad
 P^0=\sqrt{P_z^2+m^2(1+\tau)}.
 \label{eq:EFtau32}
\end{equation}
The initial and final states have the same energy $P^0$ and
longitudinal momentum $P_z$, and the momentum transfer is purely
transverse. At $P_z=0$, these kinematics reduce to the transverse
Breit frame, while the limit $P_z\to\infty$ defines the IMF.

For the normalization used below, we introduce the boost parameters
evaluated in the forward limit $t\to0$, following
Ref.~\cite{Won:2025dgc}:
\begin{equation}
 E_P=\sqrt{m^2+P_z^2},
 \qquad
 \gamma_P=\frac{E_P}{m},
 \qquad
 \beta_P=\frac{P_z}{E_P}.
 \label{eq:forwardboost32}
\end{equation}
At large $P_z$, the matrix elements of $T^{00}$, $T^{03}$, and
$T^{33}$ grow with a common kinematic prefactor. We remove this
prefactor by defining
\begin{align}
 \bigl\langle \hat T_a^{\mu\nu}\bigr\rangle_{\mathrm{EF}}
 (P_z,\bm\Delta_\perp;\sigma',\sigma)
 &\equiv
 \frac{\langle p',\sigma'|\hat T_a^{\mu\nu}(0)|p,\sigma\rangle}
 {2P^0\gamma_P}.
 \label{eq:EFmatrixelement32}
\end{align}
Here, the factor $1/(2P^0)$ follows from the covariant normalization
of the external states, while $1/\gamma_P$ removes the additional
kinematic enhancement caused by the longitudinal boost. Without the
latter factor, the corresponding spatial distributions would diverge
as $P_z\to\infty$. At $P_z=0$, $\gamma_P=1$, and
Eq.~\eqref{eq:EFmatrixelement32} reduces to the normalization used in
the transverse Breit frame.

\subsection{Matrix elements in the transverse Breit frame}
Setting $P_z=0$ in Eqs.~\eqref{eq:EFkinematics32} and
\eqref{eq:EFtau32}, we obtain
\begin{align}
 P^\mu&=\left(m\sqrt{1+\tau},\bm0_\perp,0\right),
 \qquad \Delta^\mu=(0,\bm\Delta_\perp,0).
 \label{eq:BFkinematics32}
\end{align}
These momenta satisfy $t=\Delta^2=-\bm\Delta_\perp^2$,
$P\cdot\Delta=0$, and
\begin{align}
 &P^2=m^2(1+\tau)=m^2-\frac{t}{4},
 \label{eq:BFconstraints32}
\end{align}
in agreement with the on-shell relation in
Eq.~\eqref{eq:covariantConstraints32}.

We evaluate the covariant EMT matrix element in
Eq.~\eqref{eq:covariantEMTspin32} with the canonical Rarita-Schwinger
spinors for the kinematics of Eq.~\eqref{eq:BFkinematics32}. Expanding
the $T^{00}$, $T^{03}$, and $T^{33}$ components in the multipole basis
of Sec.~\ref{sec:ME}, we obtain
\begin{subequations}
\label{eq:BF32}
\begin{align}
\left.\bigl\langle \hat T_a^{00}\bigr\rangle_{\mathrm{EF}}\right|_{P_z=0}
&=m\Big[
\mathcal E_{0,a}(t)\bm1
+4\tau\mathcal E_{2,a}(t)
\left(
Q^{33}-2Q^{ij}X_2^{ij}(\theta_\Delta)
\right)
\Big]_{\sigma'\sigma},
\label{eq:BFenergy32}\\[0.5ex]
\left.\bigl\langle \hat T_a^{03}\bigr\rangle_{\mathrm{EF}}\right|_{P_z=0}
&=2m\sqrt{\tau}\Big[
\mathcal J_{1,a}(t)
i\epsilon^{ij3}S^iX_1^j(\theta_\Delta)
+4\tau\mathcal J_{3,a}(t)\Big(
i\epsilon^{ij3}O^{i33}X_1^j(\theta_\Delta)
-4i\epsilon^{3ik}O^{ij\ell}X_3^{kj\ell}(\theta_\Delta)
\Big)
\Big]_{\sigma'\sigma},
\label{eq:BFangular32}\\[0.5ex]
\left.\bigl\langle \hat T_a^{33}\bigr\rangle_{\mathrm{EF}}\right|_{P_z=0}
&=m\Big[
\mathcal P_{0,a}(t)\bm1
+\mathcal P_{0Q,a}(t)Q^{33}
+4\tau\mathcal P_{2,a}(t)
Q^{ij}X_2^{ij}(\theta_\Delta)
\Big]_{\sigma'\sigma}.
\label{eq:BFstress32}
\end{align}
\end{subequations}
The seven multipole form factors in Eq.~\eqref{eq:BF32} are
dimensionless. The authors of Ref.~\cite{Kim:2020lrs} expressed the
same Breit-frame matrix elements in a three-dimensional basis, whereas
we employ a two-dimensional transverse one. The distinction is purely
geometrical: tracelessness is imposed in three or in two
dimensions~\cite{Kim:2022bia}. Appendix~\ref{app:intrinsicFF32} gives
the explicit relations between the two sets of form factors.

In the present convention, we relate the seven multipole form factors
to the covariant functions $F^a_{i,j}(t)$ as
\begin{subequations}
\label{eq:32FF}
\begin{align}
\mathcal E_{0,a}(t)=\frac{1}{3}\Big[&
 (1+\tau)(3+2\tau)F^a_{1,0}(t)
 +2\tau(1+\tau)^2F^a_{1,1}(t)
 +\tau(3+2\tau)F^a_{2,0}(t)
 +2\tau^2(1+\tau)F^a_{2,1}(t)
 \nonumber\\
 &+(3+2\tau)F^a_{3,0}(t)
 +2\tau(1+\tau)F^a_{3,1}(t)
 -2\tau(3+2\tau)F^a_{4,0}(t)
 -4\tau^2(1+\tau)F^a_{4,1}(t)
 \nonumber\\
 &-8\tau(1+2\tau)F^a_{5,0}(t)
 +2\tau F^a_{6,0}(t)\Big],
 \label{eq:E0spin32}\\[1pt]
\mathcal E_{2,a}(t)=\frac{1}{12}\Big[&
 (1+\tau)F^a_{1,0}(t)
 +(1+\tau)^2F^a_{1,1}(t)
 +\tau F^a_{2,0}(t)
 +\tau(1+\tau)F^a_{2,1}(t)
 +F^a_{3,0}(t)
 +(1+\tau)F^a_{3,1}(t)
 \nonumber\\
 &-2\tau F^a_{4,0}(t)
 -2\tau(1+\tau)F^a_{4,1}(t)
 -4(1+2\tau)F^a_{5,0}(t)
 +F^a_{6,0}(t)\Big],
 \label{eq:E2spin32}\\[1pt]
\mathcal J_{1,a}(t)=\frac{1}{15}\Big[&
 (5+4\tau)F^a_{4,0}(t)
 +4\tau(1+\tau)F^a_{4,1}(t)
 +20\tau F^a_{5,0}(t)
 -5F^a_{6,0}(t)\Big],
 \label{eq:J1spin32}\\[1pt]
\mathcal J_{3,a}(t)=\frac{1}{24}\Big[&
 F^a_{4,0}(t)
 +(1+\tau)F^a_{4,1}(t)\Big],
 \label{eq:J3spin32}\\[1pt]
\mathcal P_{0,a}(t)=-\frac{1}{3}\Big[&
 \tau(3+2\tau)F^a_{2,0}(t)
 +2\tau^2(1+\tau)F^a_{2,1}(t)
 +(3+2\tau)F^a_{3,0}(t)
 +2\tau(1+\tau)F^a_{3,1}(t)
 \nonumber\\
 &-8\tau(2+\tau)F^a_{5,0}(t)
 +2F^a_{6,0}(t)\Big],
 \label{eq:P0spin32}\\[1pt]
\mathcal P_{0Q,a}(t)=-\frac{1}{3}\Big[&
 \tau^2F^a_{2,0}(t)
 +\tau^2(1+\tau)F^a_{2,1}(t)
 +\tau F^a_{3,0}(t)
 +\tau(1+\tau)F^a_{3,1}(t)
 \nonumber\\
 &+4\tau(1-\tau)F^a_{5,0}(t)
 -2F^a_{6,0}(t)\Big],
 \label{eq:P0Qspin32}\\[1pt]
\mathcal P_{2,a}(t)=\frac{1}{6}\Big[&
 \tau F^a_{2,0}(t)
 +\tau(1+\tau)F^a_{2,1}(t)
 +F^a_{3,0}(t)
 +(1+\tau)F^a_{3,1}(t)
 -4(1+\tau)F^a_{5,0}(t)\Big].
 \label{eq:P2spin32}
\end{align}
\end{subequations}
For an individual quark or gluon contribution, these pressure
multipoles retain the nonconserved EMT form factors and therefore
provide a useful basis for interpreting the monopole and quadrupole
forces between the two subsystems~\cite{Kim:2025iis}.
When the partonic label $a$ is omitted, each form factor denotes the
sum over all quark and gluon contributions. Applying the total-EMT
constraints in Eq.~\eqref{eq:totalconstraints32} to the relations
above gives
\begin{equation}
 \mathcal E_0(0)=1,
 \qquad
 \mathcal J_1(0)=\frac12,
 \qquad
 \mathcal P_0(0)=\mathcal P_{0Q}(0)=0.
 \label{eq:intrinsicforwardchecks32}
\end{equation}
The first two relations express the mass and spin sum rules. Although
the baryon has spin $3/2$, the spin dependence in
Eq.~\eqref{eq:BFangular32} is carried by the operator $S^i$, so the
sum rule takes the form $2\mathcal J_1(0)S^i=S^i$ and yields
$\mathcal J_1(0)=1/2$~\cite{Kim:2020lrs}. Conservation of the total
EMT requires $\mathcal P_0(0)$ and $\mathcal P_{0Q}(0)$ to vanish.

\subsection{Longitudinal boost and Wigner rotation}
We now obtain the EF matrix elements at finite $P_z$ by applying a
longitudinal boost to the transverse Breit-frame results in
Eq.~\eqref{eq:BF32}. The boost mixes the EMT components $T^{00}$,
$T^{03}$, and $T^{33}$ and induces a Wigner rotation of the canonical
spin states. Below we derive the general transformation relating the
matrix elements in the two frames and work out its explicit multipole
form in the next subsection. The initial and final Breit-frame momenta
are written as 
\begin{equation}
 p_{\rm B}^{\mu}
 =\left(
 m\sqrt{1+\tau},
 -\frac{\bm\Delta_\perp}{2},
 0
 \right),
 \qquad
 p_{\rm B}^{\prime\mu}
 =\left(
 m\sqrt{1+\tau},
 +\frac{\bm\Delta_\perp}{2},
 0
 \right).
 \label{eq:BFmomenta32}
\end{equation}
The two momenta are on shell and have the same energy, while their
transverse components have equal magnitudes and opposite directions.
A Lorentz boost $\Lambda$ along the $z$ axis maps them to the
finite-$P_z$ EF kinematics defined in
Eq.~\eqref{eq:EFkinematics32}:
\begin{equation}
 p^\mu=\Lambda^\mu{}_\nu p_{\rm B}^\nu,
 \qquad
 p'^\mu=\Lambda^\mu{}_\nu p_{\rm B}^{\prime\nu}.
 \label{eq:boostedmomenta32}
\end{equation}
The corresponding boost matrix is given by
\begin{equation}
 \Lambda^\mu{}_\nu=
 \begin{pmatrix}
  \gamma&0&0&\gamma\beta\\
  0&1&0&0\\
  0&0&1&0\\
  \gamma\beta&0&0&\gamma
 \end{pmatrix},
 \qquad
 \gamma=\frac{1}{\sqrt{1-\beta^2}}
 =\frac{P^0}{m\sqrt{1+\tau}},
 \qquad
 \beta=\frac{P_z}{P^0}.
 \label{eq:boost32}
\end{equation}
Because $P^0$ depends on $\tau$, the boost velocity $\beta$ at
nonzero $t$ differs from the velocity $\beta_P=P_z/E_P$ in the
forward limit, defined in Eq.~\eqref{eq:forwardboost32}. In addition,
both $p_{\rm B}$ and $p'_{\rm B}$ acquire transverse components, so
the longitudinal boost is not collinear with either momentum. The
canonical spin states therefore undergo Wigner
rotations~\cite{Wigner:1939cj,Keister:1991sb}. Their explicit form
for the present EF kinematics was derived in
Refs.~\cite{Chen:2022smg,Won:2025dgc}.

The matrix $D^{(3/2)}(p_{\rm B},\Lambda)$ for the incoming state
connects the canonical spin projection $\sigma_{\rm B}$ in the
transverse Breit frame to the projection $\sigma$ in the EF. In
spin-$3/2$ space, it is given by
\begin{align}
 D^{(3/2)}(p_{\rm B},\Lambda)
 &=\exp\!\left[
 i\theta(P_z,t)\,
 \epsilon^{ij3}S^iX_1^j(\theta_\Delta)
 \right],
 \label{eq:Dspin32}
\end{align}
The Wigner angle $\theta(P_z,t)$ is determined by
\begin{equation}
 \cos\theta(P_z,t)
 =\frac{P^0+m(1+\tau)}{(P^0+m)\sqrt{1+\tau}},
 \qquad
 \sin\theta(P_z,t)
 =-\frac{\sqrt\tau\,P_z}{(P^0+m)\sqrt{1+\tau}}.
 \label{eq:Wignerangle32}
\end{equation}
At $P_z=0$, Eq.~\eqref{eq:Wignerangle32} yields $\theta(0,t)=0$, and
the rotation matrix~\eqref{eq:Dspin32} reduces to the identity. Its
compact form in the multipole basis of Sec.~\ref{sec:ME} follows once
we define
\begin{equation}
 c_n\equiv\cos(n\theta),
 \qquad
 s_n\equiv\sin(n\theta),
 \qquad
 n=\frac12,1,\frac32,3.
 \label{eq:Wignertrigshorthand32}
\end{equation}
Expressed in terms of $c_n$ and $s_n$, Eq.~\eqref{eq:Dspin32} reads
\begin{align}
 D^{(3/2)}_{\sigma'\sigma}(p_{\rm B},\Lambda)
={}&\Bigg[
 \frac12\left(
 c_{1/2}+c_{3/2}
 \right)\bm1
 +\frac15\left(
 s_{1/2}+3s_{3/2}
 \right)
 i\epsilon^{ij3}S^iX_1^j(\theta_\Delta)
 \nonumber\\
&+\frac14\left(
 c_{1/2}-c_{3/2}
 \right)
 \left\{
 Q^{33}+2Q^{ij}X_2^{ij}(\theta_\Delta)
 \right\}
 \nonumber\\
&+\frac14\left(
 3s_{1/2}-s_{3/2}
 \right)
 \left\{
 i\epsilon^{ij3}O^{i33}X_1^j(\theta_\Delta)
 +\frac43 i\epsilon^{3ik}O^{ij\ell}
 X_3^{kj\ell}(\theta_\Delta)
 \right\}
 \Bigg]_{\sigma'\sigma}.
 \label{eq:Dspin32multipole}
\end{align}

The corresponding matrix $D^{(3/2)}(p'_{\rm B},\Lambda)$ for the
outgoing state connects $\sigma'_{\rm B}$ in the transverse Breit
frame to $\sigma'$ in the EF. Since the transverse momentum of the
outgoing state is reversed, the Wigner rotation has the opposite
sense. Consequently,
\begin{equation}
 D^{(3/2)\dagger}(p'_{\rm B},\Lambda)
 =D^{(3/2)}(p_{\rm B},\Lambda).
 \label{eq:Drelation32}
\end{equation}
Combining the Wigner rotations in Eqs.~\eqref{eq:Dspin32} and
\eqref{eq:Drelation32} with the Lorentz boost in
Eq.~\eqref{eq:boost32}, we obtain the general relation between the
finite-$P_z$ EF and transverse Breit-frame matrix elements:
\begin{align}
 &\langle p',\sigma'|\hat T_a^{\mu\nu}(0)|p,\sigma\rangle
 =
 \sum_{\sigma'_{\rm B},\sigma_{\rm B}}
 D^{(3/2)}_{\sigma_{\rm B}\sigma}
 (p_{\rm B},\Lambda)
 D^{(3/2)*}_{\sigma'_{\rm B}\sigma'}
 (p'_{\rm B},\Lambda)
 \Lambda^\mu{}_{\alpha}\Lambda^\nu{}_{\beta}
 \langle p'_{\rm B},\sigma'_{\rm B}|
 \hat T_a^{\alpha\beta}(0)
 |p_{\rm B},\sigma_{\rm B}\rangle.
 \label{eq:Wignertransform32}
\end{align}

\subsection{Finite-$P_z$ matrix elements and multipole form factors}
Applying the general boost relation in
Eq.~\eqref{eq:Wignertransform32} to the EMT components $T^{00}$,
$T^{03}$, and $T^{33}$, we express their EF matrix elements at finite
$P_z$ in terms of the transverse Breit-frame results:
\begin{align}
\left.
\begin{pmatrix}
\langle \hat T_a^{00}\rangle_{\mathrm{EF}}\\[5pt]
\langle \hat T_a^{03}\rangle_{\mathrm{EF}}\\[5pt]
\langle \hat T_a^{33}\rangle_{\mathrm{EF}}
\end{pmatrix}
\right|_{P_z}
={}&\frac{\gamma}{\gamma_P}
\begin{pmatrix}
1&2\beta&\beta^2\\
\beta&1+\beta^2&\beta\\
\beta^2&2\beta&1
\end{pmatrix}
\begin{pmatrix}
D^{(3/2)}(p_{\rm B},\Lambda)
\left.\langle \hat T_a^{00}\rangle_{\mathrm{EF}}\right|_{P_z=0}
D^{(3/2)}(p_{\rm B},\Lambda)
\\[7pt]
D^{(3/2)}(p_{\rm B},\Lambda)
\left.\langle \hat T_a^{03}\rangle_{\mathrm{EF}}\right|_{P_z=0}
D^{(3/2)}(p_{\rm B},\Lambda)
\\[7pt]
D^{(3/2)}(p_{\rm B},\Lambda)
\left.\langle \hat T_a^{33}\rangle_{\mathrm{EF}}\right|_{P_z=0}
D^{(3/2)}(p_{\rm B},\Lambda)
\end{pmatrix}_{\sigma'\sigma}.
\label{eq:EFamplitudes32factorized}
\end{align}
At $P_z=0$, $\beta=\theta=0$, $\gamma=\gamma_P=1$, and
$D^{(3/2)}=\bm{1}$. Equation~\eqref{eq:EFamplitudes32factorized} then 
reduces to Eq.~\eqref{eq:BF32}. 

To verify Eq.~\eqref{eq:EFamplitudes32factorized} independently, we
evaluate its left-hand side directly from the covariant EMT matrix
element in Eq.~\eqref{eq:covariantEMTspin32}, using the EF kinematics
at finite $P_z$ in Eqs.~\eqref{eq:EFkinematics32} and
\eqref{eq:EFtau32} together with the normalization in
Eq.~\eqref{eq:EFmatrixelement32}. The result agrees with the
right-hand side obtained by boosting the transverse Breit-frame
matrix elements.

To identify the corresponding multipole form factors, we expand each
of the three matrix elements in
Eq.~\eqref{eq:EFamplitudes32factorized} in the six multipole
structures introduced in Sec.~\ref{sec:ME}:
\begin{align}
\bigl\langle \hat T_a^{\mu\nu}\bigr\rangle_{\mathrm{EF}}
={}&m\Big[
C_{0,a}^{\mu\nu}(P_z,t)\bm1
+C_{0Q,a}^{\mu\nu}(P_z,t)Q^{33}+2\sqrt\tau\,C_{1S,a}^{\mu\nu}(P_z,t)
i\epsilon^{ij3}S^iX_1^j(\theta_\Delta)
\nonumber\\[-2pt]
&\hspace{13mm}
+2\sqrt\tau\,C_{1O,a}^{\mu\nu}(P_z,t)
i\epsilon^{ij3}O^{i33}X_1^j(\theta_\Delta)
+4\tau C_{2Q,a}^{\mu\nu}(P_z,t)
Q^{ij}X_2^{ij}(\theta_\Delta)
\nonumber\\[-2pt]
&\hspace{13mm}
+8\tau^{3/2}C_{3O,a}^{\mu\nu}(P_z,t)
 i\epsilon^{3ik}O^{ij\ell}X_3^{kj\ell}(\theta_\Delta)
 \Big]_{\sigma'\sigma}.
\label{eq:EFmultipoledecomposition32}
\end{align}
Here and below, $\mu,\nu\in\{0,3\}$. The six dimensionless
functions $C_{A,a}^{\mu\nu}(P_z,t)$, with $A\in\{0,0Q,1S,1O,2Q,3O\}$,
are the EF multipole form factors for the component $T^{\mu\nu}$.
Unlike the covariant GFFs $F^a_{i,j}(t)$, they carry an explicit
$P_z$ dependence, which stems solely from the Lorentz mixing of the
EMT components and the Wigner rotation of the canonical spin states.

Their explicit expressions in terms of the seven transverse
Breit-frame multipole form factors are given in
Appendix~\ref{app:DMD32}. At $P_z=0$, they reduce to the multipole
form factors in Eq.~\eqref{eq:BF32}, while their $P_z\to\infty$ limits
give the IMF multipole form factors discussed next.

\subsection{Multipole form factors in the IMF}
We obtain the IMF matrix elements and multipole form factors by
taking the $P_z\to+\infty$ limit of the EF relation in
Eq.~\eqref{eq:EFamplitudes32factorized}. At fixed $t$, the
definitions in Eqs.~\eqref{eq:EFtau32}, \eqref{eq:forwardboost32},
and \eqref{eq:boost32} lead to 
\begin{equation}
 \beta\to1,
 \qquad
 \gamma_P\to\infty,
 \qquad
 \frac{\gamma}{\gamma_P}
 =\frac{P^0}{E_P\sqrt{1+\tau}}
 \longrightarrow\frac{1}{\sqrt{1+\tau}}.
 \label{eq:IMFboostlimits32}
\end{equation}
The $P_z$ dependence of the Wigner matrices in
Eq.~\eqref{eq:EFamplitudes32factorized} enters through
$\theta(P_z,t)$, for which Eq.~\eqref{eq:Wignerangle32} yields 
\begin{equation}
 \lim_{P_z\to+\infty}\cos\theta(P_z,t)
 =\frac{1}{\sqrt{1+\tau}},
 \qquad
 \lim_{P_z\to+\infty}\sin\theta(P_z,t)
 =-\frac{\sqrt\tau}{\sqrt{1+\tau}}.
 \label{eq:WignerIMF32}
\end{equation}

We define the IMF matrix element for each component
$\mu,\nu\in\{0,3\}$ by
\[
 \bigl\langle\hat T_a^{\mu\nu}\bigr\rangle_{\mathrm{IMF}}
 \equiv
 \lim_{P_z\to+\infty}
 \bigl\langle\hat T_a^{\mu\nu}\bigr\rangle_{\mathrm{EF}}.
\]
In this limit, the three rows of the boost matrix in
Eq.~\eqref{eq:EFamplitudes32factorized} become identical. Therefore,
\begin{equation}
 \bigl\langle\hat T_a^{00}\bigr\rangle_{\mathrm{IMF}}
 =
 \bigl\langle\hat T_a^{03}\bigr\rangle_{\mathrm{IMF}}
 =
 \bigl\langle\hat T_a^{33}\bigr\rangle_{\mathrm{IMF}}.
 \label{eq:IMFcomponentlimit32}
\end{equation}
The common matrix element is written in the six multipole structures
of Eq.~\eqref{eq:EFmultipoledecomposition32} as
\begin{align}
 \bigl\langle\hat T_a^{\mu\nu}\bigr\rangle_{\mathrm{IMF}}
 &=m\Big[
 C_{0,a}^{\mathrm{IMF}}(t)\bm1
 +C_{0Q,a}^{\mathrm{IMF}}(t)Q^{33}
 +2\sqrt\tau\,C_{1S,a}^{\mathrm{IMF}}(t)
 i\epsilon^{ij3}S^iX_1^j(\theta_\Delta)
 +2\sqrt\tau\,C_{1O,a}^{\mathrm{IMF}}(t)
 i\epsilon^{ij3}O^{i33}X_1^j(\theta_\Delta)
 \nonumber\\[-2pt]
 &\hspace{13mm}
 +4\tau C_{2Q,a}^{\mathrm{IMF}}(t)
 Q^{ij}X_2^{ij}(\theta_\Delta)
 +8\tau^{3/2}C_{3O,a}^{\mathrm{IMF}}(t)
 i\epsilon^{3ik}O^{ij\ell}X_3^{kj\ell}(\theta_\Delta)
 \Big]_{\sigma'\sigma},
 \qquad
\mu,\nu\in\{0,3\}.
 \label{eq:IMFmatrixelement32}
\end{align}
The six functions $C_{A,a}^{\mathrm{IMF}}(t)$ appearing in
Eq.~\eqref{eq:IMFmatrixelement32} are the IMF multipole form factors.
In Eq.~\eqref{eq:EFmultipoledecomposition32}, the three EF matrix
elements are decomposed in terms of the same six structures. Since
they become identical in the IMF, as shown in
Eq.~\eqref{eq:IMFcomponentlimit32}, the form factors for each
structure must approach a common limit in all three components:
\begin{equation}
\begin{aligned}
 C_{A,a}^{\mathrm{IMF}}(t)
 &\equiv
 \lim_{P_z\to+\infty}C_{A,a}^{00}(P_z,t)
 \\
 &=
 \lim_{P_z\to+\infty}C_{A,a}^{03}(P_z,t)
 \\
 &=
 \lim_{P_z\to+\infty}C_{A,a}^{33}(P_z,t),
\end{aligned}
\qquad
 A\in\{0,0Q,1S,1O,2Q,3O\}.
\label{eq:EFmultipoleFFIMFlimit32}
\end{equation}
The six IMF multipole form factors are given by
\begin{widetext}
\begin{subequations}
\label{eq:IMFformfactors32}
\begin{align}
 C_{0,a}^{\mathrm{IMF}}(t)
={}&\frac{1}{(1+\tau)^2}\Big[
 (1-\tau)(\mathcal E_{0,a}(t)+\mathcal P_{0,a}(t))
 +\tau\mathcal P_{0Q,a}(t)
 +2\tau(5-\tau)\mathcal J_{1,a}(t)
 \nonumber\\[-2pt]
 &\hspace{25mm}
 -8\tau^2\mathcal E_{2,a}(t)
 +6\tau^2\mathcal P_{2,a}(t)
 -\frac{96}{5}\tau^3\mathcal J_{3,a}(t)
 \Big],
 \label{eq:IMFC0spin32}\\[2pt]
 C_{0Q,a}^{\mathrm{IMF}}(t)
={}&\frac{1}{(1+\tau)^2}\Big[
 \tau(\mathcal E_{0,a}(t)+\mathcal P_{0,a}(t))
 +\frac{2+\tau}{2}\mathcal P_{0Q,a}(t)
 +2\tau(\tau-2)\mathcal J_{1,a}(t)
 \nonumber\\[-2pt]
 &\hspace{25mm}
 +4\tau(1+2\tau)\mathcal E_{2,a}(t)
 -3\tau^2\mathcal P_{2,a}(t)
 +\frac{48}{5}\tau^2(1+2\tau)\mathcal J_{3,a}(t)
 \Big],
 \label{eq:IMFC0Qspin32}\\[2pt]
 C_{1S,a}^{\mathrm{IMF}}(t)
={}&\frac{1}{(1+\tau)^2}\Big[
 \frac{\tau-5}{5}(\mathcal E_{0,a}(t)+\mathcal P_{0,a}(t))
 +\frac{2-\tau}{5}\mathcal P_{0Q,a}(t)
 +\frac{2}{5}(5-13\tau)\mathcal J_{1,a}(t)
 \nonumber\\[-2pt]
 &\hspace{25mm}
 +\frac{8}{5}\tau(\tau-2)\mathcal E_{2,a}(t)
 +\frac{6}{5}\tau(2-\tau)\mathcal P_{2,a}(t)
 -\frac{288}{25}\tau^2\mathcal J_{3,a}(t)
 \Big],
 \label{eq:IMFC1Sspin32}\\[2pt]
 C_{1O,a}^{\mathrm{IMF}}(t)
={}&\frac{1}{(1+\tau)^2}\Big[
 -\frac{\tau}{2}(\mathcal E_{0,a}(t)+\mathcal P_{0,a}(t))
 -\frac{4+3\tau}{4}\mathcal P_{0Q,a}(t)
 +3\tau\mathcal J_{1,a}(t)
 \nonumber\\[-2pt]
 &\hspace{25mm}
 -2\tau(1+2\tau)\mathcal E_{2,a}(t)
 +\frac{\tau(\tau-2)}{2}\mathcal P_{2,a}(t)
 +\frac{8}{5}\tau(5+8\tau)\mathcal J_{3,a}(t)
 \Big],
 \label{eq:IMFC1Ospin32}\\[2pt]
 C_{2Q,a}^{\mathrm{IMF}}(t)
={}&\frac{1}{(1+\tau)^2}\Big[
 \frac12(\mathcal E_{0,a}(t)+\mathcal P_{0,a}(t))
 -\frac14\mathcal P_{0Q,a}(t)
 +(\tau-2)\mathcal J_{1,a}(t)
 \nonumber\\[-2pt]
 &\hspace{25mm}
 -2\mathcal E_{2,a}(t)
 +\left(1-\frac{\tau}{2}\right)\mathcal P_{2,a}(t)
 -\frac{8}{5}\tau(7+4\tau)\mathcal J_{3,a}(t)
 \Big],
 \label{eq:IMFC2Qspin32}\\[2pt]
 C_{3O,a}^{\mathrm{IMF}}(t)
={}&\frac{1}{(1+\tau)^2}\Big[
 -\frac16(\mathcal E_{0,a}(t)+\mathcal P_{0,a}(t))
 +\frac1{12}\mathcal P_{0Q,a}(t)
 +\mathcal J_{1,a}(t)
 \nonumber\\[-2pt]
 &\hspace{25mm}
 +\left(2+\frac{4\tau}{3}\right)\mathcal E_{2,a}(t)
 -\frac{2+\tau}{2}\mathcal P_{2,a}(t)
 -\frac{8}{5}(5+4\tau)\mathcal J_{3,a}(t)
 \Big].
 \label{eq:IMFC3Ospin32}
\end{align}
\end{subequations}
\end{widetext}
We obtain Eqs.~\eqref{eq:IMFC0spin32}--\eqref{eq:IMFC3Ospin32} by
taking the $P_z\to+\infty$ limit of the finite-$P_z$ EF form factors
listed in Appendix~\ref{app:DMD32}.

At $t=0$, the total-EMT conditions in
Eq.~\eqref{eq:intrinsicforwardchecks32} fix four of the six IMF
multipole form factors:
\begin{align}
 C_0^{\mathrm{IMF}}(0)
 =1, \qquad
 C_{0Q}^{\mathrm{IMF}}(0)
=0, \qquad
 C_{1S}^{\mathrm{IMF}}(0)
 =0, \qquad
 C_{1O}^{\mathrm{IMF}}(0)
 =0.
 \label{eq:IMFconstrainedforward32}
\end{align}
No symmetry relation fixes the remaining two form factors:
\begin{align}
 C_{2Q}^{\mathrm{IMF}}(0)
 =-\frac12-2\mathcal E_2(0)+\mathcal P_2(0), \qquad
 C_{3O}^{\mathrm{IMF}}(0)
 =\frac13+2\mathcal E_2(0)-\mathcal P_2(0)
 -8\mathcal J_3(0).
 \label{eq:IMFremainingforward32}
\end{align}
They correspond to the quadrupole and octupole moments of the IMF
EMT distributions. The constants $-1/2$ and $1/3$ follow from the
mass and spin sum rules: the Melosh rotation transfers
$\mathcal E_0(0)=1$ and $\mathcal J_1(0)=1/2$ into the $2Q$ and
$3O$ channels. The dynamical input consists of $\mathcal E_2(0)$,
$\mathcal P_2(0)$, and $\mathcal J_3(0)$.

\section{Transverse distributions in EFs}
\label{sec:spatial32}
In this section, we construct transverse spatial distributions of
energy, longitudinal momentum, and longitudinal momentum flux at fixed
$P_z$. We analyze these distributions for baryon states polarized
along the longitudinal $z$ axis or the transverse $x$ axis.

\subsection{Multipole expansion in coordinate space}
In the EF, the momentum transfer remains purely transverse,
$\Delta^\mu=(0,\bm\Delta_\perp,0)$. The transverse spatial EMT
distribution at fixed $P_z$ is therefore defined as the
two-dimensional Fourier transform of the EF matrix element in
Eq.~\eqref{eq:EFmatrixelement32} with respect to
$\bm\Delta_\perp$~\cite{Burkardt:2000za,Lorce:2020onh,
Kim:2021jjf,Lorce:2022cle}:
\begin{align}
 T_{a,\mathrm{EF}}^{\mu\nu}
 (\bm x_\perp;P_z;\sigma',\sigma)
 &=
 \int\frac{d^2\bm\Delta_\perp}{(2\pi)^2}
 e^{-i\bm\Delta_\perp\cdot\bm x_\perp}
 \bigl\langle\hat T_a^{\mu\nu}\bigr\rangle_{\mathrm{EF}}
 (P_z,\bm\Delta_\perp;\sigma',\sigma),
 \label{eq:EFdistribution32}
\end{align}
where $\bm x_\perp$ denotes the transverse position conjugate to
$\bm\Delta_\perp$. The $1/\gamma_P$ factor in
Eq.~\eqref{eq:EFmatrixelement32} is independent of $\bm\Delta_\perp$
and can therefore be taken outside the integral in
Eq.~\eqref{eq:EFdistribution32}. This factor affects the overall
normalization of the distribution at fixed $P_z$ but not its
dependence on $\bm x_\perp$.

We define $\hat{\bm x}_\perp\equiv\bm x_\perp/x_\perp
=(\cos\theta_x,\sin\theta_x)=(b_x,b_y)/x_\perp$. The coordinate-space tensors
$X_n(\theta_x)$ are obtained from Eq.~\eqref{eq:Xspin32} by replacing
$\hat{\bm\Delta}_\perp$ with $\hat{\bm x}_\perp$. Since the $X_n$
form an irreducible multipole basis, each rank transforms
independently under the two-dimensional Fourier transform. Inserting
Eq.~\eqref{eq:EFmultipoledecomposition32} into
Eq.~\eqref{eq:EFdistribution32} yields
\begin{align}
 T_{a,\mathrm{EF}}^{\mu\nu}
 (\bm x_\perp;P_z;\sigma',\sigma)
={}&\Big[
 \mathcal C_{0,a}^{\mu\nu}(x_\perp;P_z)\bm1
 +\mathcal C_{0Q,a}^{\mu\nu}(x_\perp;P_z)Q^{33}
 \nonumber\\
 &+\mathcal C_{1S,a}^{\mu\nu}(x_\perp;P_z)
 \epsilon^{ij3}S^iX_1^j(\theta_x)
 +\mathcal C_{1O,a}^{\mu\nu}(x_\perp;P_z)
 \epsilon^{ij3}O^{i33}X_1^j(\theta_x)
 \nonumber\\
 &+\mathcal C_{2Q,a}^{\mu\nu}(x_\perp;P_z)
 Q^{ij}X_2^{ij}(\theta_x)
 +\mathcal C_{3O,a}^{\mu\nu}(x_\perp;P_z)
 \epsilon^{3ik}O^{ij\ell}X_3^{kj\ell}(\theta_x)
 \Big]_{\sigma'\sigma}.
 \label{eq:coordinateC32}
\end{align}
The radial functions $\mathcal C_{A,a}^{\mu\nu}(x_\perp;P_z)$ are
expressed in terms of the two-dimensional Fourier transforms of the
multipole form factors,
\begin{align}
 \widetilde C_{A,a}^{\mu\nu}(x_\perp;P_z)
 &=
 \int\frac{d^2\bm\Delta_\perp}{(2\pi)^2}
 e^{-i\bm\Delta_\perp\cdot\bm x_\perp}
 C_{A,a}^{\mu\nu}(P_z,t).
 \label{eq:Ctilde32}
\end{align}
In coordinate space, the factor $|\bm{\Delta}_\perp|^n$ accompanying
a rank-$n$ transverse tensor in
Eq.~\eqref{eq:EFmultipoledecomposition32} turns into an $n$th-order
radial differential operator acting on
$\widetilde C_{A,a}^{\mu\nu}$. The radial functions in
Eq.~\eqref{eq:coordinateC32} then read
\begin{subequations}
\label{eq:calCdefinitions32}
\begin{align}
 \mathcal C_{0,a}^{\mu\nu}(x_\perp;P_z)
 &=
 m\widetilde C_{0,a}^{\mu\nu}(x_\perp;P_z),
 &&\mathcal C_{0Q,a}^{\mu\nu}(x_\perp;P_z)
 =
 m\widetilde C_{0Q,a}^{\mu\nu}(x_\perp;P_z),
 \\[1ex]
 \mathcal C_{1S,a}^{\mu\nu}(x_\perp;P_z)
 &=
 -\frac{d}{dx_\perp}
 \widetilde C_{1S,a}^{\mu\nu}(x_\perp;P_z),
 &&\mathcal C_{1O,a}^{\mu\nu}(x_\perp;P_z)
 =
 -\frac{d}{dx_\perp}
 \widetilde C_{1O,a}^{\mu\nu}(x_\perp;P_z),
 \\
 \mathcal C_{2Q,a}^{\mu\nu}(x_\perp;P_z)
 &=
 -\frac{x_\perp}{m}\frac{d}{dx_\perp}
 \frac{1}{x_\perp}\frac{d}{dx_\perp}
 \widetilde C_{2Q,a}^{\mu\nu}(x_\perp;P_z),
 \\
 \mathcal C_{3O,a}^{\mu\nu}(x_\perp;P_z)
 &=
 \frac{x_\perp^2}{m^2}\frac{d}{dx_\perp}
 \frac{1}{x_\perp}\frac{d}{dx_\perp}
 \frac{1}{x_\perp}\frac{d}{dx_\perp}
 \widetilde C_{3O,a}^{\mu\nu}(x_\perp;P_z).
\end{align}
\end{subequations}
The signs in Eq.~\eqref{eq:calCdefinitions32} are fixed by the
Fourier phase convention in Eq.~\eqref{eq:EFdistribution32}.

The forward limit of each multipole form factor of rank $n$ is
obtained by integrating the corresponding radial function in
Eq.~\eqref{eq:calCdefinitions32} over the transverse plane with the
weight $x_\perp^n$:
\begin{subequations}
\label{eq:EFmultipoleMoments32}
\begin{align}
 C_{0}^{\mu\nu}(P_z,0)
 &=\frac{1}{m}\int d^2\bm x_\perp\,
 \mathcal C_{0}^{\mu\nu}(x_\perp;P_z),
 \label{eq:EFmonopoleMoment32}\\
 C_{0Q}^{\mu\nu}(P_z,0)
 &=\frac{1}{m}\int d^2\bm x_\perp\,
 \mathcal C_{0Q}^{\mu\nu}(x_\perp;P_z),
 \label{eq:EFspinMonopoleMoment32}\\
 C_{1S}^{\mu\nu}(P_z,0)
 &=\frac12\int d^2\bm x_\perp\,x_\perp
 \mathcal C_{1S}^{\mu\nu}(x_\perp;P_z),
 \label{eq:EF1SMoment32}\\
 C_{1O}^{\mu\nu}(P_z,0)
 &=\frac12\int d^2\bm x_\perp\,x_\perp
 \mathcal C_{1O}^{\mu\nu}(x_\perp;P_z),
 \label{eq:EF1OMoment32}\\
 C_{2Q}^{\mu\nu}(P_z,0)
 &=-\frac{m}{8}\int d^2\bm x_\perp\,x_\perp^2
 \mathcal C_{2Q}^{\mu\nu}(x_\perp;P_z),
 \label{eq:EFquadrupoleMoment32}\\
 C_{3O}^{\mu\nu}(P_z,0)
 &=-\frac{m^2}{48}\int d^2\bm x_\perp\,x_\perp^3
 \mathcal C_{3O}^{\mu\nu}(x_\perp;P_z).
 \label{eq:EFoctupoleMoment32}
\end{align}
\end{subequations}
The first two relations fix the monopole normalizations, while the
integrals weighted by $x_\perp$, $x_\perp^2$, and $x_\perp^3$ define
the dipole, quadrupole, and octupole moments, respectively. The
physical meaning of each moment becomes explicit in the polarized
distributions discussed in the next subsection.

\subsection{Polarized distributions}
We evaluate Eq.~\eqref{eq:coordinateC32} for definite baryon
polarization using the canonical spin states. Along the $z$ axis,
these form the basis
\begin{align}
 \left\{
 |\sigma\rangle\;\middle|\;
 \sigma=\frac32,\frac12,-\frac12,-\frac32
 \right\}.
 \label{eq:LSpinStates32}
\end{align}
Along the $x$ axis, we denote the spin projection by $s_x$.
Following Ref.~\cite{Alexandrou:2009hs}, the two transverse
polarization states considered here are written in the $z$-quantized
basis of Eq.~\eqref{eq:LSpinStates32} as
\begin{subequations}
\label{eq:transverseSpinStates32}
\begin{align}
 |s_x=3/2\rangle
 &=\frac{1}{\sqrt8}\left(
 |3/2\rangle+\sqrt3|1/2\rangle
 +\sqrt3|-1/2\rangle+|-3/2\rangle
 \right),
 \\
 |s_x=1/2\rangle
 &=\frac{1}{\sqrt8}\left(
 \sqrt3|3/2\rangle+|1/2\rangle
 -|-1/2\rangle-\sqrt3|-3/2\rangle
 \right).
\end{align}
\end{subequations}
The longitudinally and transversely polarized EF distributions are
then defined by
\begin{subequations}
\label{eq:polarizedEFmatrixelements32}
\begin{align}
 T_{a,\mathrm{EF}}^{\mu\nu,L}
 (\bm x_\perp;P_z;\sigma)
 &\equiv
 T_{a,\mathrm{EF}}^{\mu\nu}
 (\bm x_\perp;P_z;\sigma,\sigma),
 \label{eq:longitudinalEFmatrixelement32}\\[1ex]
 T_{a,\mathrm{EF}}^{\mu\nu,T}
 (\bm x_\perp;P_z;s_x)
 &\equiv
 \sum_{\sigma',\sigma}
 \langle s_x|\sigma'\rangle
 T_{a,\mathrm{EF}}^{\mu\nu}
 (\bm x_\perp;P_z;\sigma',\sigma)
 \langle\sigma|s_x\rangle.
 \label{eq:transverseEFmatrixelement32}
\end{align}
\end{subequations}

The polarization dependence becomes explicit in the diagonal matrix
elements of the six spin structures in Eq.~\eqref{eq:coordinateC32}.
With the shorthand $X_n\equiv X_n(\theta_x)$, they read
\begin{widetext}
\begingroup
\renewcommand{\arraystretch}{2.5}
\setlength{\arraycolsep}{10pt}
\begin{equation}
\begin{array}{c|cccccc}
 \text{state}
 &\bm1
 &Q^{33}
 &\epsilon^{ij3}S^iX_1^j
 &\epsilon^{ij3}O^{i33}X_1^j
 &Q^{ij}X_2^{ij}
 &\epsilon^{3ik}O^{ij\ell}X_3^{kj\ell}
 \\
 & (0)
 & (0Q)
 & (1S)
 & (1O)
 & (2Q)
 & (3O)
 \\
\hline
 \sigma=\dfrac32
 &1&1&0&0&0&0\\
 \sigma=\dfrac12
 &1&-1&0&0&0&0\\
 s_x=\dfrac32
 &1&-\dfrac12
 &\dfrac32\sin\theta_x
 &-\dfrac{3}{20}\sin\theta_x
 &\dfrac34\cos 2\theta_x
 &\dfrac{3}{16}\sin 3\theta_x\\
 s_x=\dfrac12
 &1&\dfrac12
 &\dfrac12\sin\theta_x
 &\dfrac{9}{20}\sin\theta_x
 &-\dfrac34\cos 2\theta_x
 &-\dfrac{9}{16}\sin 3\theta_x
\end{array}
\label{eq:spinMultipoleExpectations32}
\end{equation}
\endgroup
\end{widetext}
The complete $4\times4$ matrices of these structures in the
$z$-quantized basis are collected in
Appendix~\ref{app:spinMultipoleMatrices32}.

Combining the entries in Eq.~\eqref{eq:spinMultipoleExpectations32}
with the radial functions in Eq.~\eqref{eq:coordinateC32} yields the
polarized distributions. The three independent EMT components
$\mu,\nu\in\{0,3\}$ take the form
\begin{subequations}
\label{eq:explicitPolarizedEFdistributions32}
\begin{align}
 T_{a,\mathrm{EF}}^{\mu\nu,L}
 (\bm x_\perp;P_z;3/2)
 ={}&
 \mathcal C_{0,a}^{\mu\nu}(x_\perp;P_z)
 +\mathcal C_{0Q,a}^{\mu\nu}(x_\perp;P_z),
 \label{eq:explicitPolarizedEFdistributions32a}\\
 T_{a,\mathrm{EF}}^{\mu\nu,L}
 (\bm x_\perp;P_z;1/2)
 ={}&
 \mathcal C_{0,a}^{\mu\nu}(x_\perp;P_z)
 -\mathcal C_{0Q,a}^{\mu\nu}(x_\perp;P_z),
 \label{eq:explicitPolarizedEFdistributions32b}\\
 T_{a,\mathrm{EF}}^{\mu\nu,T}
 (\bm x_\perp;P_z;3/2)
 ={}&
 \mathcal C_{0,a}^{\mu\nu}(x_\perp;P_z)
 -\frac12\mathcal C_{0Q,a}^{\mu\nu}(x_\perp;P_z)
 \nonumber\\
 &+\sin\theta_x\left[
 \frac32\mathcal C_{1S,a}^{\mu\nu}(x_\perp;P_z)
 -\frac{3}{20}\mathcal C_{1O,a}^{\mu\nu}(x_\perp;P_z)
 \right]
 \nonumber\\
 &+\frac34\cos 2\theta_x\,
 \mathcal C_{2Q,a}^{\mu\nu}(x_\perp;P_z)
 +\frac{3}{16}\sin 3\theta_x\,
 \mathcal C_{3O,a}^{\mu\nu}(x_\perp;P_z),
 \label{eq:explicitPolarizedEFdistributions32c}\\
 T_{a,\mathrm{EF}}^{\mu\nu,T}
 (\bm x_\perp;P_z;1/2)
 ={}&
 \mathcal C_{0,a}^{\mu\nu}(x_\perp;P_z)
 +\frac12\mathcal C_{0Q,a}^{\mu\nu}(x_\perp;P_z)
 \nonumber\\
 &+\sin\theta_x\left[
 \frac12\mathcal C_{1S,a}^{\mu\nu}(x_\perp;P_z)
 +\frac{9}{20}\mathcal C_{1O,a}^{\mu\nu}(x_\perp;P_z)
 \right]
 \nonumber\\
 &-\frac34\cos 2\theta_x\,
 \mathcal C_{2Q,a}^{\mu\nu}(x_\perp;P_z)
 -\frac{9}{16}\sin 3\theta_x\,
 \mathcal C_{3O,a}^{\mu\nu}(x_\perp;P_z).
 \label{eq:explicitPolarizedEFdistributions32d}
\end{align}
\end{subequations}
These relations hold separately for each quark and gluon contribution
$a$ and, after summing over $a$, for the total EMT. 

The longitudinal distributions in
Eqs.~\eqref{eq:explicitPolarizedEFdistributions32a} and
\eqref{eq:explicitPolarizedEFdistributions32b} contain only the
monopole terms $\mathcal C_{0,a}^{\mu\nu}$ and
$\mathcal C_{0Q,a}^{\mu\nu}$. The transverse distributions in
Eqs.~\eqref{eq:explicitPolarizedEFdistributions32c} and
\eqref{eq:explicitPolarizedEFdistributions32d} involve, in addition,
the dipole terms $\mathcal C_{1S,a}^{\mu\nu}$ and
$\mathcal C_{1O,a}^{\mu\nu}$ with angular dependence $\sin\theta_x$,
the quadrupole term $\mathcal C_{2Q,a}^{\mu\nu}$ with
$\cos 2\theta_x$, and the octupole term $\mathcal C_{3O,a}^{\mu\nu}$
with $\sin 3\theta_x$. The moment relations in
Eq.~\eqref{eq:EFmultipoleMoments32} characterize the overall
strengths of these patterns: the first two fix the $0$ and $0Q$
monopole strengths, the next two define the $1S$ and $1O$ dipole
moments quantifying the lateral asymmetry, and the final two the
quadrupole and octupole moments quantifying the rank-$2$ and
rank-$3$ transverse deformations.

We now identify the physical distributions associated with the three
EMT components. Denoting the polarization argument by $s_\alpha$
with $s_L=\sigma$, $s_T=s_x$, and $\alpha=L,T$, we define
\begin{subequations}
\label{eq:physicalEFdistributions32}
\begin{align}
 \rho_a^{\alpha}(\bm x_\perp;P_z;s_\alpha)
 &\equiv
 T_{a,\mathrm{EF}}^{00,\alpha}
 (\bm x_\perp;P_z;s_\alpha),
 \label{eq:energyEFdistribution32}\\
 \mathcal P_a^{z,\alpha}(\bm x_\perp;P_z;s_\alpha)
 &\equiv
 T_{a,\mathrm{EF}}^{03,\alpha}
 (\bm x_\perp;P_z;s_\alpha),
 \label{eq:momentumEFdistribution32}\\
 \mathcal I_a^{z,\alpha}(\bm x_\perp;P_z;s_\alpha)
 &\equiv
 T_{a,\mathrm{EF}}^{30,\alpha}
 (\bm x_\perp;P_z;s_\alpha),
 \label{eq:energyfluxEFdistribution32}\\
 \Pi_a^{zz,\alpha}(\bm x_\perp;P_z;s_\alpha)
 &\equiv
 T_{a,\mathrm{EF}}^{33,\alpha}
 (\bm x_\perp;P_z;s_\alpha),
 \label{eq:momentumfluxEFdistribution32}
\end{align}
\end{subequations}
where $\rho_a^\alpha$, $\mathcal P_a^{z,\alpha}$,
$\mathcal I_a^{z,\alpha}$, and $\Pi_a^{zz,\alpha}$ denote the
distributions of energy, longitudinal momentum, longitudinal energy
flux, and longitudinal momentum flux, respectively.
Since the Belinfante EMT is symmetric ($T_a^{03}=T_a^{30}$), the
longitudinal momentum density and longitudinal energy flux coincide: 
\begin{equation}
 \mathcal I_a^{z,\alpha}(\bm x_\perp;P_z;s_\alpha)
 =
 \mathcal P_a^{z,\alpha}(\bm x_\perp;P_z;s_\alpha),
 \qquad \alpha=L,T.
 \label{eq:momentumEnergyFluxEquality32}
\end{equation}
We therefore consider the three independent distributions
$\rho_a^{\alpha}$, $\mathcal P_a^{z,\alpha}$, and
$\Pi_a^{zz,\alpha}$. The same three EMT components describe the
polarized spin-$1/2$ distributions in Ref.~\cite{Won:2025dgc}. In the
present spin-$3/2$ case, the six multipole structures in
Eq.~\eqref{eq:coordinateC32} include quadrupole and octupole terms
that are absent for a spin-$1/2$ target.

The corresponding distributions for the total EMT are expressed as 
\begin{equation}
 \rho^{\alpha}=\sum_a\rho_a^{\alpha},
 \qquad
 \mathcal P^{z,\alpha}=\sum_a\mathcal P_a^{z,\alpha},
 \qquad
 \Pi^{zz,\alpha}=\sum_a\Pi_a^{zz,\alpha}.
 \label{eq:totalEFdistributions32}
\end{equation}
Integration over the transverse position projects onto the forward
matrix element at $\bm\Delta_\perp=0$.
Equations~\eqref{eq:EFmatrixelement32} and
\eqref{eq:totalconstraints32} then lead to
\begin{equation}
 \begin{aligned}
 \int d^2\bm x_\perp\,
 \rho^{\alpha}(\bm x_\perp;P_z;s_\alpha)
 &=m,
 \\
 \int d^2\bm x_\perp\,
 \mathcal P^{z,\alpha}(\bm x_\perp;P_z;s_\alpha)
 &=m\beta_P,
 \\
 \int d^2\bm x_\perp\,
 \Pi^{zz,\alpha}(\bm x_\perp;P_z;s_\alpha)
 &=m\beta_P^2,
 \qquad \alpha=L,T.
 \end{aligned}
 \label{eq:EFdistributionnormalizations32}
\end{equation}
Equation~\eqref{eq:EFdistributionnormalizations32} holds at any
$P_z$. The boost velocity $\beta_P$ vanishes in the transverse Breit
frame and approaches unity in the IMF, so the corresponding
normalizations are
\begin{align}
 \int d^2\bm x_\perp
 \left(
 \rho^{\alpha},
 \mathcal P^{z,\alpha},
 \Pi^{zz,\alpha}
 \right)
 \bigg|_{P_z=0}
 &=(m,0,0),
 \nonumber\\
 \lim_{P_z\to\infty}\int d^2\bm x_\perp
 \left(
 \rho^{\alpha},
 \mathcal P^{z,\alpha},
 \Pi^{zz,\alpha}
 \right)
 &=(m,m,m),
 \qquad \alpha=L,T.
 \label{eq:BFIMFdistributionnormalizations32}
\end{align}
Beyond these integrated normalizations,
Eq.~\eqref{eq:EFmultipoleFFIMFlimit32} shows that the $00$, $03$,
and $33$ multipole form factors have the same IMF limit. The
distributions of energy, longitudinal momentum, and longitudinal
momentum flux at fixed polarization thus become identical in the
IMF.

\section{LF multipole form factors and the IMF limit}
\label{sec:LFspin32}
In this section, we calculate the spin-$3/2$ EMT matrix element
directly on the light front using LF Rarita--Schwinger spinors and
determine its multipole form factors. We then verify that this LF
result reproduces the IMF limit of the boosted EF matrix element.

\subsection{Symmetric Drell-Yan kinematics}

We work in the front form of relativistic dynamics introduced by
Dirac~\cite{Dirac:1949cp} and use the LF convention
of Refs.~\cite{Kogut:1969xa,Soper:1971wn,Brodsky:1997de}:
\begin{equation}
 a^\pm=\frac{a^0\pm a^3}{\sqrt{2}},
 \qquad
 a\cdot b=a^+b^-+a^-b^+-\bm a_\perp\cdot\bm b_\perp,
 \label{eq:LFconvention32}
\end{equation}
and write $a^\mu=(a^+,a^-,\bm a_\perp)$.
In the symmetric Drell--Yan frame~\cite{Drell:1969km,West:1970av},
we choose $\bm P_\perp=0$ and $\Delta^+=0$. For $P^+>0$, the
on-shell conditions $p^2=p^{\prime 2}=m^2$ then give
\begin{equation}
 \Delta^-=0,
 \qquad
 t=\Delta^2=-\bm\Delta_\perp^2,
 \qquad
 P^-=\frac{m^2(1+\tau)}{2P^+}.
 \label{eq:LFDYonshell32}
\end{equation}
The average momentum and momentum transfer are therefore
\begin{equation}
 P^\mu
 =\left(P^+,\frac{m^2(1+\tau)}{2P^+},\bm 0_\perp\right),
 \qquad
 \Delta^\mu=(0,0,\bm\Delta_\perp).
 \label{eq:LFDYkinematics32}
\end{equation}
We use the LF state
normalization~\cite{Soper:1971wn,Brodsky:1997de}
\begin{align}
 {}_{\mathrm{LF}}\langle p',\lambda'|p,\lambda\rangle_{\mathrm{LF}}
 &=
 2p^+(2\pi)^3
 \delta(p^{\prime +}-p^+)
 \delta^{(2)}(\bm p'_\perp-\bm p_\perp)
 \delta_{\lambda'\lambda}.
 \label{eq:LFstateNormalization32}
\end{align}
Since $p^+=p^{\prime +}=P^+$, we define the normalized LF EMT matrix
element as
\begin{equation}
 \bigl\langle \hat T_a^{\mu\nu}\bigr\rangle_{\mathrm{LF}}
 (P^+,\bm\Delta_\perp;\lambda',\lambda)
 \equiv
 \frac{{}_{\mathrm{LF}}\langle p',\lambda'|
 \hat T_a^{\mu\nu}(0)|p,\lambda\rangle_{\mathrm{LF}}}{2P^+},
 \label{eq:LFmatrixelement32}
\end{equation}
where $\lambda$ and $\lambda'$ denote the LF helicities of the
incoming and outgoing baryons, respectively. The LF Dirac and
Rarita--Schwinger spinors used in the calculation are given in
Eqs.~\eqref{eq:LFDiracSpinors32}--\eqref{eq:LFRSSpinor32}. With the
LF boost and phase conventions of Appendix~\ref{app:canonicalRS32},
the $P_z\to\infty$ limit of the Wigner matrix
$D^{(3/2)}(p_{\rm B},\Lambda)$ in Eq.~\eqref{eq:Dspin32} yields the
spin-$3/2$ Melosh rotation, which relates the canonical spin basis
in the transverse Breit frame to the LF helicity
basis~\cite{Soper:1971wn,Melosh:1974cu,Ahluwalia:1993ff,
Keister:1991sb,Carlson:2003je}.

\subsection{LF multipole decomposition}
The symmetric EMT has three independent LF components with
$\mu,\nu\in\{+,-\}$, namely $T_a^{++}$, $T_a^{+-}=T_a^{-+}$, and
$T_a^{--}$. The matrix element
defined in Eq.~\eqref{eq:LFmatrixelement32} is decomposed in the
same six multipole structures used in the EF:
\begin{align}
\bigl\langle \hat T_a^{\mu\nu}\bigr\rangle_{\mathrm{LF}}
={}&m\Big[
C_{0,a,\mathrm{LF}}^{\mu \nu}\bm1
+C_{0Q,a,\mathrm{LF}}^{\mu \nu} Q^{33}
+2\sqrt\tau\,C_{1S,a,\mathrm{LF}}^{\mu \nu}
i\epsilon^{ij3}S^iX_1^j(\theta_\Delta)
\nonumber\\
&+2\sqrt\tau\,C_{1O,a,\mathrm{LF}}^{\mu \nu}
i\epsilon^{ij3}O^{i33}X_1^j(\theta_\Delta)
+4\tau C_{2Q,a,\mathrm{LF}}^{\mu \nu}
Q^{ij}X_2^{ij}(\theta_\Delta)
\nonumber\\
&+8\tau^{3/2}C_{3O,a,\mathrm{LF}}^{\mu \nu}
i\epsilon^{3ik}O^{ij\ell}X_3^{kj\ell}(\theta_\Delta)
\Big]_{\lambda'\lambda}.
\label{eq:LFmultipoledecomposition32}
\end{align}
The functions $C_{A,a,\mathrm{LF}}^{\mu\nu}(P^+,t)$ are the LF multipole
form factors.

We calculate the LF multipole form factors from the covariant EMT
matrix element in Eq.~\eqref{eq:covariantEMTspin32} and express them
in terms of the seven transverse Breit-frame functions defined in
Eqs.~\eqref{eq:E0spin32}--\eqref{eq:P2spin32}; this form allows a
direct comparison with the $P_z\to\infty$ limit of the EF matrix
elements. The six form factors for $T_a^{++}$ read
\begin{widetext}
\begin{subequations}
\label{eq:directLFformfactors32}
\begin{align}
 C_{0,a,\mathrm{LF}}^{++}(P^+,t)
={}&\frac{P^+}{m(1+\tau)^2}\Big[
 (1-\tau)(\mathcal E_{0,a}(t)+\mathcal P_{0,a}(t))
 +\tau\mathcal P_{0Q,a}(t)
 +2\tau(5-\tau)\mathcal J_{1,a}(t)
 \nonumber\\[-2pt]
 &\hspace{25mm}
 -8\tau^2\mathcal E_{2,a}(t)
 +6\tau^2\mathcal P_{2,a}(t)
 -\frac{96}{5}\tau^3\mathcal J_{3,a}(t)
 \Big],
 \label{eq:LFC0spin32}\\[2pt]
 C_{0Q,a,\mathrm{LF}}^{++}(P^+,t)
={}&\frac{P^+}{m(1+\tau)^2}\Big[
 \tau(\mathcal E_{0,a}(t)+\mathcal P_{0,a}(t))
 +\frac{2+\tau}{2}\mathcal P_{0Q,a}(t)
 +2\tau(\tau-2)\mathcal J_{1,a}(t)
 \nonumber\\[-2pt]
 &\hspace{25mm}
 +4\tau(1+2\tau)\mathcal E_{2,a}(t)
 -3\tau^2\mathcal P_{2,a}(t)
 +\frac{48}{5}\tau^2(1+2\tau)\mathcal J_{3,a}(t)
 \Big],
 \label{eq:LFC0Qspin32}\\[2pt]
 C_{1S,a,\mathrm{LF}}^{++}(P^+,t)
={}&\frac{P^+}{m(1+\tau)^2}\Big[
 \frac{\tau-5}{5}(\mathcal E_{0,a}(t)+\mathcal P_{0,a}(t))
 +\frac{2-\tau}{5}\mathcal P_{0Q,a}(t)
 +\frac{2}{5}(5-13\tau)\mathcal J_{1,a}(t)
 \nonumber\\[-2pt]
 &\hspace{25mm}
 +\frac{8}{5}\tau(\tau-2)\mathcal E_{2,a}(t)
 +\frac{6}{5}\tau(2-\tau)\mathcal P_{2,a}(t)
 -\frac{288}{25}\tau^2\mathcal J_{3,a}(t)
 \Big],
 \label{eq:LFC1Sspin32}\\[2pt]
 C_{1O,a,\mathrm{LF}}^{++}(P^+,t)
={}&\frac{P^+}{m(1+\tau)^2}\Big[
 -\frac{\tau}{2}(\mathcal E_{0,a}(t)+\mathcal P_{0,a}(t))
 -\frac{4+3\tau}{4}\mathcal P_{0Q,a}(t)
 +3\tau\mathcal J_{1,a}(t)
 \nonumber\\[-2pt]
 &\hspace{25mm}
 -2\tau(1+2\tau)\mathcal E_{2,a}(t)
 +\frac{\tau(\tau-2)}{2}\mathcal P_{2,a}(t)
 +\frac{8}{5}\tau(5+8\tau)\mathcal J_{3,a}(t)
 \Big],
 \label{eq:LFC1Ospin32}\\[2pt]
 C_{2Q,a,\mathrm{LF}}^{++}(P^+,t)
={}&\frac{P^+}{m(1+\tau)^2}\Big[
 \frac12(\mathcal E_{0,a}(t)+\mathcal P_{0,a}(t))
 -\frac14\mathcal P_{0Q,a}(t)
 +(\tau-2)\mathcal J_{1,a}(t)
 \nonumber\\[-2pt]
 &\hspace{25mm}
 -2\mathcal E_{2,a}(t)
 +\left(1-\frac{\tau}{2}\right)\mathcal P_{2,a}(t)
 -\frac{8}{5}\tau(7+4\tau)\mathcal J_{3,a}(t)
 \Big],
 \label{eq:LFC2Qspin32}\\[2pt]
 C_{3O,a,\mathrm{LF}}^{++}(P^+,t)
={}&\frac{P^+}{m(1+\tau)^2}\Big[
 -\frac16(\mathcal E_{0,a}(t)+\mathcal P_{0,a}(t))
 +\frac1{12}\mathcal P_{0Q,a}(t)
 +\mathcal J_{1,a}(t)
 \nonumber\\[-2pt]
 &\hspace{25mm}
 +\left(2+\frac{4\tau}{3}\right)\mathcal E_{2,a}(t)
 -\frac{2+\tau}{2}\mathcal P_{2,a}(t)
 -\frac{8}{5}(5+4\tau)\mathcal J_{3,a}(t)
 \Big].
 \label{eq:LFC3Ospin32}
\end{align}
\end{subequations}
\end{widetext}
The corresponding functions $C_{A,a,\mathrm{LF}}^{+-}(P^+,t)$ and
$C_{A,a,\mathrm{LF}}^{--}(P^+,t)$ are given in
Appendix~\ref{app:directLF32}.

\subsection{LF matrix elements and the IMF limit of EF matrix elements}
\label{subsec:EFLFmatching32}
The average momentum $P^\mu$ and momentum transfer $\Delta^\mu$ in
the EF take the symmetric Drell--Yan LF form for any $P_z$. This
kinematic relation alone does not imply equality of the EMT matrix
elements: the EF calculation uses canonical spin states, whereas the
LF one is based on LF helicity states. We first rewrite the EF
kinematics in LF coordinates and then compare the two sets of matrix
elements in the IMF. In this limit, the Wigner rotation becomes the
Melosh rotation between the two spin bases, and the leading term of
each EF matrix element reproduces its LF counterpart.

In LF coordinates, the average momentum of the EF kinematics in
Eq.~\eqref{eq:EFkinematics32} has the components
\begin{equation}
 P^+
 =\frac{P^0+P_z}{\sqrt2},
 \qquad
 P^-
 =\frac{P^0-P_z}{\sqrt2}
 =\frac{m^2(1+\tau)}{2P^+}.
 \label{eq:EFLFmomenta32}
\end{equation}
The second expression for $P^-$ follows from
$(P^0)^2-P_z^2=m^2(1+\tau)$. Since $\Delta^0=\Delta^3=0$, the
momentum transfer also satisfies $\Delta^+=\Delta^-=0$. The EF
momenta therefore reproduce the symmetric Drell--Yan kinematics in
Eq.~\eqref{eq:LFDYkinematics32} at any $P_z$; only the comparison of
the matrix elements in the two spin bases requires the IMF limit.

We compare the EF and LF matrix elements themselves, without the
normalization factors introduced in Eqs.~\eqref{eq:EFmatrixelement32}
and \eqref{eq:LFmatrixelement32}. The EF components with
$\mu,\nu\in\{+,-\}$ are related to the temporal and longitudinal
components $T_a^{00}$, $T_a^{03}$, and $T_a^{33}$ by
\begin{subequations}
\label{eq:EFlightfrontmatrixelements32}
\begin{align}
 \langle p'|\hat T_a^{++}(0)|p\rangle
 &=
 \frac12\langle p'|
 \bigl(\hat T_a^{00}(0)+2\hat T_a^{03}(0)+\hat T_a^{33}(0)\bigr)
 |p\rangle,
 \label{eq:EFplusplusmatrixelement32}\\
 \langle p'|\hat T_a^{+-}(0)|p\rangle
 &=
 \frac12\langle p'|
 \bigl(\hat T_a^{00}(0)-\hat T_a^{33}(0)\bigr)
 |p\rangle,
 \label{eq:EFplusminusmatrixelement32}\\
 \langle p'|\hat T_a^{--}(0)|p\rangle
 &=
 \frac12\langle p'|
 \bigl(\hat T_a^{00}(0)-2\hat T_a^{03}(0)+\hat T_a^{33}(0)\bigr)
 |p\rangle,
 \label{eq:EFminusminusmatrixelement32}
\end{align}
\end{subequations}
where the canonical spin labels $\sigma'$ and $\sigma$ are omitted.
Although the momenta agree at finite $P_z$, the matrix elements in
Eq.~\eqref{eq:EFlightfrontmatrixelements32} are written in the
canonical spin basis and cannot yet be identified with their LF
helicity counterparts. In the IMF, the Wigner angle approaches the
limit in Eq.~\eqref{eq:WignerIMF32}, and the rotation reduces to the
Melosh rotation between the two bases. Relating the spin indices
through this rotation, we obtain
\begin{equation}
 \left.
 \langle p',\sigma'|\hat T_a^{\mu\nu}(0)|p,\sigma\rangle
 \right|_{\mathrm{leading\ IMF}}
 =
 {}_{\mathrm{LF}}\langle p',\lambda'|
 \hat T_a^{\mu\nu}(0)|p,\lambda\rangle_{\mathrm{LF}}
 \,,
 \qquad
\mu,\nu\in\{+,-\}.
 \label{eq:IMFLFmatrixrelation32}
\end{equation}
The three EMT components scale with different powers of $P_z$. The
subscript ``leading IMF'' denotes, for each component, the leading
nonvanishing term in its large-$P_z$ expansion. With $P^+$ related
to $P_z$ through Eq.~\eqref{eq:EFLFmomenta32},
Eq.~\eqref{eq:IMFLFmatrixrelation32} equates this leading EF term
with the LF matrix element calculated directly on the light front.
The corresponding LF multipole form factors for $T_a^{+-}$ and
$T_a^{--}$ are listed in Appendix~\ref{app:directLF32}. 

\section{\texorpdfstring{Numerical results for the $\Delta$-baryon EMT
under longitudinal boosts}{Numerical results for the Delta-baryon EMT
under longitudinal boosts}}
\label{sec:results32}
We now turn to a spin-$3/2$ target within the EF formalism. Our aim
is to visualize its transverse EMT distributions and examine how the
individual multipoles change under longitudinal boosts. As a concrete
example, we take the $\Delta$ baryon, with its EMT form factors from
the Skyrme model of Ref.~\cite{Kim:2020lrs}. A multipole
parametrization of their $t$ dependence serves as input for the six
EF multipole form factors as functions of $P_z$ and $t$. Their
two-dimensional Fourier transforms yield the distributions of
energy, longitudinal momentum, and longitudinal momentum flux for
longitudinally and transversely polarized $\Delta$ states.
\subsection{\texorpdfstring{Form factors from the Skyrme model and their multipole
parametrization}{Form factors from the Skyrme model and their
multipole parametrization}}
The numerical calculation uses the total $\Delta$-baryon EMT form
factors obtained in the Skyrme model of Ref.~\cite{Kim:2020lrs}.
Since only the total EMT is considered, the label $a$ is omitted. We
parametrize the spacelike $t$ dependence as
\begin{equation}
 F_{i,j}(t)
 =\frac{F_{i,j}(0)}{\left(1-t/M_{i,j}^2\right)^p}
 =\frac{F_{i,j}(0)}{\left(1+Q^2/M_{i,j}^2\right)^p},
 \qquad Q^2=-t,\qquad p=6.
\label{eq:poleParametrization32}
\end{equation}
The power $p=6$ ensures numerically stable transverse Fourier
transforms. The normalizations $F_{i,j}(0)$ are those of
Ref.~\cite{Kim:2020lrs}, while the pole masses $M_{i,j}$ are
determined by fitting the model results over
$0\leq Q^2\leq1\;\mathrm{GeV}^2$. The resulting parameters are
listed in Table~\ref{tab:poleInput32}. Throughout, we set
$m=1.232\;\mathrm{GeV}$ in $\tau$ and the boost factors.
\begin{table}[ht]
  \caption{Parameters of the multipole parametrization in
Eq.~\eqref{eq:poleParametrization32} with $p=6$. The normalizations
$F_{i,j}(0)$ are taken from Ref.~\cite{Kim:2020lrs}, and the pole
masses $M_{i,j}$ are fitted over $0\leq Q^2\leq1\;\mathrm{GeV}^2$.}
\label{tab:poleInput32}
\begin{ruledtabular}
\begin{tabular}{cccc}
Form factor & $F_{i,j}(0)$ & $p$ & $M_{i,j}\;[\mathrm{GeV}]$\\
\hline
$F_{1,0}$ & $ 1.00$ & $6$ & $2.05$\\
$F_{1,1}$ & $-3.60$ & $6$ & $1.53$\\
$F_{2,0}$ & $-4.30$ & $6$ & $1.26$\\
$F_{2,1}$ & $ 2.40$ & $6$ & $0.78$\\
$F_{4,0}$ & $ 1.50$ & $6$ & $1.88$\\
$F_{4,1}$ & $-1.50$ & $6$ & $1.88$\\
$F_{5,0}$ & $-0.15$ & $6$ & $1.88$\\
\end{tabular}
\end{ruledtabular}
\end{table}
The conserved total EMT satisfies
$F_{3,0}(t)=F_{3,1}(t)=F_{6,0}(t)=0$, as required by
Eq.~\eqref{eq:totalconstraints32}. Table~\ref{tab:BFmultipoleInput32}
lists the forward values of the seven transverse Breit-frame
multipole form factors, obtained from
Eqs.~\eqref{eq:E0spin32}--\eqref{eq:P2spin32} with the parameters in
Table~\ref{tab:poleInput32}.
\begin{table}[ht]
  \caption{Forward values of the seven transverse Breit-frame
multipole form factors evaluated with the parameters in
Table~\ref{tab:poleInput32}. All values are dimensionless.}
\label{tab:BFmultipoleInput32}
\begin{ruledtabular}
\begin{tabular}{cc}
Multipole form factor & Value at $t=0$\\
\hline
$\mathcal E_0$     & $ 1.00$\\
$\mathcal E_2$     & $-0.17$\\
$\mathcal J_1$     & $ 0.50$\\
$\mathcal J_3$     & $ 0.00$\\
$\mathcal P_0$     & $ 0.00$\\
$\mathcal P_{0Q}$  & $ 0.00$\\
$\mathcal P_2$     & $ 0.10$\\
\end{tabular}
\end{ruledtabular}
\end{table}

\begin{figure}[htp]
\centering
\includegraphics[width=\textwidth]{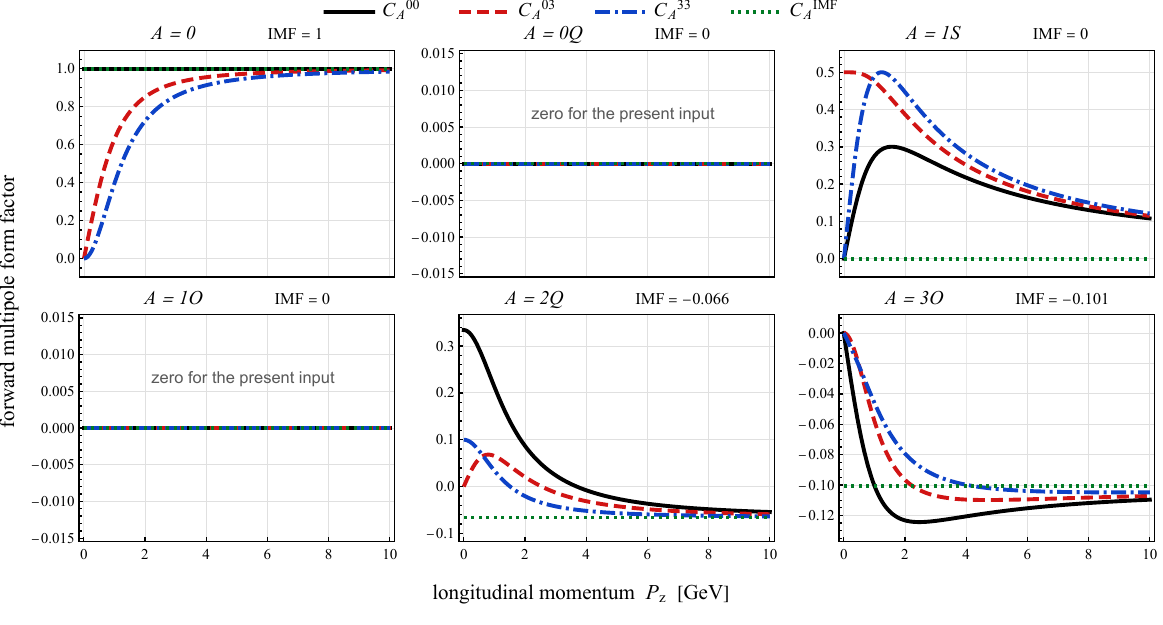}
\caption{Forward limits of the six EF multipole form factors as
functions of $P_z$, evaluated with the input in
Table~\ref{tab:BFmultipoleInput32}. The solid (black), dashed (red),
and dot-dashed (blue) curves show $C_A^{00}$, $C_A^{03}$, and
$C_A^{33}$, respectively. The horizontal dotted (green) lines and
the values quoted above the panels indicate $C_A^{\mathrm{IMF}}$.
For the present input, the forward limits of the $0Q$ and $1O$
multipoles vanish for all $P_z$.}
\label{fig:2}
\end{figure}

The same input also allows us to display the frame dependence before
the Fourier transform. Figure~\ref{fig:2} shows the forward limits of
$C_A^{00}$, $C_A^{03}$, and $C_A^{33}$ for the six multipoles,
obtained from Eqs.~\eqref{eq:appC0spin32}--\eqref{eq:appC3Ospin32}.
At $P_z=0$, they coincide with the multipole form factors in the
transverse Breit frame. As $P_z$ increases, the three components
approach the common IMF values in
Eq.~\eqref{eq:EFmultipoleFFIMFlimit32}.
The $A=0$ components approach the mass-normalized value
$C_0^{\mathrm{IMF}}(0)=1$. At $t=0$, conservation of the total EMT
requires $\mathcal P_0(0)=\mathcal P_{0Q}(0)=0$. In the $0Q$ and
$1O$ sectors, the terms surviving at $t=0$ are proportional to
$\mathcal P_{0Q}(0)$, so
$C_{0Q}^{\mu\nu}(P_z,0)=C_{1O}^{\mu\nu}(P_z,0)=0$ for all $P_z$.
The $1S$ components show different $P_z$ dependence, but their
common IMF limit vanishes through the mass and spin sum rules.
Setting $t=0$ in Eq.~\eqref{eq:IMFC1Sspin32} yields
\begin{align}
 C_{1S}^{\mathrm{IMF}}(0)
 &=-\mathcal E_0(0)-\mathcal P_0(0)
 +\frac{2}{5}\mathcal P_{0Q}(0)+2\mathcal J_1(0)
 \nonumber\\
 &=-\mathcal E_0(0)+2\mathcal J_1(0)=0,
 \label{eq:C1SforwardCancellation32}
\end{align}
where the second line uses EMT conservation together with
$\mathcal E_0(0)=1$ and $\mathcal J_1(0)=1/2$ from the mass and spin
sum rules. In contrast, the $2Q$ and $3O$ components approach the nonzero
values $C_{2Q}^{\mathrm{IMF}}(0)=-0.066$ and
$C_{3O}^{\mathrm{IMF}}(0)=-0.101$, respectively, obtained from
Eq.~\eqref{eq:IMFremainingforward32} with the input in
Table~\ref{tab:BFmultipoleInput32}. The zeros at $t=0$ constrain
only the corresponding multipole moments; the $0Q$, $1S$, and $1O$
distributions themselves need not vanish, since their Fourier
transforms involve the form factors over the full spacelike $t$
range.

\subsection{Longitudinally polarized distributions}
When the target is polarized along the $z$ axis, the distributions
in Eqs.~\eqref{eq:explicitPolarizedEFdistributions32a} and
\eqref{eq:explicitPolarizedEFdistributions32b} contain only the
azimuthally symmetric $0$ and $0Q$ multipoles.
Figures~\ref{fig:3}--\ref{fig:5} show one-dimensional profiles of
the corresponding energy, longitudinal momentum, and longitudinal
momentum flux distributions along a transverse axis through the
baryon center.
\subsubsection{Energy distribution}
\begin{figure}[htb!]
\begin{subfigure}{.4\textwidth}
    \centering
    \includegraphics[width=1.\linewidth]{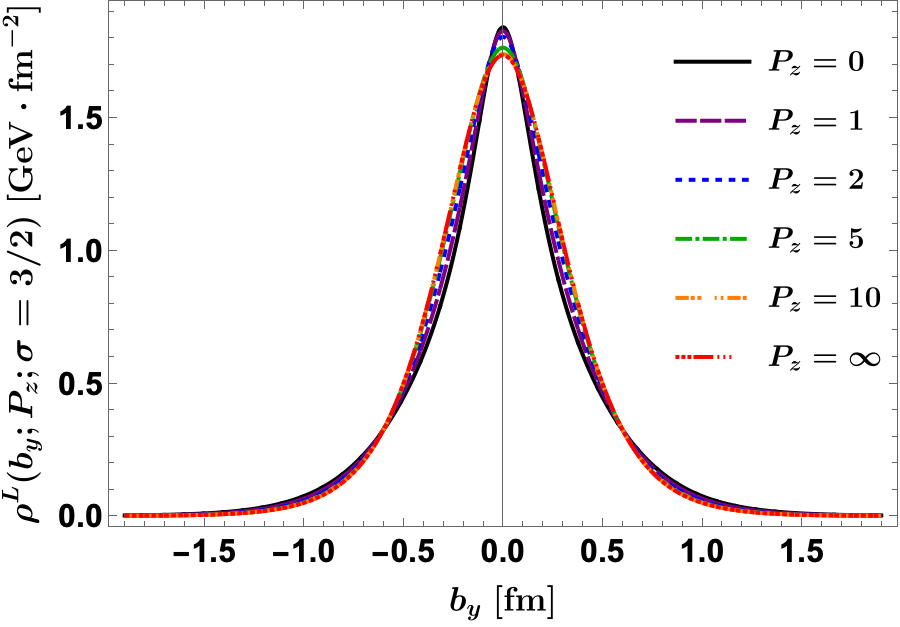}
    \caption{$s_z=3/2$}
    \label{fig:3a}
\end{subfigure}
\begin{subfigure}{.4\textwidth}
    \centering
    \includegraphics[width=1.\linewidth]{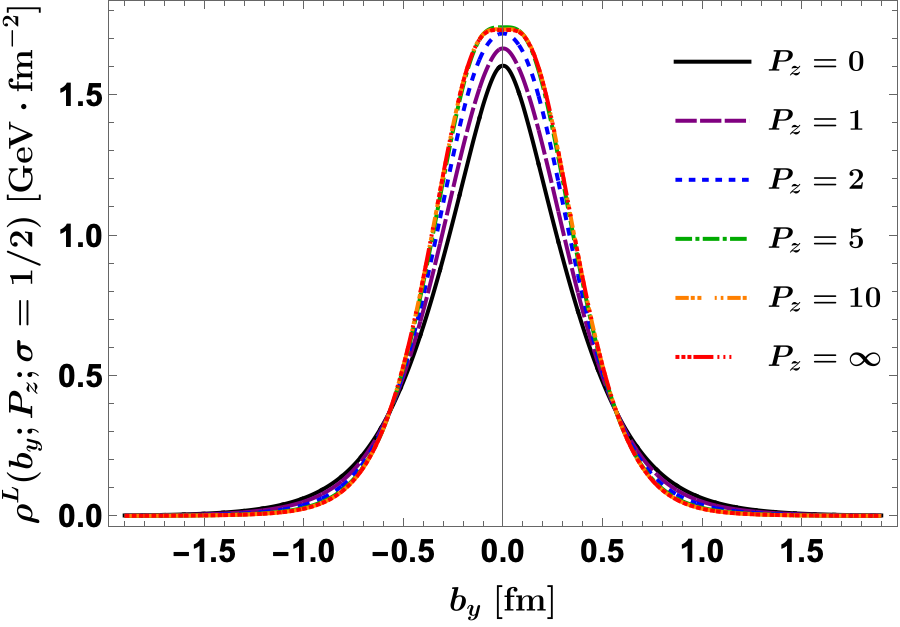}
    \caption{$s_z=1/2$}
    \label{fig:3b}
  \end{subfigure}
  \caption{Energy distribution $\rho^L(x_\perp;P_z;s_z)$ along a
transverse axis for (a) $s_z=3/2$ and (b) $s_z=1/2$. The curves
correspond to $P_z=0,1,2,5,10\;\mathrm{GeV}$ and to the IMF limit
$P_z\to\infty$.}
\label{fig:3}
\end{figure}

The left and right panels of Fig.~\ref{fig:3} show the energy
distributions for $s_z=3/2$ and $s_z=1/2$, respectively. Their
dependence on $P_z$ is modest: as $P_z$ increases, the central value
at $x_\perp=0$ decreases slightly for $s_z=3/2$ but increases for
$s_z=1/2$, and the curves for $P_z\gtrsim5$ GeV are nearly
indistinguishable from the IMF limit. In the transverse Breit frame,
Eq.~\eqref{eq:BFenergy32} contains the $0$ monopole $\mathcal E_0$
and the $0Q$ monopole $4\tau\mathcal E_2$. The values
$\mathcal E_0(0)=1$ and $\mathcal E_2(0)=-0.167$ in
Table~\ref{tab:BFmultipoleInput32}, together with the explicit
factor of $\tau$, suppress the $0Q$ term at low $|t|$, so the $0$
monopole dominates both profiles. The $0Q$ term enters with opposite
signs because the diagonal matrix element of $Q^{33}$ is $+1$ for
$s_z=3/2$ and $-1$ for $s_z=1/2$; see
Eq.~\eqref{eq:spinMonopoleMatrices32}. This sign difference explains
the opposite trends of the two central values. 
Equation~\eqref{eq:EFdistributionnormalizations32} fixes the
transverse integral to the baryon mass at every $P_z$ for both spin
projections. The changes near the center are therefore compensated
in the outer region: the $s_z=3/2$ distribution broadens slightly
with $P_z$, whereas the $s_z=1/2$ distribution becomes more compact.
\subsubsection{\texorpdfstring{Distribution of longitudinal momentum}{Distribution
of longitudinal momentum}}
\begin{figure}[htb!]
\begin{subfigure}{.4\textwidth}
    \centering
    \includegraphics[width=1.\linewidth]{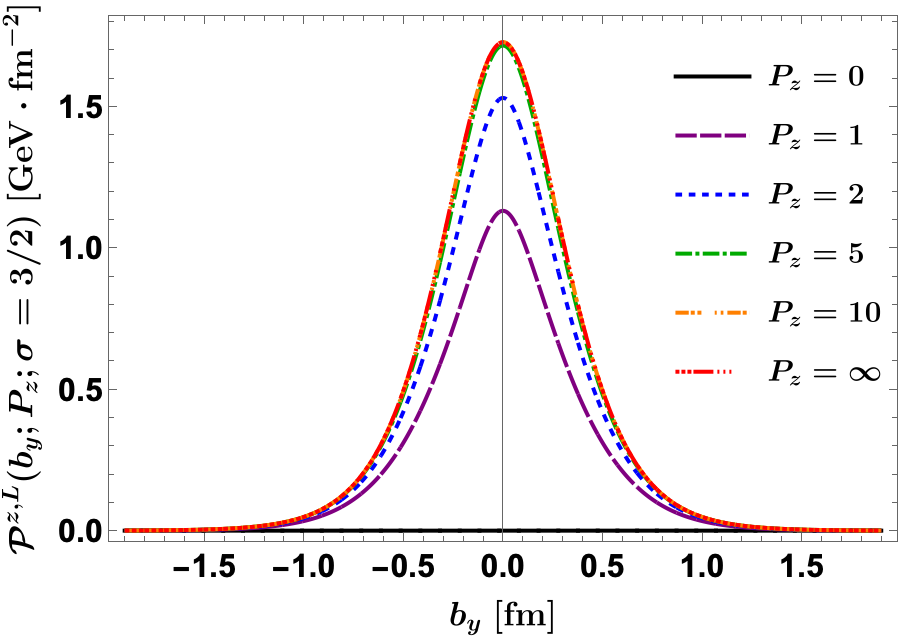}
    \caption{$s_z=3/2$}
    \label{fig:4a}
\end{subfigure}
\begin{subfigure}{.4\textwidth}
    \centering
    \includegraphics[width=1.\linewidth]{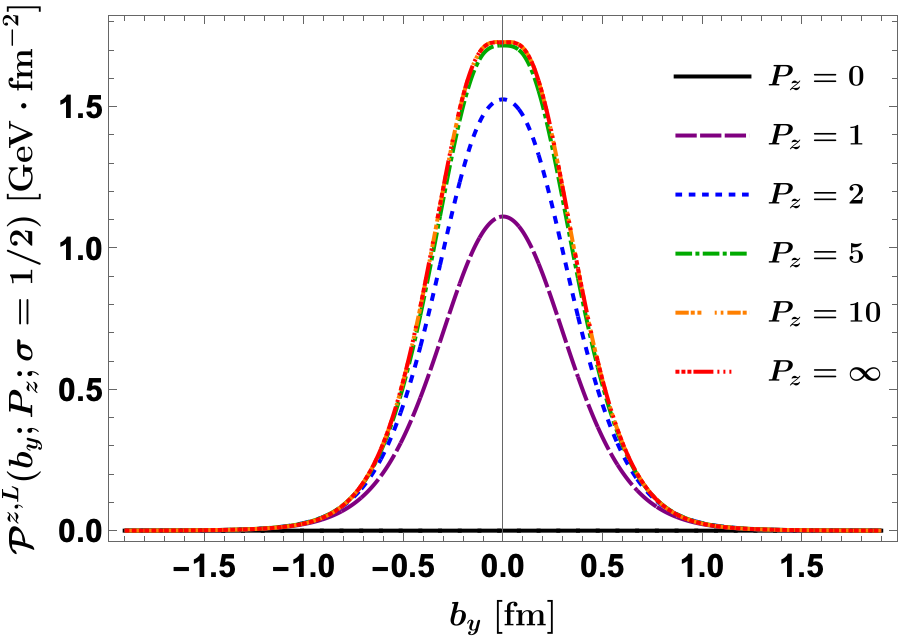}
    \caption{$s_z=1/2$}
    \label{fig:4b}
  \end{subfigure}
  \caption{Longitudinal momentum distribution
$\mathcal P^{z,L}(x_\perp;P_z;s_z)$. The notation is the same as in
Fig.~\ref{fig:3}.}
\label{fig:4}
\end{figure}

The left and right panels of Fig.~\ref{fig:4} show the longitudinal
momentum distributions for $s_z=3/2$ and $s_z=1/2$, respectively. At
$P_z=0$, both profiles vanish for all $x_\perp$. In
Eq.~\eqref{eq:BFangular32}, $\mathcal J_1$ multiplies
$i\epsilon^{ij3}S^iX_1^j$, while $\mathcal J_3$ multiplies the
octupole operators, and the expectation values of both vanish in
longitudinal spin states. The transverse Breit-frame matrix element
of $T^{03}$ therefore carries no longitudinal momentum density. At
finite $P_z$, the middle row of
Eq.~\eqref{eq:EFamplitudes32factorized} combines the transverse
Breit-frame matrix elements with weights $\beta$, $1+\beta^2$, and
$\beta$ for $T^{00}$, $T^{03}$, and $T^{33}$, respectively, each
accompanied by the Wigner rotation. Consequently, all three
Breit-frame components contribute to the $0$ and $0Q$ multipoles of
the longitudinally polarized EF matrix element of $T^{03}$, in
agreement with Eqs.~\eqref{eq:explicitPolarizedEFdistributions32a}
and \eqref{eq:explicitPolarizedEFdistributions32b}.

At $t=0$, the Wigner angle vanishes, the Breit-frame matrix element
of $T^{03}$ is zero for longitudinal polarization, and the stress
monopoles obey $\mathcal P_0(0)=\mathcal P_{0Q}(0)=0$, whereas the
energy monopole retains $\mathcal E_0(0)=1$; see
Table~\ref{tab:BFmultipoleInput32}. Hence only the $\beta T^{00}$
term survives in the forward limit and fixes the transverse integral
of either profile to $m\beta_P$, as required by
Eq.~\eqref{eq:EFdistributionnormalizations32}.
Since $\beta_P=P_z/E_P$ increases from $0$ to $1$, the monopole
normalization grows from zero at $P_z=0$ to $m$ in the IMF, which
explains the overall rise of the longitudinal momentum profiles in
Fig.~\ref{fig:4} and their saturation for
$P_z\gtrsim5\;\mathrm{GeV}$. In the IMF,
Eq.~\eqref{eq:EFmultipoleFFIMFlimit32} makes them identical to the
energy distributions in Fig.~\ref{fig:3} for the same spin
projection. The $0Q$ contribution changes sign between the two spin
projections, but its effect is barely visible in Fig.~\ref{fig:4}.
\subsubsection{\texorpdfstring{Distribution of longitudinal momentum flux}{Distribution of longitudinal momentum flux}}
\begin{figure}[htb!]
\begin{subfigure}{.4\textwidth}
    \centering
    \includegraphics[width=1.\linewidth]{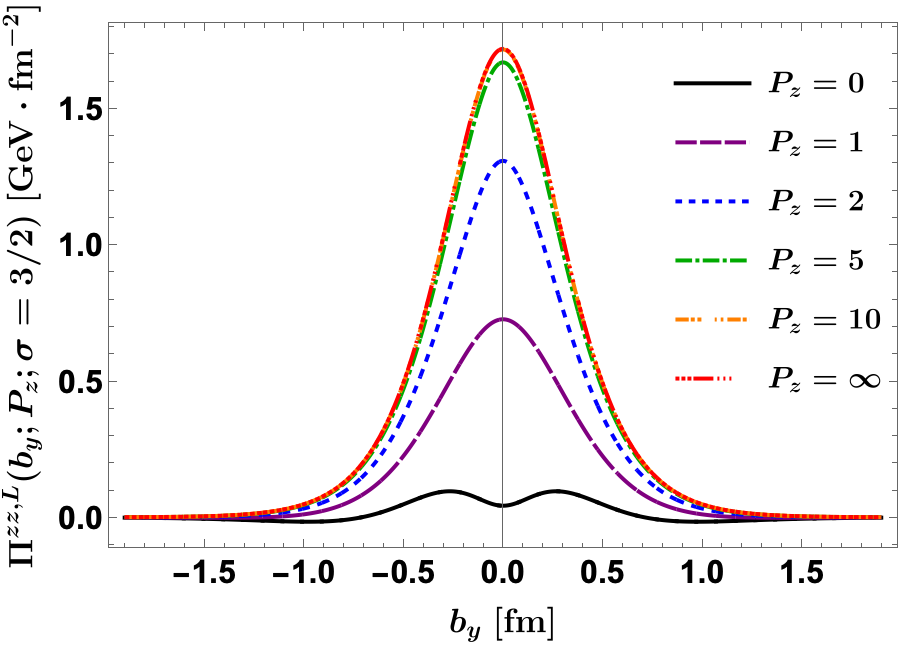}
    \caption{$s_z=3/2$}
    \label{fig:5a}
\end{subfigure}
\begin{subfigure}{.4\textwidth}
    \centering
    \includegraphics[width=1.\linewidth]{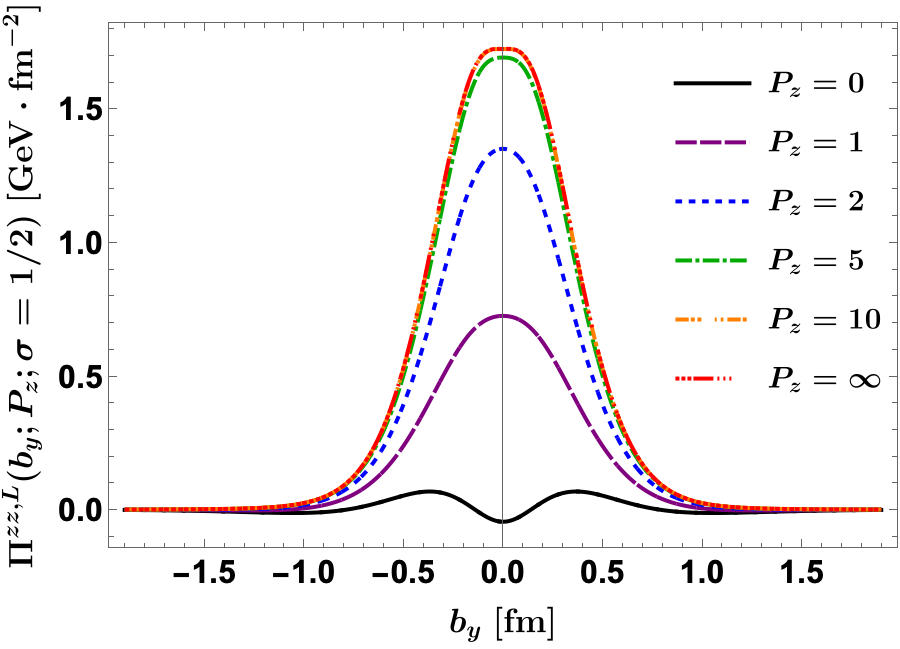}
    \caption{$s_z=1/2$}
    \label{fig:5b}
  \end{subfigure}
  \caption{Longitudinal momentum flux distribution
$\Pi^{zz,L}(x_\perp;P_z;s_z)$. The notation is the same as in
Fig.~\ref{fig:3}.}
\label{fig:5}
\end{figure}

The left and right panels of Fig.~\ref{fig:5} show the longitudinal
momentum flux distributions for $s_z=3/2$ and $s_z=1/2$,
respectively. Unlike the longitudinal momentum distributions in
Fig.~\ref{fig:4}, the momentum flux distributions do not vanish
identically at $P_z=0$, owing to the nonvanishing expectation values of the $\bm{1}$
and $Q^{33}$ terms in Eq.~\eqref{eq:BFstress32}. Their transverse
integrals, however, vanish because
$\mathcal P_0(0)=\mathcal P_{0Q}(0)=0$; see
Table~\ref{tab:BFmultipoleInput32}. The resulting cancellation between
positive and negative regions is the two-dimensional von Laue
condition for mechanical stability~\cite{vonLaue:1911,Polyakov:2002yz,Polyakov:2018zvc,
Kim:2021jjf,Lorce:2022cle}. The node structure differs between the
two spin projections: the profile is negative at the center for
$s_z=1/2$ but remains positive there for $s_z=3/2$, reflecting the
opposite signs of the $Q^{33}$ expectation values.

At finite $P_z$, the third row of
Eq.~\eqref{eq:EFamplitudes32factorized} combines the Breit-frame
components with weights $\beta^2$, $2\beta$, and $1$ for $T^{00}$,
$T^{03}$, and $T^{33}$, respectively. After the Wigner rotation, all
three contribute to the $0$ and $0Q$ multipoles, and the profiles
turn into single positive peaks. At $t=0$, only the $\beta^2T^{00}$
term survives, so the transverse integral is $m\beta_P^2$; see
Eq.~\eqref{eq:EFdistributionnormalizations32}. Since
$\beta_P^2<\beta_P$ for $0<\beta_P<1$, the momentum flux
normalization rises more slowly than that of the longitudinal
momentum, but both approach $m$ in the IMF. The factor $m\beta_P^2$
explains the overall rise of the profiles in Fig.~\ref{fig:5} and
their saturation for $P_z\gtrsim5\;\mathrm{GeV}$. In the IMF,
Eq.~\eqref{eq:EFmultipoleFFIMFlimit32} makes the momentum flux
identical to the energy and longitudinal momentum densities in
Figs.~\ref{fig:3} and \ref{fig:4} for the same spin projection.
\subsection{Transversely polarized multipole distributions}
The states with transverse spin in
Eq.~\eqref{eq:transverseSpinStates32} receive contributions from all
six multipole structures. In Figs.~\ref{fig:6}--\ref{fig:11}, the
$0$ and $0Q$ terms are combined into the monopole, the $1S$ and
$1O$ terms into the dipole, and the $2Q$ and $3O$ terms constitute
the quadrupole and octupole, respectively. Each figure displays
these four multipole contributions and their sum. The baryon is
polarized along $+\hat x$, and we show one-dimensional cuts along
the transverse $y$ axis, perpendicular to the spin, with the signed
coordinate $b_y$. The monopole and quadrupole are even under
$b_y\to-b_y$, whereas the dipole and octupole are odd.
\subsubsection{Energy distribution}
\begin{figure}[htb!]
\begin{subfigure}{.4\textwidth}
    \centering
    \signedcutgraphic[width=1.\linewidth]{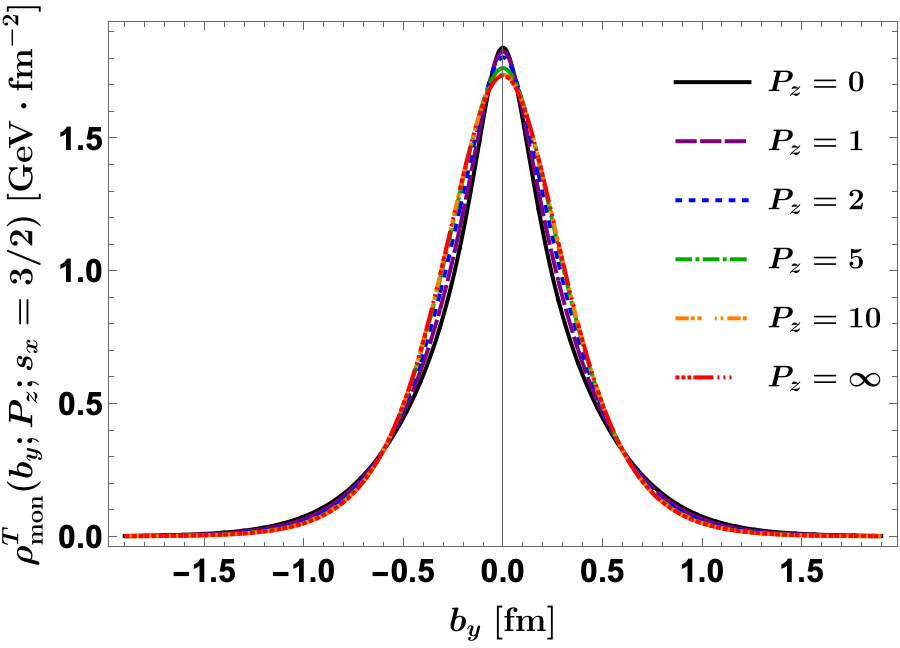}
    \caption{Monopole}
    \label{fig:6a}
\end{subfigure}
\begin{subfigure}{.4\textwidth}
    \centering
    \signedcutgraphic[width=1.\linewidth]{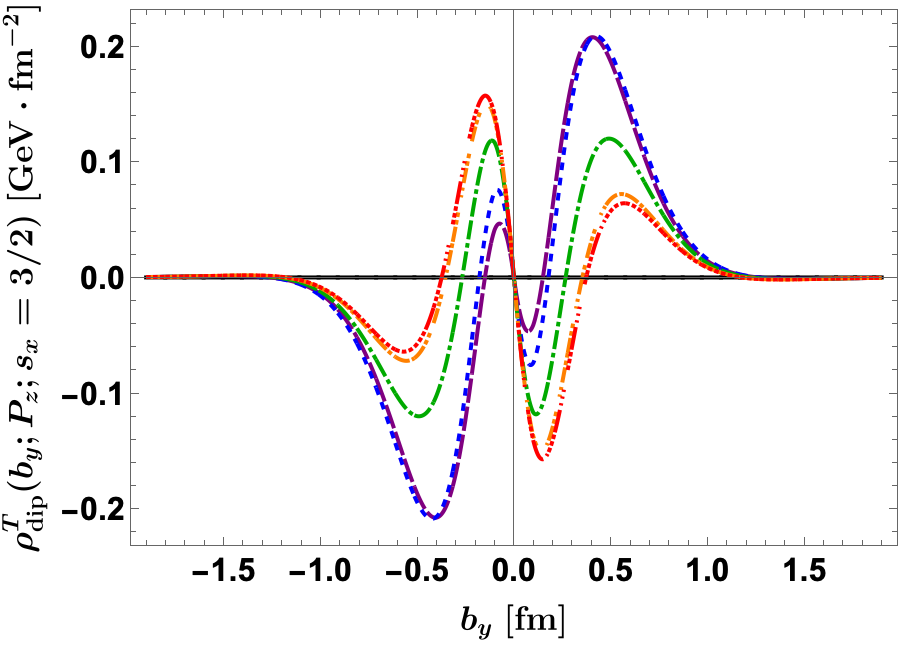}
    \caption{Dipole}
    \label{fig:6b}
\end{subfigure}
\begin{subfigure}{.4\textwidth}
    \centering
    \signedcutgraphic[width=1.\linewidth]{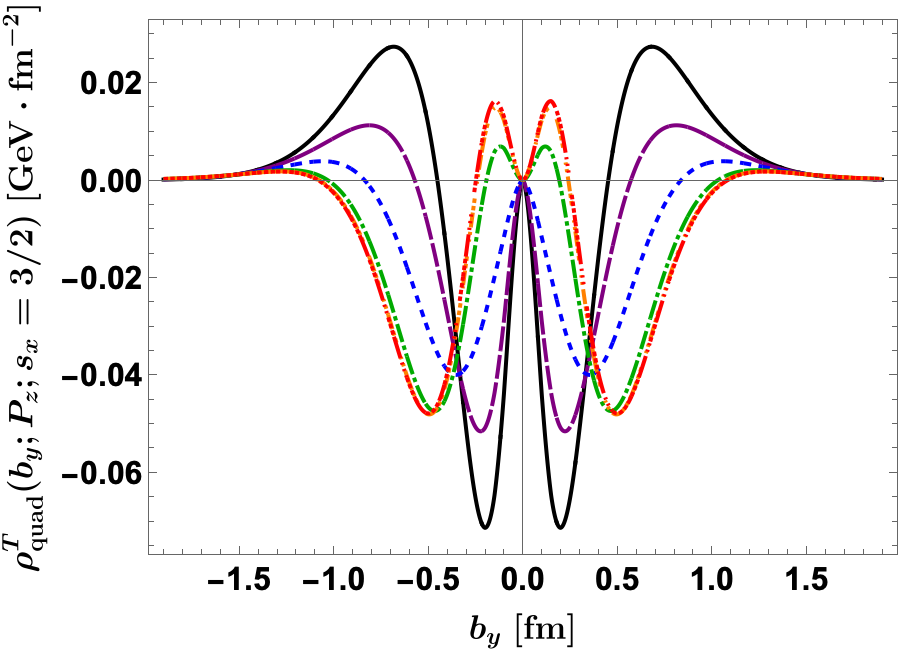}
    \caption{Quadrupole}
    \label{fig:6c}
\end{subfigure}
\begin{subfigure}{.4\textwidth}
    \centering
    \signedcutgraphic[width=1.\linewidth]{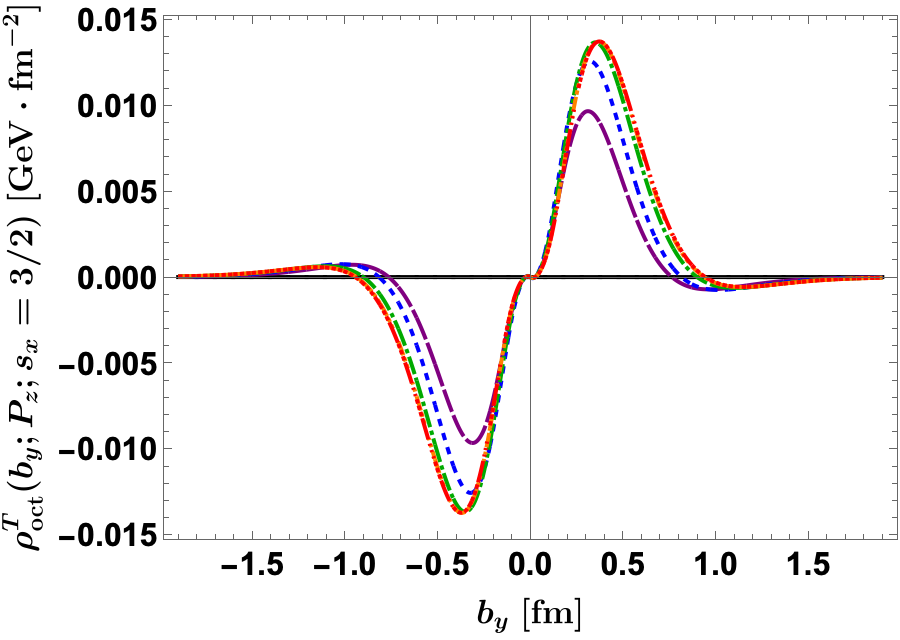}
    \caption{Octupole}
    \label{fig:6d}
\end{subfigure}
\begin{subfigure}{.4\textwidth}
    \centering
    \signedcutgraphic[width=1.\linewidth]{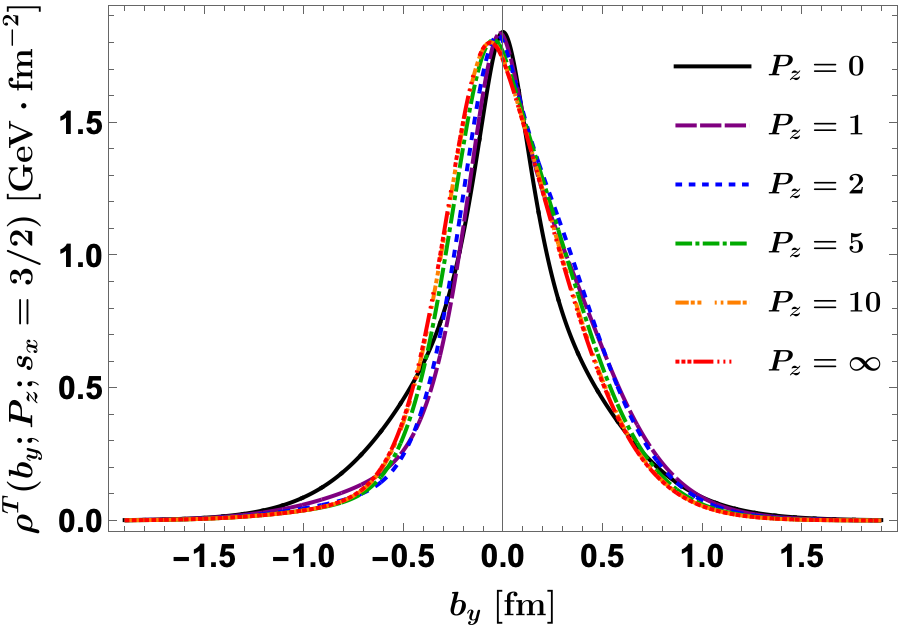}
    \caption{Total}
    \label{fig:6e}
  \end{subfigure}
\caption{Monopole (a), dipole (b), quadrupole (c), octupole (d), and
total (e) contributions to the energy distribution
$\rho^T(b_y;P_z;s_x)$ for $s_x=3/2$, shown along the $b_y$ axis
perpendicular to the spin. The monopole combines the $0$ and $0Q$
structures, the dipole the $1S$ and $1O$ structures, and the
quadrupole and octupole correspond to $2Q$ and $3O$, respectively.
The line styles are the same as in Fig.~\ref{fig:3}.}
\label{fig:6}
\end{figure}

\begin{figure}[htb!]
\begin{subfigure}{.4\textwidth}
    \centering
    \signedcutgraphic[width=1.\linewidth]{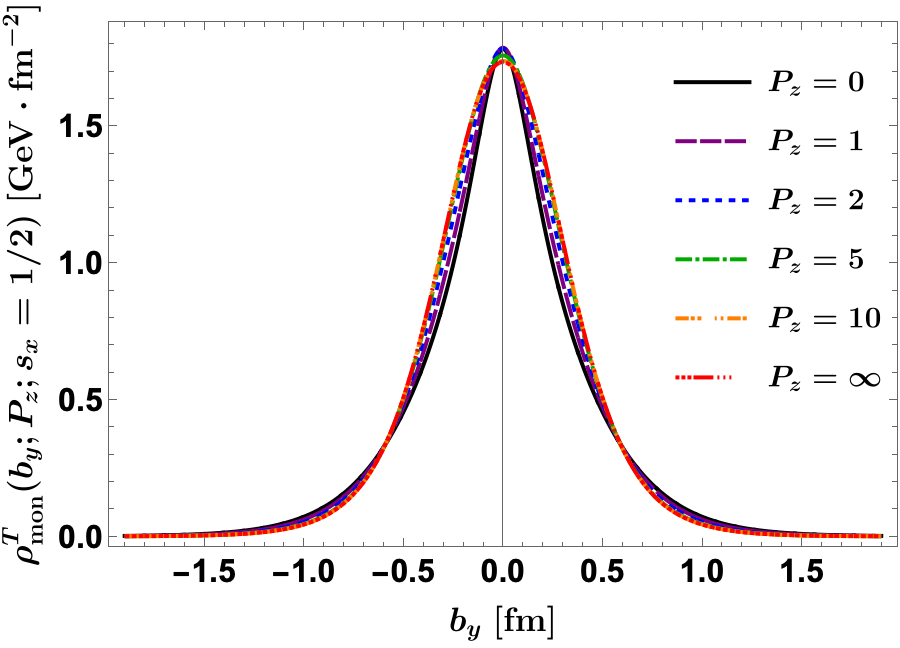}
    \caption{Monopole}
    \label{fig:7a}
\end{subfigure}
\begin{subfigure}{.4\textwidth}
    \centering
    \signedcutgraphic[width=1.\linewidth]{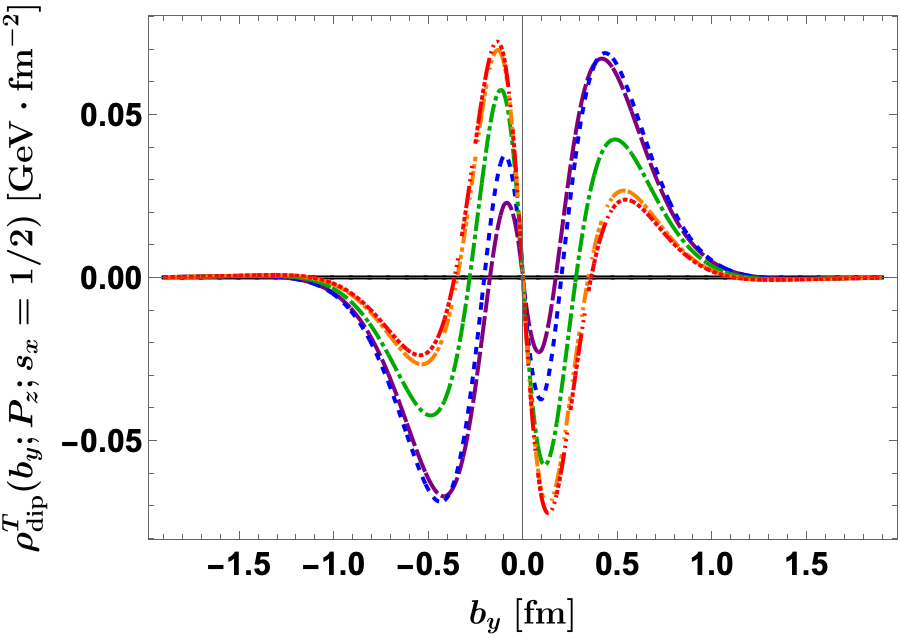}
    \caption{Dipole}
    \label{fig:7b}
\end{subfigure}
\begin{subfigure}{.4\textwidth}
    \centering
    \signedcutgraphic[width=1.\linewidth]{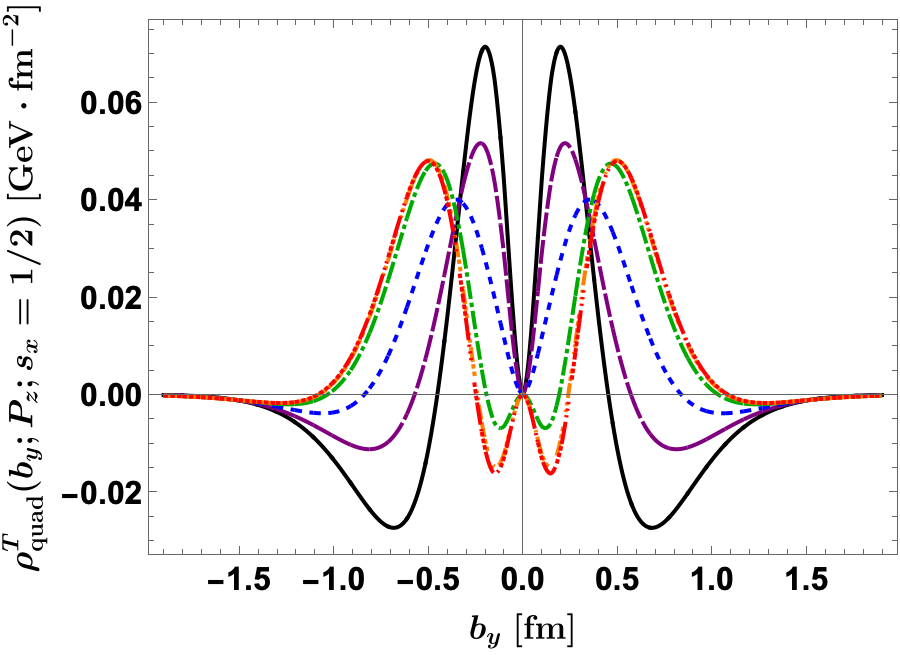}
    \caption{Quadrupole}
    \label{fig:7c}
\end{subfigure}
\begin{subfigure}{.4\textwidth}
    \centering
    \signedcutgraphic[width=1.\linewidth]{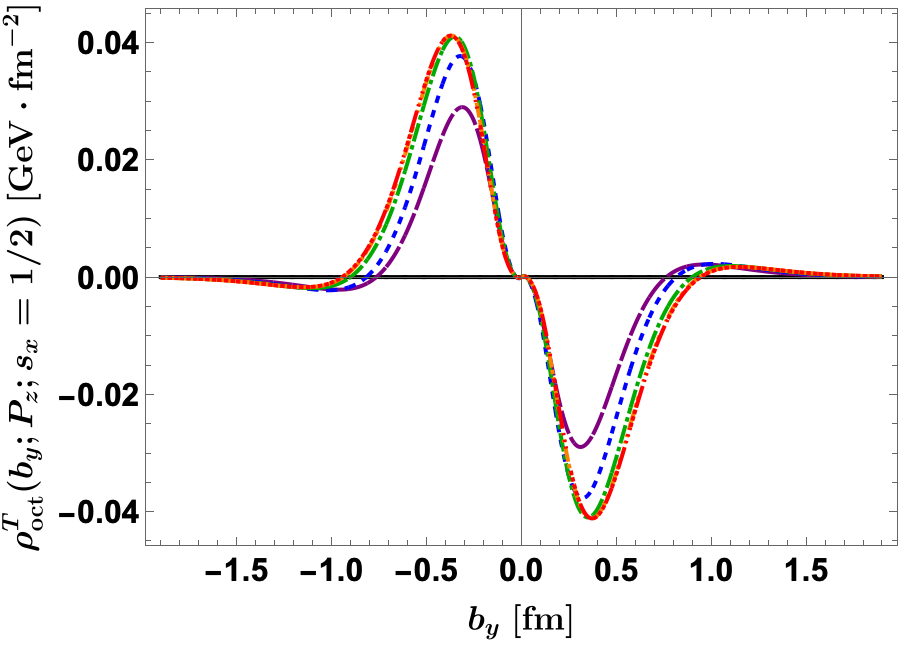}
    \caption{Octupole}
    \label{fig:7d}
\end{subfigure}
\begin{subfigure}{.4\textwidth}
    \centering
    \signedcutgraphic[width=1.\linewidth]{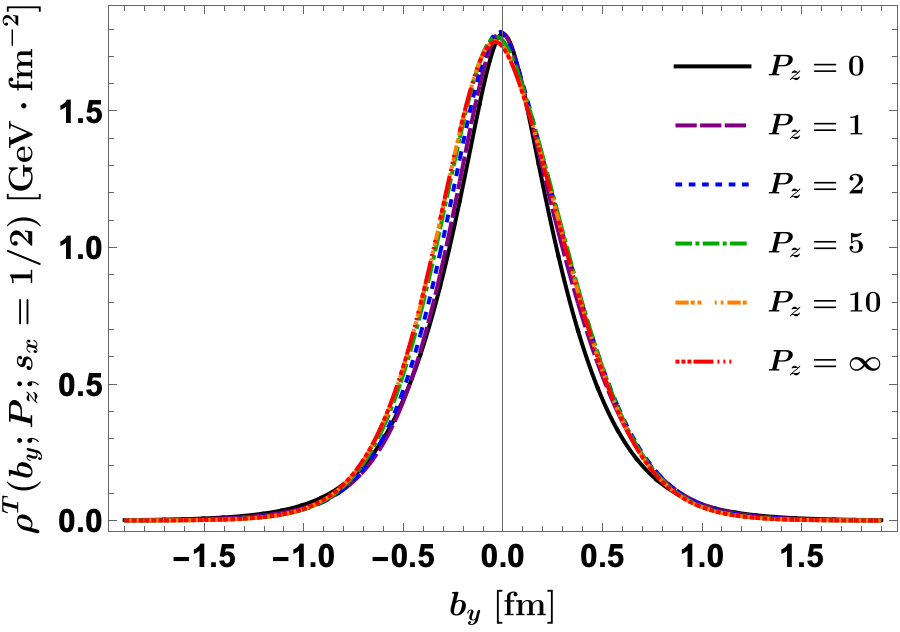}
    \caption{Total}
    \label{fig:7e}
  \end{subfigure}
\caption{Monopole (a), dipole (b), quadrupole (c), octupole (d), and
total (e) contributions to the energy distribution
$\rho^T(b_y;P_z;s_x)$ for $s_x=1/2$. The notation is the same as in
Fig.~\ref{fig:6}.}
\label{fig:7}
\end{figure}

Figures~\ref{fig:6} and \ref{fig:7} show the multipole contributions
to the transversely polarized energy distributions for $s_x=3/2$ and
$s_x=1/2$, respectively. In both cases, the monopole is much larger
than the higher multipoles and sets the overall size of the total
distribution. At $P_z=0$, $\beta=\theta=0$, and the EF energy matrix
element reduces to Eq.~\eqref{eq:BFenergy32}, which contains only
the monopole and quadrupole; the dipole and octupole therefore
vanish.

At finite $P_z$, the first row of
Eq.~\eqref{eq:EFamplitudes32factorized} combines the Breit-frame
components as $T^{00}+2\beta T^{03}+\beta^2T^{33}$.
The $2\beta T^{03}$ term introduces the Breit-frame dipole and
octupole into the EF energy matrix element, and the Wigner rotation
of the spin states generates additional odd-multipole contributions.
The dipole dominates over the octupole and breaks the $b_y\to-b_y$
symmetry. Its orientation depends on $P_z$: for
$P_z\simeq1$--$2\;\mathrm{GeV}$ the dipole enhances the energy density
at $b_y>0$, whereas at larger $P_z$ a structure of the opposite sign
develops near the center and moves the peak of the total distribution
to $b_y<0$ in the IMF. The dipole extrema are largest for
$P_z\simeq1$--$2\;\mathrm{GeV}$ and then decrease toward the IMF,
consistent with the $1S$ form factor in Fig.~\ref{fig:2}.
\subsubsection{\texorpdfstring{Distribution of longitudinal momentum}{Distribution
of longitudinal momentum}}
\begin{figure}[htb!]
\begin{subfigure}{.4\textwidth}
    \centering
    \signedcutgraphic[width=1.\linewidth]{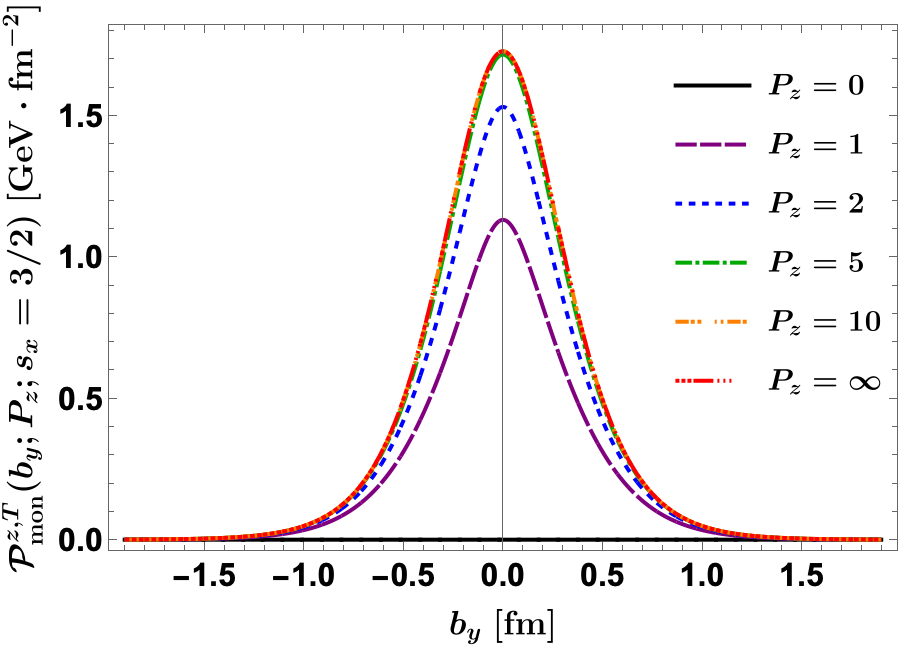}
    \caption{Monopole}
    \label{fig:8a}
\end{subfigure}
\begin{subfigure}{.4\textwidth}
    \centering
    \signedcutgraphic[width=1.\linewidth]{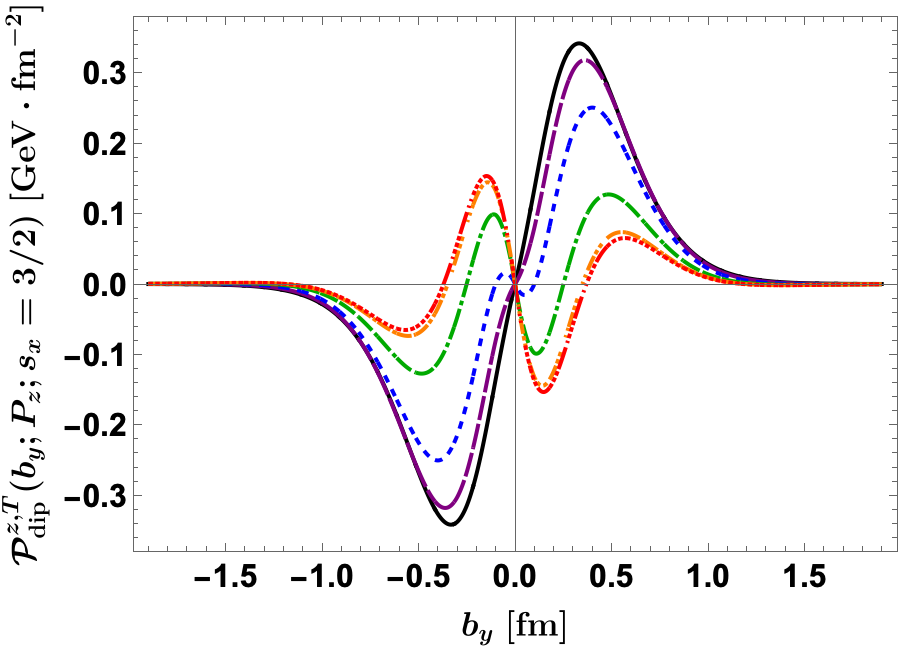}
    \caption{Dipole}
    \label{fig:8b}
\end{subfigure}
\begin{subfigure}{.4\textwidth}
    \centering
    \signedcutgraphic[width=1.\linewidth]{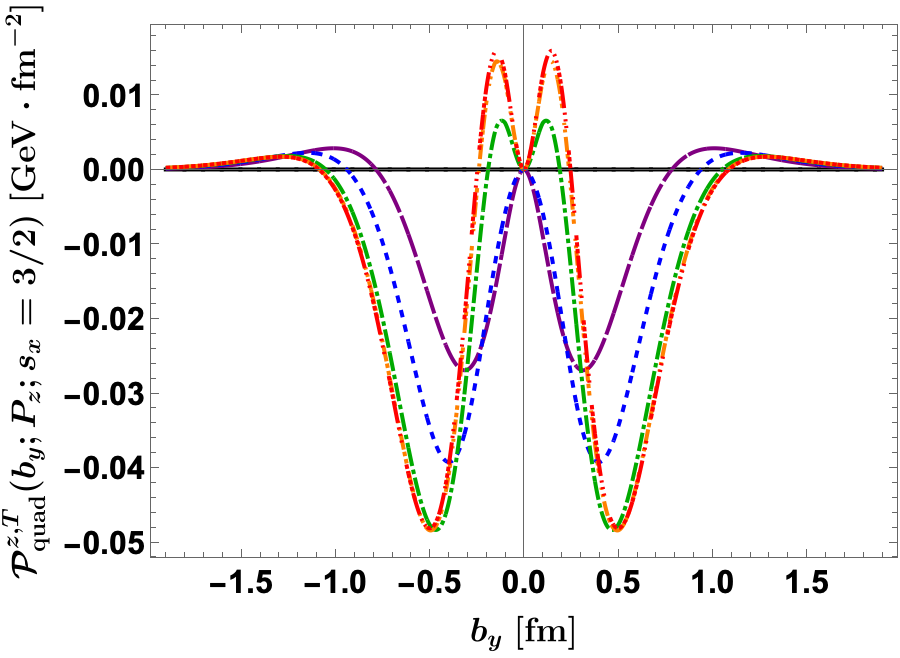}
    \caption{Quadrupole}
    \label{fig:8c}
\end{subfigure}
\begin{subfigure}{.4\textwidth}
    \centering
    \signedcutgraphic[width=1.\linewidth]{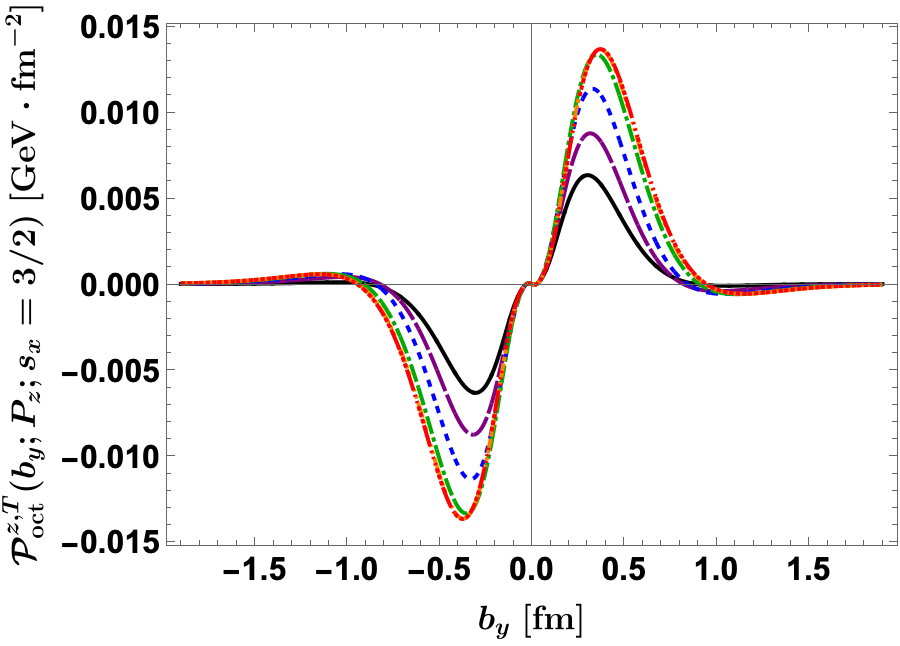}
    \caption{Octupole}
    \label{fig:8d}
\end{subfigure}
\begin{subfigure}{.4\textwidth}
    \centering
    \signedcutgraphic[width=1.\linewidth]{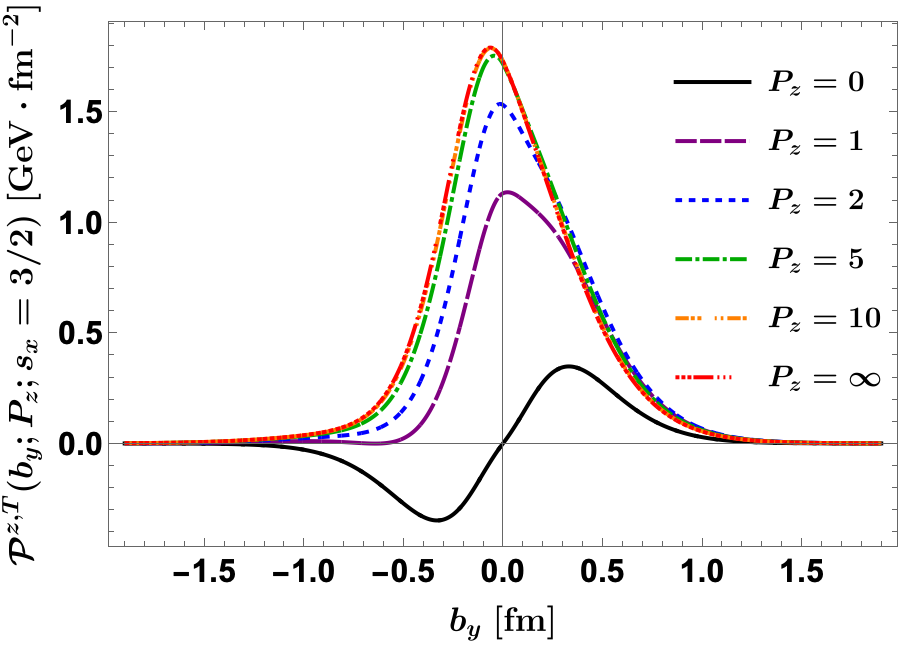}
    \caption{Total}
    \label{fig:8e}
  \end{subfigure}
  \caption{Monopole (a), dipole (b), quadrupole (c), octupole (d), and
total (e) contributions to the longitudinal momentum distribution
$\mathcal P^{z,T}(b_y;P_z;s_x)$ for $s_x=3/2$. The notation is the
same as in Fig.~\ref{fig:6}.}
\label{fig:8}
\end{figure}

\begin{figure}[htb!]
\begin{subfigure}{.4\textwidth}
    \centering
    \signedcutgraphic[width=1.\linewidth]{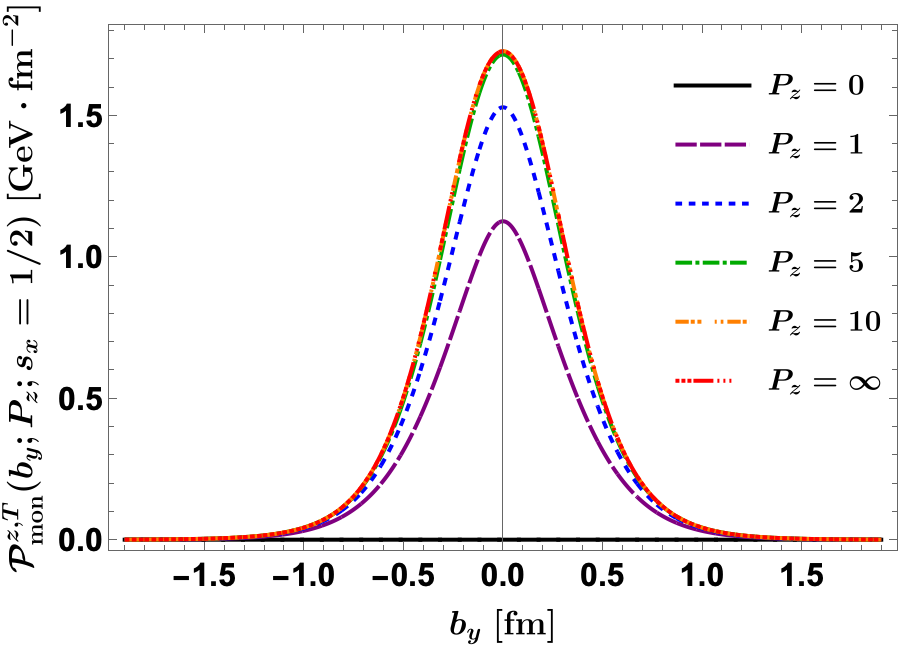}
    \caption{Monopole}
    \label{fig:9a}
\end{subfigure}
\begin{subfigure}{.4\textwidth}
    \centering
    \signedcutgraphic[width=1.\linewidth]{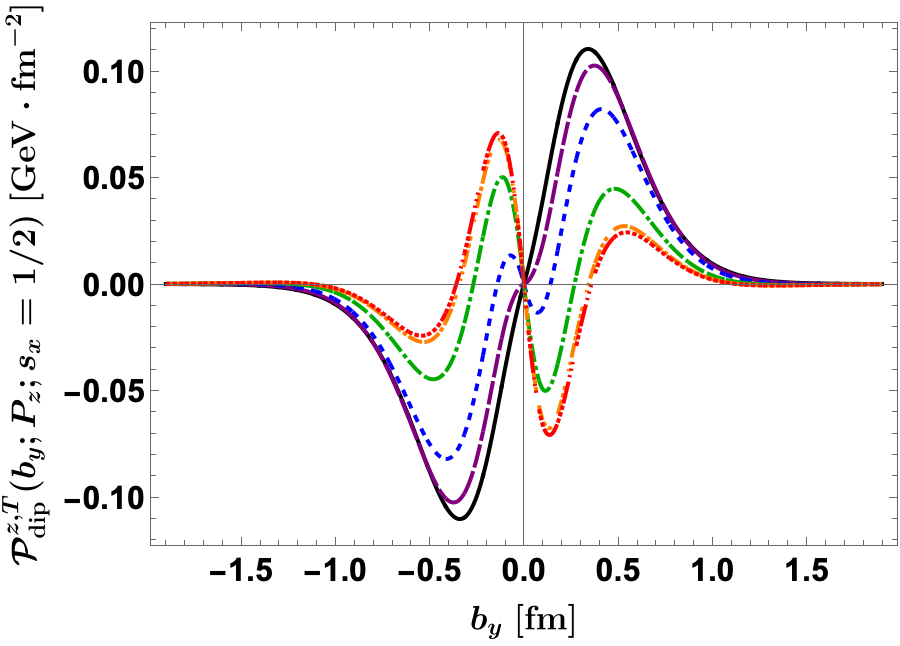}
    \caption{Dipole}
    \label{fig:9b}
\end{subfigure}
\begin{subfigure}{.4\textwidth}
    \centering
    \signedcutgraphic[width=1.\linewidth]{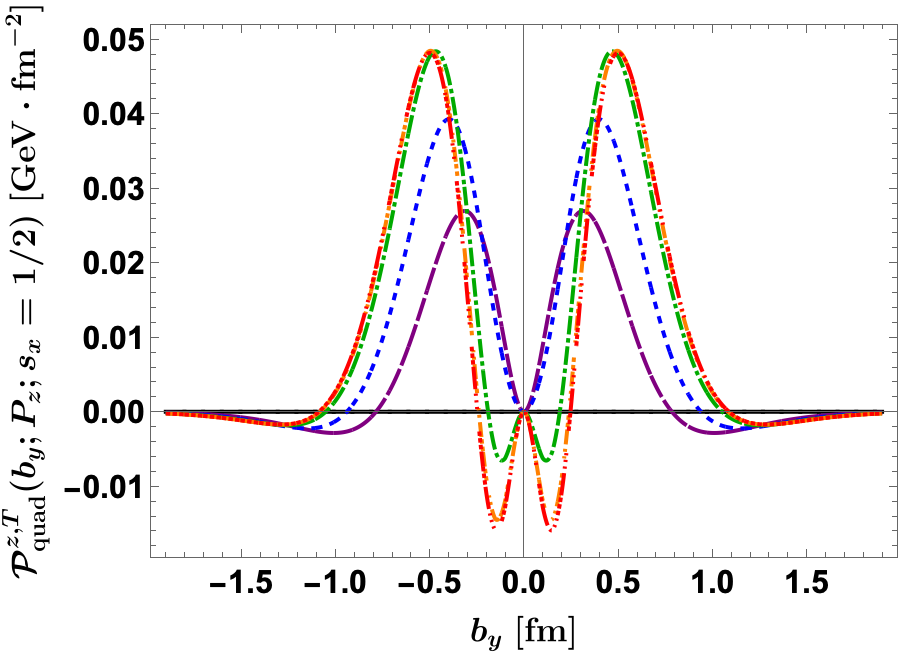}
    \caption{Quadrupole}
    \label{fig:9c}
\end{subfigure}
\begin{subfigure}{.4\textwidth}
    \centering
    \signedcutgraphic[width=1.\linewidth]{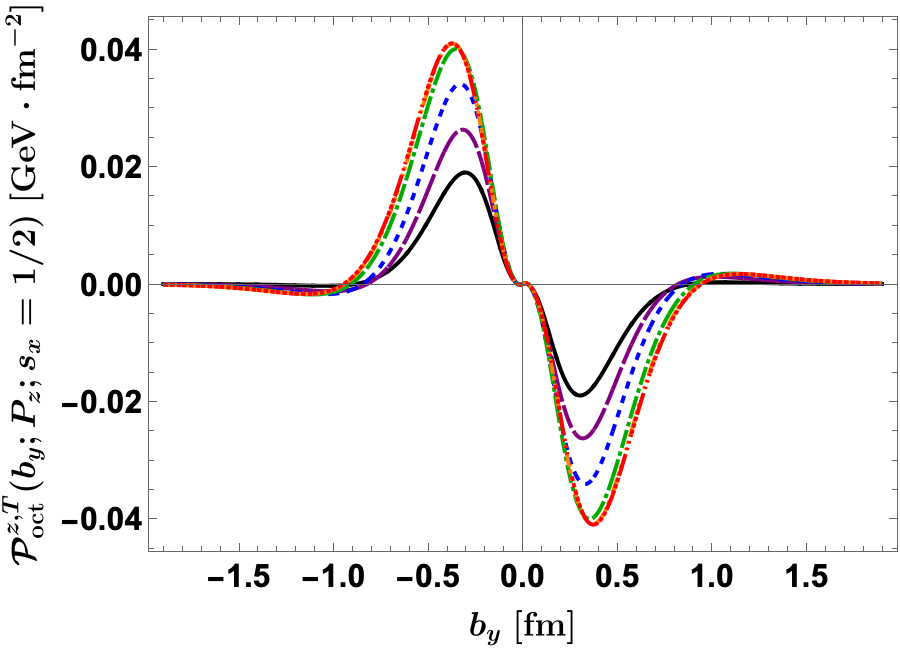}
    \caption{Octupole}
    \label{fig:9d}
\end{subfigure}
\begin{subfigure}{.4\textwidth}
    \centering
    \signedcutgraphic[width=1.\linewidth]{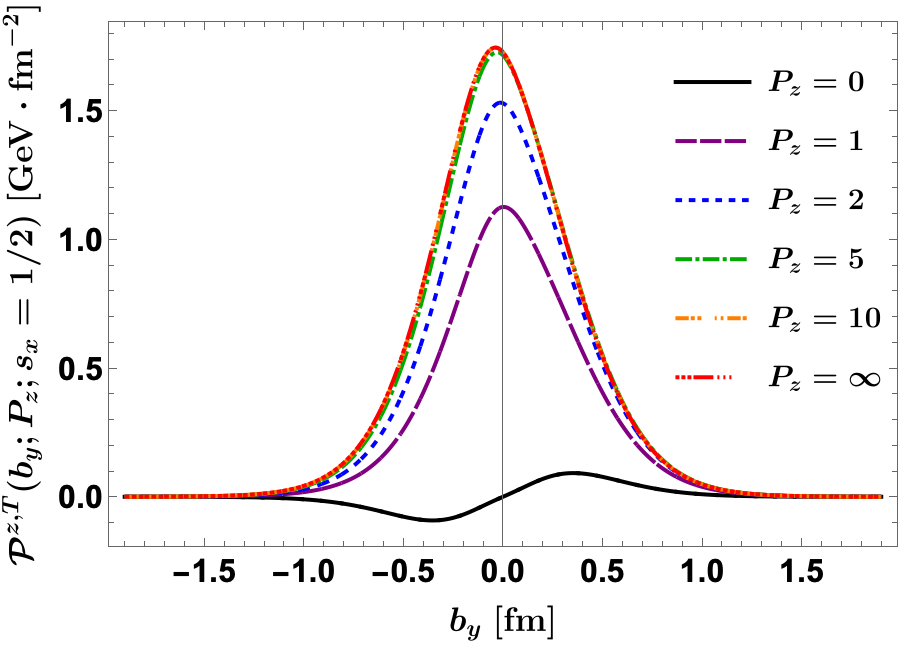}
    \caption{Total}
    \label{fig:9e}
  \end{subfigure}
\caption{Monopole (a), dipole (b), quadrupole (c), octupole (d), and
total (e) contributions to the longitudinal momentum distribution
$\mathcal P^{z,T}(b_y;P_z;s_x)$ for $s_x=1/2$. The notation is the
same as in Fig.~\ref{fig:6}.}
\label{fig:9}
\end{figure}

While the energy distributions in Figs.~\ref{fig:6} and \ref{fig:7}
contain the even multipoles already at $P_z=0$, the longitudinal
momentum distributions in Figs.~\ref{fig:8} and \ref{fig:9} are
purely odd in the transverse Breit frame. At $P_z=0$, the matrix
element of $T^{03}$ in Eq.~\eqref{eq:BFangular32} contains only the
dipole and octupole, so the profile changes sign under
$b_y\to-b_y$ and its transverse integral vanishes. At finite $P_z$,
the boost effects discussed for the energy distributions generate a
positive monopole, which rapidly overwhelms the Breit-frame odd
pattern. The dipole itself behaves oppositely to the energy case: it
is largest at $P_z=0$ and decreases with $P_z$. The total
distribution is therefore a single positive peak tilted by the
dipole, with smaller angular deformations from the quadrupole and
octupole.
\subsubsection{\texorpdfstring{Distribution of longitudinal momentum flux}{Distribution of longitudinal momentum flux}}
\begin{figure}[htb!]
\begin{subfigure}{.4\textwidth}
    \centering
    \signedcutgraphic[width=1.\linewidth]{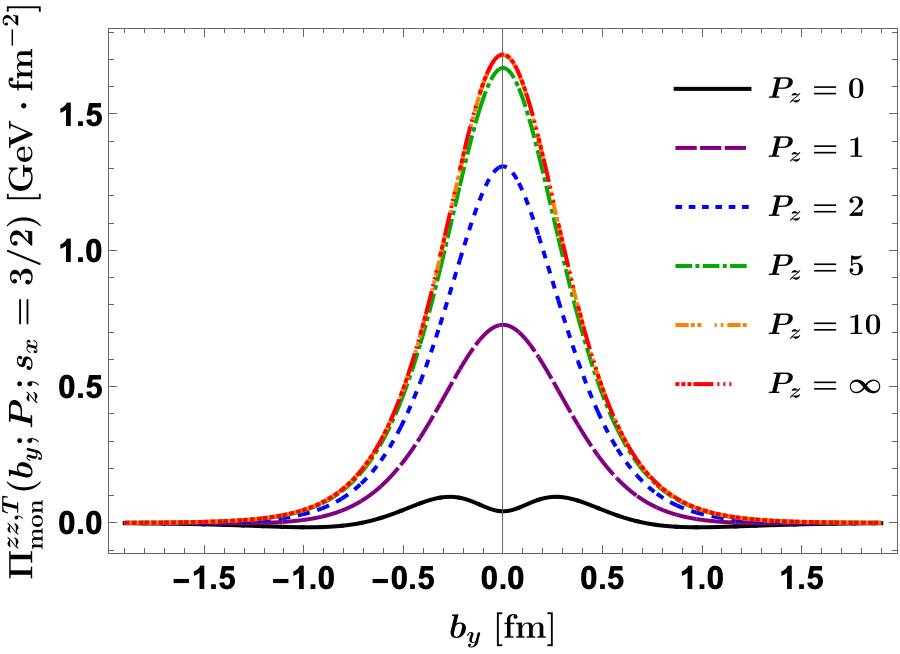}
    \caption{Monopole}
    \label{fig:10a}
\end{subfigure}
\begin{subfigure}{.4\textwidth}
    \centering
    \signedcutgraphic[width=1.\linewidth]{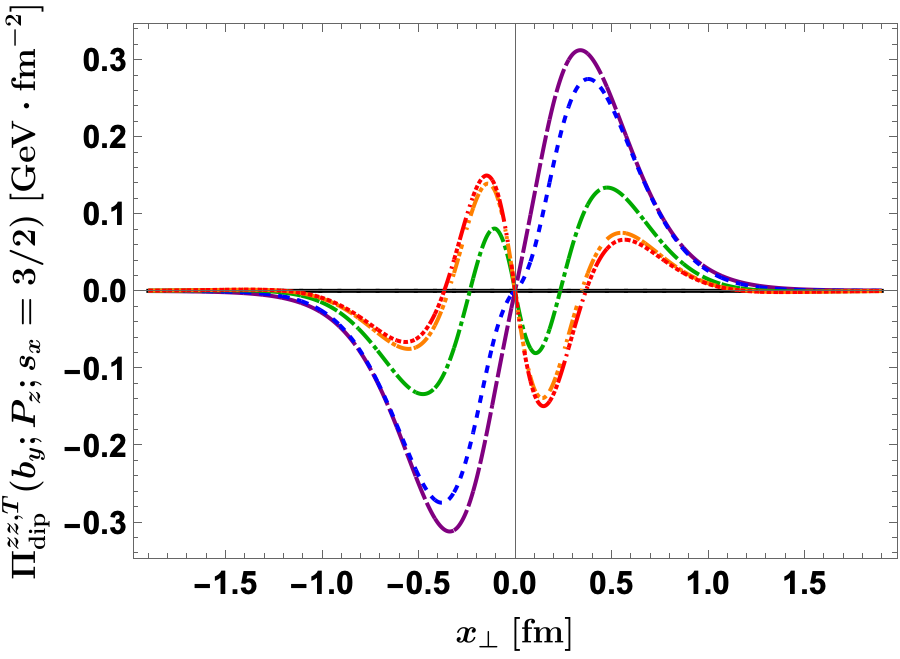}
    \caption{Dipole}
    \label{fig:10b}
\end{subfigure}
\begin{subfigure}{.4\textwidth}
    \centering
    \signedcutgraphic[width=1.\linewidth]{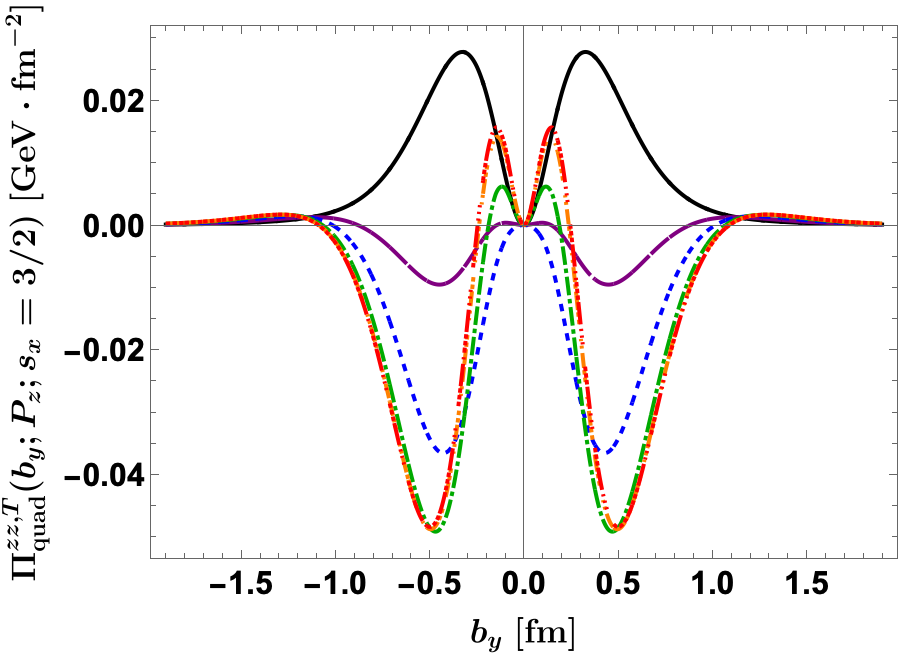}
    \caption{Quadrupole}
    \label{fig:10c}
\end{subfigure}
\begin{subfigure}{.4\textwidth}
    \centering
    \signedcutgraphic[width=1.\linewidth]{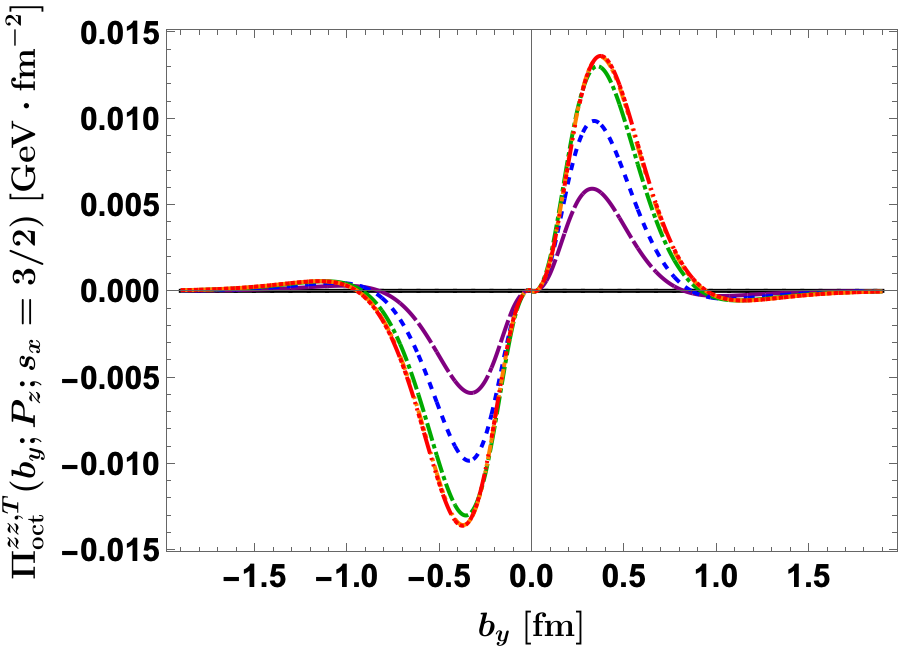}
    \caption{Octupole}
    \label{fig:10d}
\end{subfigure}
\begin{subfigure}{.4\textwidth}
    \centering
    \signedcutgraphic[width=1.\linewidth]{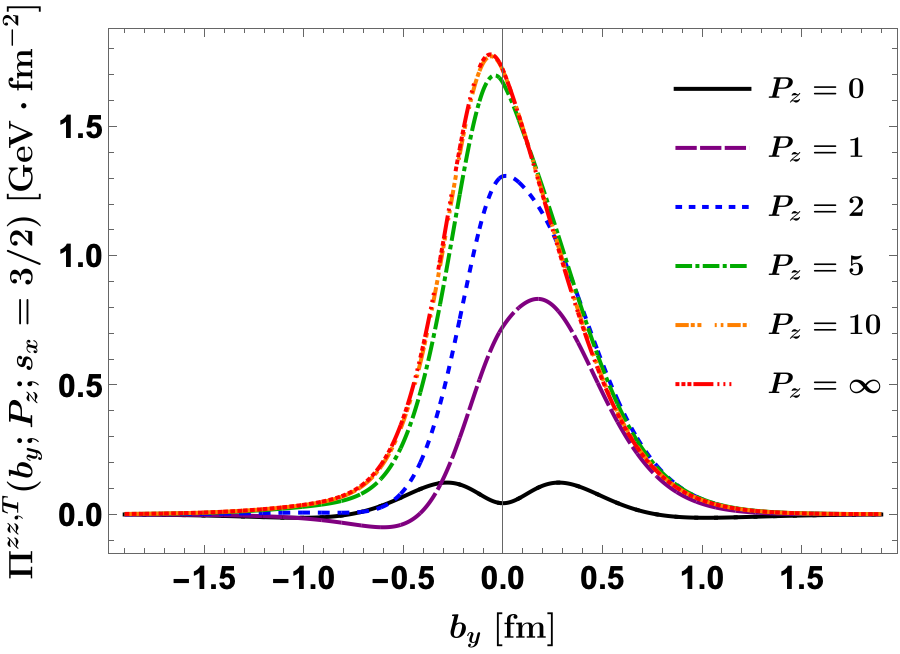}
    \caption{Total}
    \label{fig:10e}
  \end{subfigure}
\caption{Monopole (a), dipole (b), quadrupole (c), octupole (d), and
total (e) contributions to the longitudinal momentum flux
distribution $\Pi^{zz,T}(b_y;P_z;s_x)$ for $s_x=3/2$. The notation
is the same as in Fig.~\ref{fig:6}.}
\label{fig:10}
\end{figure}

\begin{figure}[htb!]
\begin{subfigure}{.4\textwidth}
    \centering
    \signedcutgraphic[width=1.\linewidth]{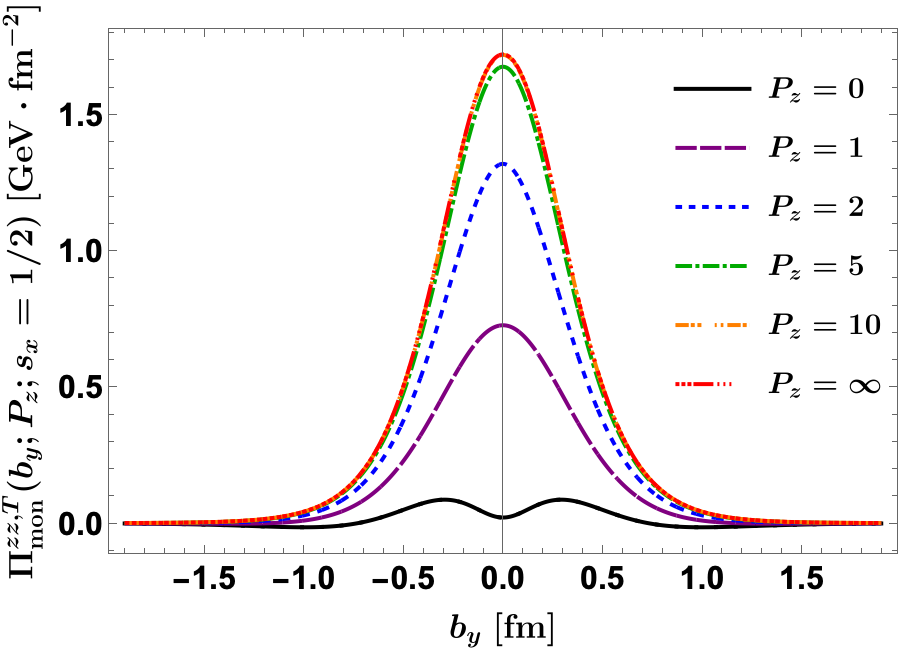}
    \caption{Monopole}
    \label{fig:11a}
\end{subfigure}
\begin{subfigure}{.4\textwidth}
    \centering
    \signedcutgraphic[width=1.\linewidth]{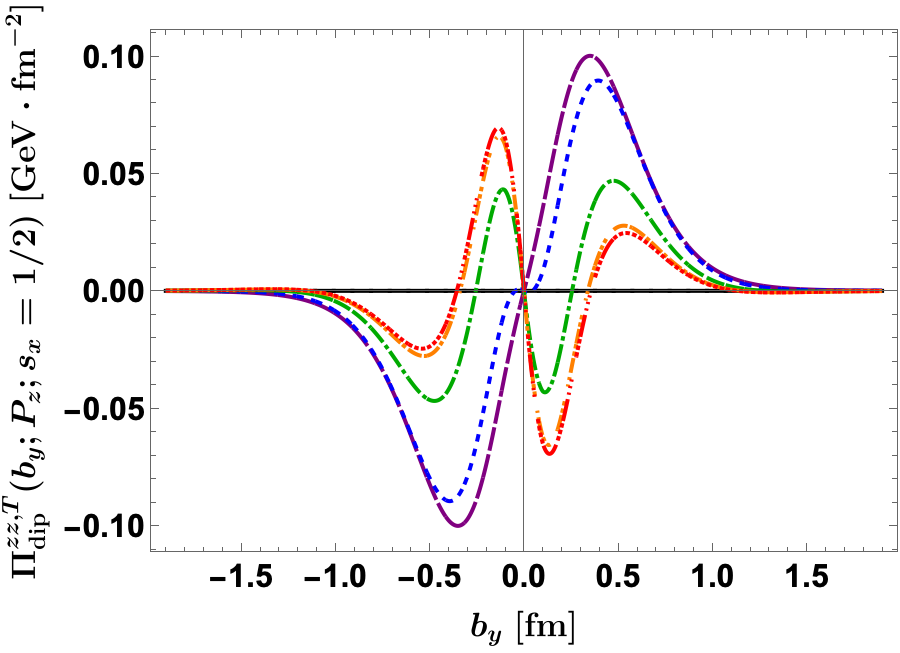}
    \caption{Dipole}
    \label{fig:11b}
\end{subfigure}
\begin{subfigure}{.4\textwidth}
    \centering
    \signedcutgraphic[width=1.\linewidth]{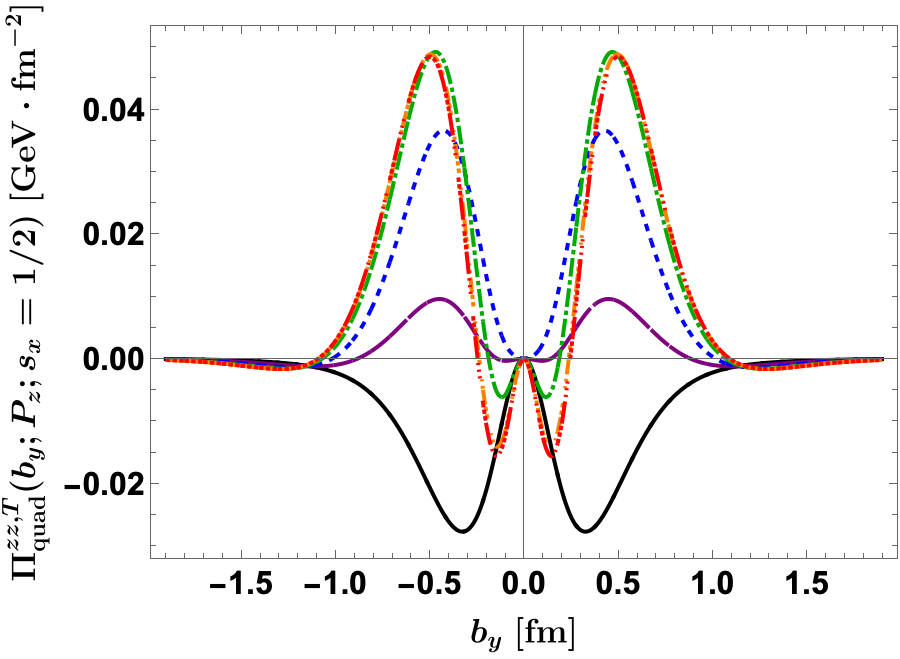}
    \caption{Quadrupole}
    \label{fig:11c}
\end{subfigure}
\begin{subfigure}{.4\textwidth}
    \centering
    \signedcutgraphic[width=1.\linewidth]{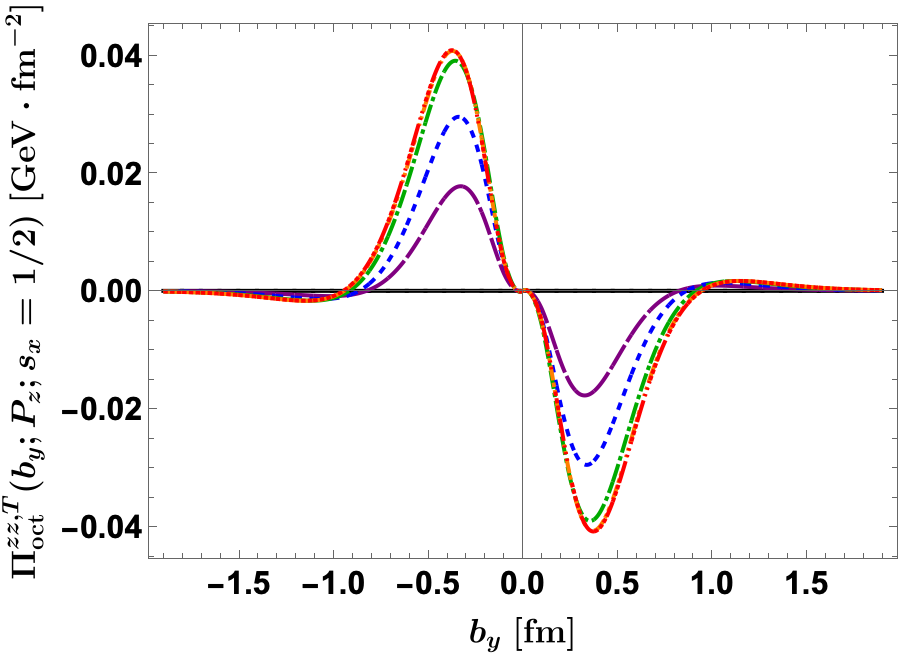}
    \caption{Octupole}
    \label{fig:11d}
\end{subfigure}
\begin{subfigure}{.4\textwidth}
    \centering
    \signedcutgraphic[width=1.\linewidth]{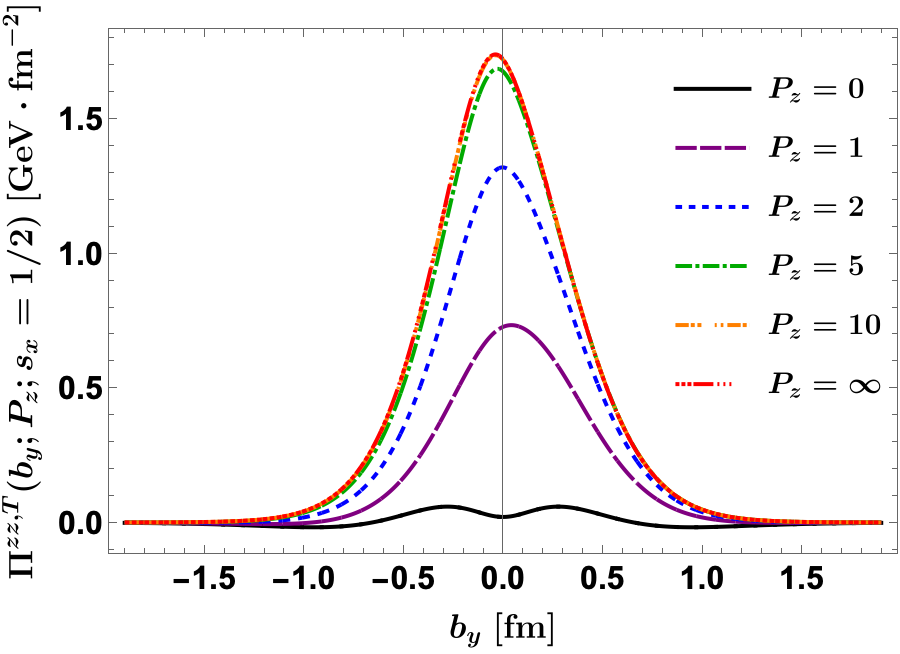}
    \caption{Total}
    \label{fig:11e}
  \end{subfigure}
\caption{Monopole (a), dipole (b), quadrupole (c), octupole (d), and
total (e) contributions to the longitudinal momentum flux
distribution $\Pi^{zz,T}(b_y;P_z;s_x)$ for $s_x=1/2$. The notation
is the same as in Fig.~\ref{fig:6}.}
\label{fig:11}
\end{figure}

The longitudinal momentum flux in Figs.~\ref{fig:10} and
\ref{fig:11} differs from the longitudinal momentum in that it does
not vanish in the transverse Breit frame. In that frame, the
monopole and quadrupole produce a sign-changing profile with zero
transverse integral. A longitudinal boost changes this balance in
two ways: odd multipoles appear, and $\beta^2T^{00}$ builds a
positive central monopole. Once the monopole dominates, the angular
deformations survive only as small modulations of the total
distribution. At $P_z\to\infty$, the momentum flux matches the
energy and longitudinal momentum distributions for the same spin
projection.
\FloatBarrier
\subsection{Dependence of the transverse spatial
distributions on longitudinal momentum}
The one-dimensional cuts in Figs.~\ref{fig:6}--\ref{fig:11} compare
the magnitude and sign of each multipole.
Figures~\ref{fig:12}--\ref{fig:14} display the total distributions
over the transverse plane for $P_z=0$, $1$, $2\;\mathrm{GeV}$, and
the IMF, with the spin along $+\hat x$. At $P_z=0$, the $T^{00}$ and
$T^{33}$ maps are symmetric under $y\to-y$, whereas the $T^{03}$ map
is antisymmetric because it already contains dipole and octupole
contributions. More generally, these maps complement the
one-dimensional cuts by revealing the angular deformations associated
with the quadrupole and octupole over the full transverse plane.
\subsubsection{Energy distribution}
\begin{figure}[htb!]
\begin{subfigure}{0.24\textwidth}
    \centering
    \includegraphics[width=1.\linewidth]{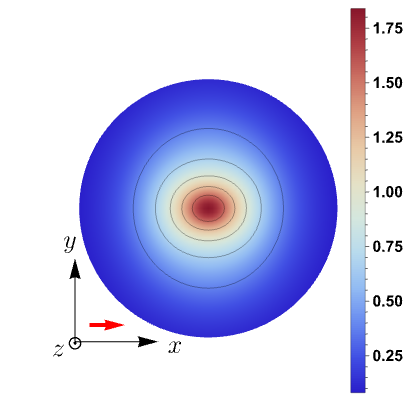}
    \caption{$P_z=0$}
    \label{fig:12a}
\end{subfigure}
\begin{subfigure}{.24\textwidth}
    \centering
    \includegraphics[width=1.\linewidth]{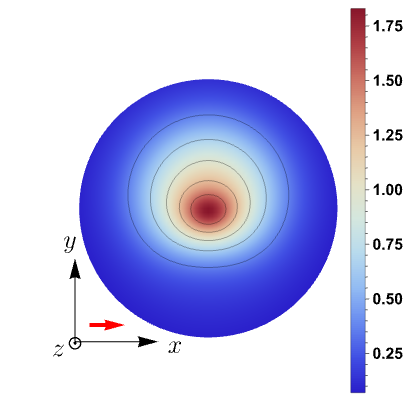}
    \caption{$P_z=1$}
    \label{fig:12b}
\end{subfigure}
\begin{subfigure}{.24\textwidth}
    \centering
    \includegraphics[width=1.\linewidth]{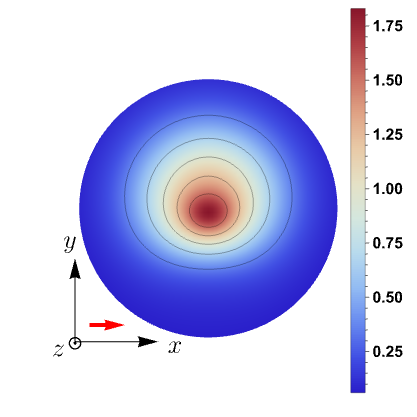}
    \caption{$P_z=2$}
    \label{fig:12c}
\end{subfigure}
\begin{subfigure}{.24\textwidth}
    \centering
    \includegraphics[width=1.\linewidth]{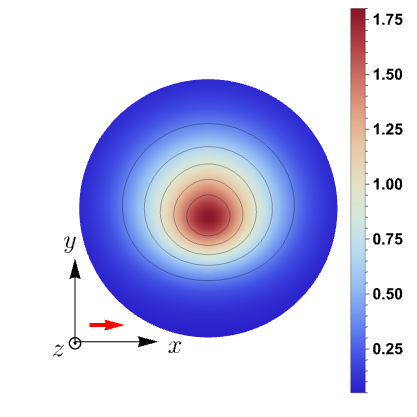}
    \caption{$P_z=\infty$}
    \label{fig:12d}
\end{subfigure}
\begin{subfigure}{.24\textwidth}
    \centering
    \includegraphics[width=1.\linewidth]{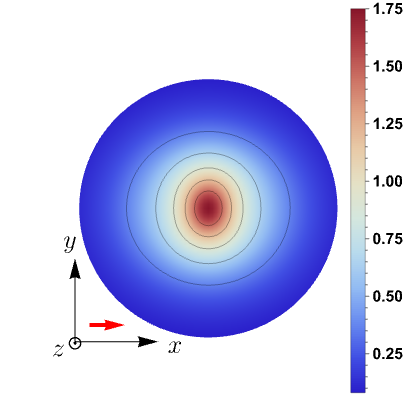}
    \caption{$P_z=0$}
    \label{fig:12e}
\end{subfigure}
\begin{subfigure}{.24\textwidth}
    \centering
    \includegraphics[width=1.\linewidth]{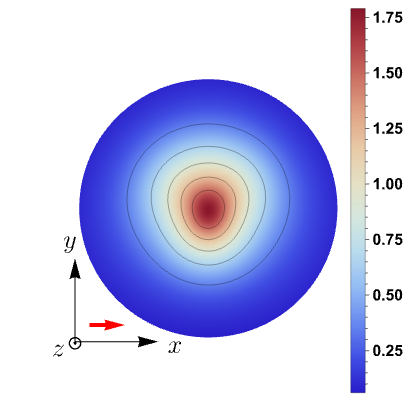}
    \caption{$P_z=1$}
    \label{fig:12f}
\end{subfigure}
\begin{subfigure}{.24\textwidth}
    \centering
    \includegraphics[width=1.\linewidth]{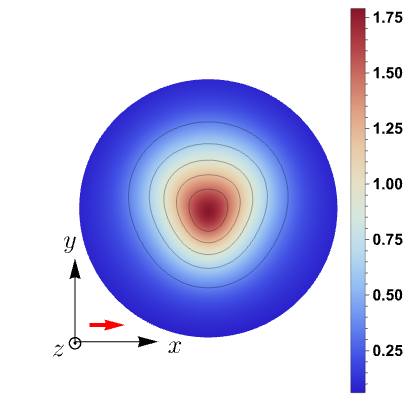}
    \caption{$P_z=2$}
    \label{fig:12g}
\end{subfigure}
\begin{subfigure}{.24\textwidth}
    \centering
    \includegraphics[width=1.\linewidth]{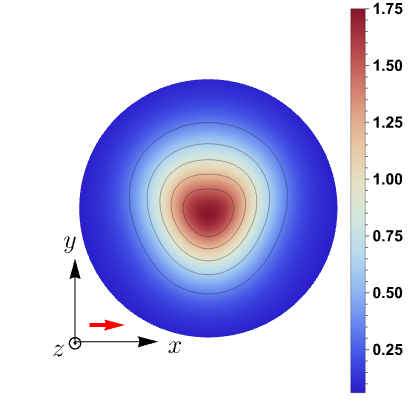}
    \caption{$P_z=\infty$}
    \label{fig:12h}
  \end{subfigure}
  \caption{Energy distribution $\rho^T(\bm x_\perp;P_z;s_x)$ over the
transverse plane. The upper and lower rows correspond to $s_x=3/2$
and $s_x=1/2$, respectively, and the columns to
$P_z=0,1,2\;\mathrm{GeV}$ and the IMF limit $P_z\to\infty$. The red
arrow indicates the polarization direction, and the color bars are
in units of $\mathrm{GeV}\,\mathrm{fm}^{-2}$.}
\label{fig:12}
\end{figure}

Figure~\ref{fig:12} shows that the energy distribution remains
positive and dominated by the central monopole over the full range
of $P_z$. At $P_z=0$, only the even multipoles contribute, and the
maps are symmetric under $y\to-y$, i.e., under reflection about the
spin ($x$) axis. The quadrupole deforms the inner contours in
opposite ways for the two projections: they are elongated along the
spin axis for $s_x=3/2$ and perpendicular to it for $s_x=1/2$. At
nonzero $P_z$, the dipole shifts the maximum slightly toward
$-\hat y$ but remains small relative to the monopole, as shown by
the separate multipole contributions in Figs.~\ref{fig:6} and
\ref{fig:7}.

The opposite orientations follow from
Eq.~\eqref{eq:spinMultipoleExpectations32}: the expectation value of
the $2Q$ structure equals $\pm\frac34\cos2\theta_x$ for $s_x=3/2$
and $s_x=1/2$, respectively, so the same radial function
$\mathcal C_{2Q}^{00}$ deforms the two densities in perpendicular
directions with equal strength. Since $\mathcal E_2(0)=-0.167$, the
coefficient $-8\tau\mathcal E_{2,a}(t)$ of the $2Q$ structure in 
Eq.~\eqref{eq:BFenergy32} is positive. This sign fixes
$\mathcal C_{2Q}^{00}>0$ near the center and determines the
orientations seen in Fig.~\ref{fig:12}.
\subsubsection{\texorpdfstring{Distribution of longitudinal momentum}{Distribution
of longitudinal momentum}}
\begin{figure}[htb!]
\begin{subfigure}{.24\textwidth}
    \centering
    \includegraphics[width=1.\linewidth]{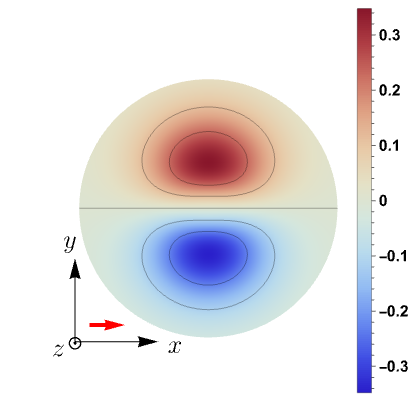}
    \caption{$P_z=0$}
    \label{fig:13a}
\end{subfigure}
\begin{subfigure}{.24\textwidth}
    \centering
    \includegraphics[width=1.\linewidth]{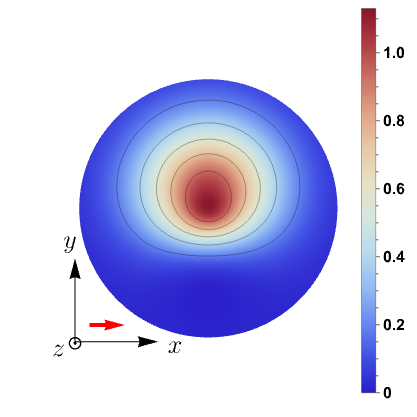}
    \caption{$P_z=1$}
    \label{fig:13b}
\end{subfigure}
\begin{subfigure}{.24\textwidth}
    \centering
    \includegraphics[width=1.\linewidth]{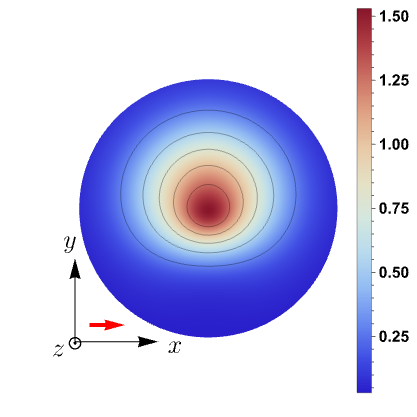}
    \caption{$P_z=2$}
    \label{fig:13c}
\end{subfigure}
\begin{subfigure}{.24\textwidth}
    \centering
    \includegraphics[width=1.\linewidth]{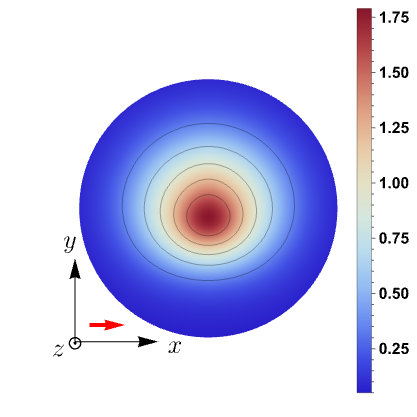}
    \caption{$P_z=\infty$}
    \label{fig:13d}
\end{subfigure}
\begin{subfigure}{.24\textwidth}
    \centering
    \includegraphics[width=1.\linewidth]{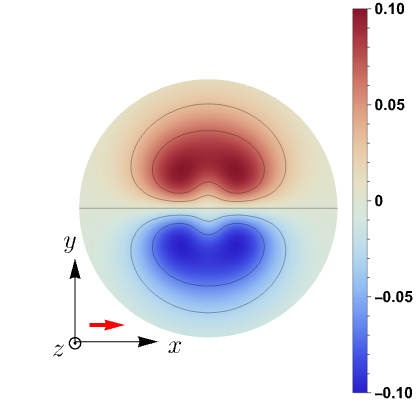}
    \caption{$P_z=0$}
    \label{fig:13e}
\end{subfigure}
\begin{subfigure}{.24\textwidth}
    \centering
    \includegraphics[width=1.\linewidth]{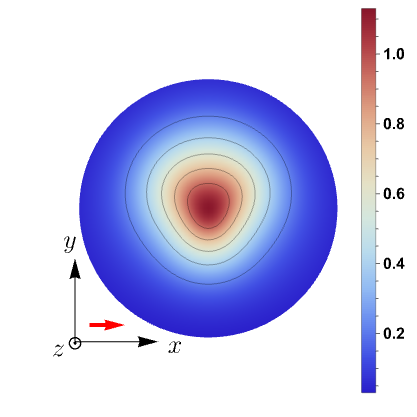}
    \caption{$P_z=1$}
    \label{fig:13f}
\end{subfigure}
\begin{subfigure}{.24\textwidth}
    \centering
    \includegraphics[width=1.\linewidth]{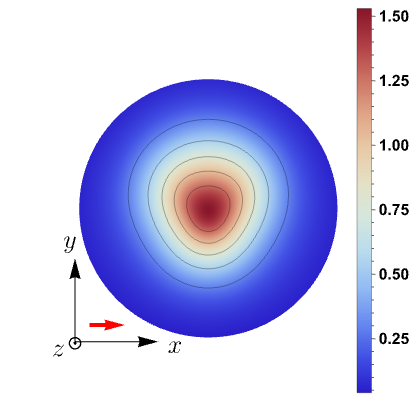}
    \caption{$P_z=2$}
    \label{fig:13g}
\end{subfigure}
\begin{subfigure}{.24\textwidth}
    \centering
    \includegraphics[width=1.\linewidth]{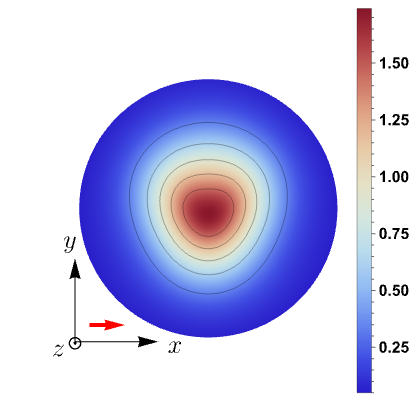}
    \caption{$P_z=\infty$}
    \label{fig:13h}
  \end{subfigure}
  \caption{Longitudinal momentum distribution
$\mathcal P^{z,T}(\bm x_\perp;P_z;s_x)$ over the transverse plane.
The notation is the same as in Fig.~\ref{fig:12}.}
\label{fig:13}
\end{figure}

Figure~\ref{fig:13} shows the two-dimensional maps of the
longitudinal momentum distributions analyzed in Figs.~\ref{fig:8}
and \ref{fig:9}. At $P_z=0$, the maps exhibit a pure dipole pattern
with the positive region at $+\hat y$. This pattern originates from
the $1S$ structure in Eq.~\eqref{eq:BFangular32}: with
$\mathcal J_3(0)=0$ for the present input, the expectation values
$\frac32\sin\theta_x$ and $\frac12\sin\theta_x$ in
Eq.~\eqref{eq:spinMultipoleExpectations32} dominate, and their ratio
explains the factor of about three between the amplitudes for
$s_x=3/2$ and $s_x=1/2$. As $P_z$ increases, the boost mixing
generates a positive monopole that grows with $\beta$ and exceeds the
Breit-frame dipole already at $P_z=1\;\mathrm{GeV}$.
Since $T^{03}$ in Eq.~\eqref{eq:BFangular32} contains no even
structures, the quadrupole deformation at finite $P_z$ arises entirely
from the admixture of the boosted $T^{00}$ and $T^{33}$. At
$P_z=1$--$2\;\mathrm{GeV}$, the maximum of the total distribution is
displaced toward $+\hat y$, opposite to the energy distribution in
Fig.~\ref{fig:12}. This displacement reverses as the IMF is approached,
where the two distributions coincide.
\FloatBarrier
\subsubsection{\texorpdfstring{Distribution of longitudinal momentum flux}
{Distribution of longitudinal momentum flux}}
\begin{figure}[htb!]
\begin{subfigure}{.24\textwidth}
    \centering
    \includegraphics[width=1.\linewidth]{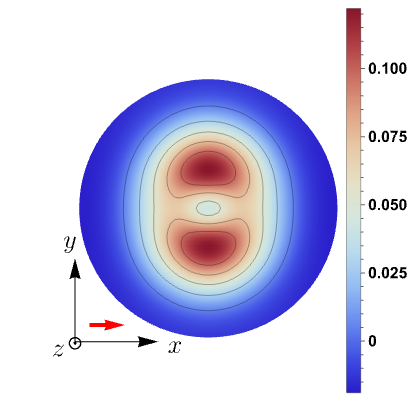}
    \caption{$P_z=0$}
    \label{fig:14a}
\end{subfigure}
\begin{subfigure}{.24\textwidth}
    \centering
    \includegraphics[width=1.\linewidth]{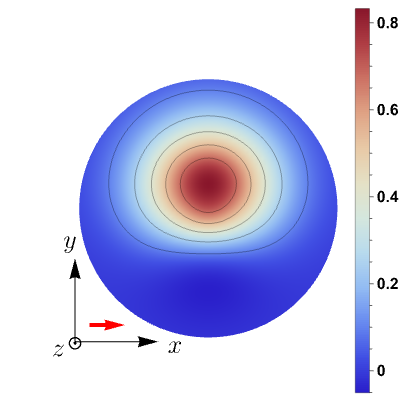}
    \caption{$P_z=1$}
    \label{fig:14b}
\end{subfigure}
\begin{subfigure}{.24\textwidth}
    \centering
    \includegraphics[width=1.\linewidth]{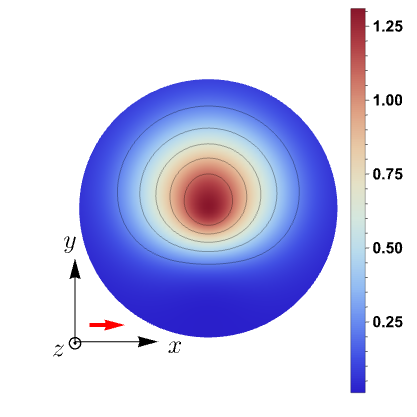}
    \caption{$P_z=2$}
    \label{fig:14c}
\end{subfigure}
\begin{subfigure}{.24\textwidth}
    \centering
    \includegraphics[width=1.\linewidth]{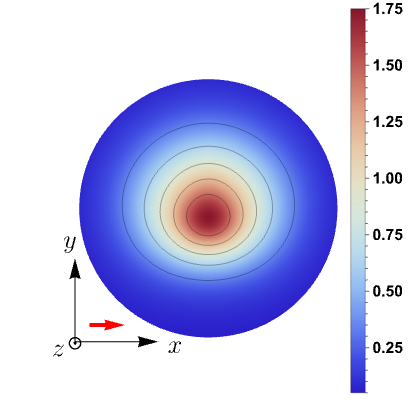}
    \caption{$P_z=\infty$}
    \label{fig:14d}
\end{subfigure}
\begin{subfigure}{.24\textwidth}
    \centering
    \includegraphics[width=1.\linewidth]{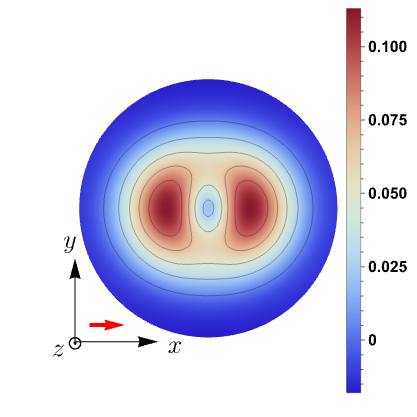}
    \caption{$P_z=0$}
    \label{fig:14e}
\end{subfigure}
\begin{subfigure}{.24\textwidth}
    \centering
    \includegraphics[width=1.\linewidth]{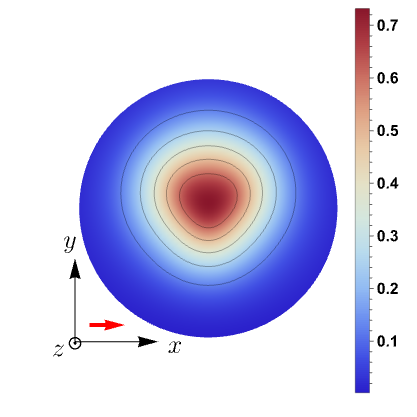}
    \caption{$P_z=1$}
    \label{fig:14f}
\end{subfigure}
\begin{subfigure}{.24\textwidth}
    \centering
    \includegraphics[width=1.\linewidth]{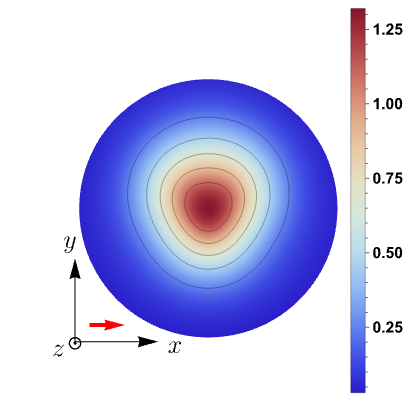}
    \caption{$P_z=2$}
    \label{fig:14g}
\end{subfigure}
\begin{subfigure}{.24\textwidth}
    \centering
    \includegraphics[width=1.\linewidth]{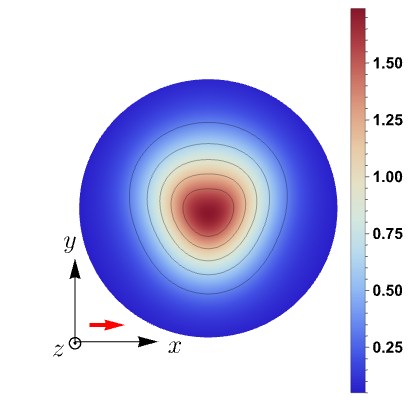}
    \caption{$P_z=\infty$}
    \label{fig:14h}
  \end{subfigure}
  \caption{Longitudinal momentum flux distribution
$\Pi^{zz,T}(\bm x_\perp;P_z;s_x)$ over the transverse plane. The
notation is the same as in Fig.~\ref{fig:12}.}
\label{fig:14}
\end{figure}

Figure~\ref{fig:14} shows the maps of the longitudinal momentum
flux. At $P_z=0$, only the even multipoles contribute, in accordance
with Eq.~\eqref{eq:BFstress32}. The monopole radial function changes
sign in $x_\perp$, as required by
$\mathcal P_0(0)=\mathcal P_{0Q}(0)=0$, and together with the
quadrupole produces two maxima: along $\pm\hat y$ for $s_x=3/2$ and
along $\pm\hat x$ for $s_x=1/2$. As in the energy case, the opposite
orientations of the two projections follow from the expectation
values $\pm\frac34\cos2\theta_x$ of the $2Q$ structure. At finite
$P_z$, the $\beta^2T^{00}$ admixture adds a positive monopole, and
the maps turn into single peaks displaced toward $+\hat y$ at
$P_z=1$--$2\;\mathrm{GeV}$. In the IMF, the maps coincide with those
of the energy and longitudinal momentum in Figs.~\ref{fig:12} and
\ref{fig:13}, as required by Eq.~\eqref{eq:EFmultipoleFFIMFlimit32}. 

\section{Summary and outlook}
\label{sec:summary32}
In the present work, we aimed at describing how the transverse
distributions of the energy--momentum tensor~(EMT) of a spin-$3/2$
baryon change under longitudinal boosts. To this end, we derived a
multipole representation of the EMT matrix elements in elastic
frames~(EFs) with purely transverse momentum transfer. In the
transverse Breit frame, the matrix elements of $T^{00}$, $T^{03}$,
and $T^{33}$ are expressed through seven multipole form factors for
energy, angular momentum, and stress. A longitudinal boost mixes
these three EMT components and rotates the canonical spin states
through a Wigner rotation. Equation~\eqref{eq:EFamplitudes32factorized}
separates the two effects and determines the EF matrix elements at
any $P_z$. Each component is described by six multipole structures:
two monopoles, two dipoles, one quadrupole, and one octupole. A
two-dimensional Fourier transform yields the transverse
distributions of energy, longitudinal momentum, and longitudinal
momentum flux. Longitudinal polarization retains only the two
monopoles, whereas transverse polarization involves all six
structures. In the infinite-momentum frame~(IMF), the multipole form
factors of $T^{00}$, $T^{03}$, and $T^{33}$ approach common values,
as shown in Eq.~\eqref{eq:EFmultipoleFFIMFlimit32}, so the three
distributions become identical at fixed spin projection.

We verified the IMF relation by calculating $T^{++}$,
$T^{+-}=T^{-+}$, and $T^{--}$ directly on the light front~(LF) with
LF Rarita--Schwinger spinors. The EF momenta have the symmetric
Drell--Yan form at any $P_z$, but the EF and LF calculations use
canonical spin and LF helicity, respectively. As $P_z\to\infty$, the
Wigner rotation becomes the Melosh rotation between these two spin
bases, and the leading large-$P_z$ term of each EF matrix element
agrees with the corresponding direct LF result, as stated in
Eq.~\eqref{eq:IMFLFmatrixrelation32}.

The numerical analysis used the total $\Delta$-baryon EMT form
factors obtained in the Skyrme model of Ref.~\cite{Kim:2020lrs},
parametrized in the momentum transfer. The energy distribution
remains positive and dominated by its central monopole, and all
distributions saturate for $P_z\gtrsim5\;\mathrm{GeV}$. Under
longitudinal polarization, the longitudinal momentum distribution
vanishes at $P_z=0$, whereas the momentum flux takes both signs with
zero transverse integral, the two-dimensional von Laue condition;
at finite $P_z$, both develop a positive central monopole. Under
transverse polarization, the quadrupole deforms the energy distribution
in perpendicular directions for the two spin projections, elongated
along the spin axis for $s_x=3/2$ and perpendicular to it for
$s_x=1/2$, as dictated by the expectation values
$\pm\frac34\cos2\theta_x$ of the $2Q$ structure. The dipole
displaces the maxima of the total distributions, toward $-\hat y$
for the energy and toward $+\hat y$ for the longitudinal momentum
and its flux at moderate $P_z$, until all three merge into the
common IMF distribution.

The present formalism isolates the kinematical effects of
longitudinal boosts, while the spatial distributions themselves are
determined by the gravitational form factors~(GFFs). Concrete
conclusions about the internal structure of spin-$3/2$ baryons
therefore require dynamical input from lattice quantum
chromodynamics~(QCD) and other models. Comparing their predictions
for the individual multipole form factors and distributions would
permit a quantitative discussion of the deformation and spin
dependence of spin-$3/2$ baryons, as has been done for the proton.

\section*{Acknowledgments}
H.-Ch. K. and J.-Y. K. thank C\'{e}dric Lorc\'{e} for valuable
discussions and for his hospitality during their visit to the Centre
de Physique Th\'{e}orique (CPHT) at \'{E}cole polytechnique, where part
of this work was carried out. This work was supported by the Basic
Science Research Program through the National Research Foundation of
Korea (NRF), funded by the Korean government through the Ministry of
Education, Science and Technology (MEST), under Grant
No.~RS-2025-00513982 (H.-Ch. K. and H.-J. L.), and by an NRF grant funded by the
Korean government under Grant No.~RS-2025-02634319 (K.-H. H.).
\clearpage
\appendix
\section{Canonical-spin and LF-helicity Rarita-Schwinger spinors}
\label{app:canonicalRS32}

\subsection{Canonical-spin representation}
The canonical Rarita--Schwinger spinor for the transverse Breit
frame and the EF is obtained by coupling a spin-1 polarization
vector to a Dirac spinor~\cite{Rarita:1941mf}:
\begin{equation}
 u^\mu(p,\sigma)=
 \sum_{\lambda,s}C_{1\lambda\,\frac12s}^{\frac32\sigma}
 u_s(p)\epsilon_\lambda^\mu(p),
 \qquad
 u_s(p)=\sqrt{m+p^0}
 \begin{pmatrix}
 \phi_s\\[2pt]
 \dfrac{\bm\sigma\cdot\bm p}{m+p^0}\phi_s
 \end{pmatrix}.
 \label{eq:RSspinor32}
\end{equation}
The canonical spin-1 polarization vector is
\begin{equation}
 \epsilon_\lambda^\mu(p)=
 \left(
 \frac{\bm e_\lambda\cdot\bm p}{m},
 \bm e_\lambda+
 \frac{\bm p\,(\bm e_\lambda\cdot\bm p)}{m(m+p^0)}
 \right),
 \label{eq:polarization32}
\end{equation}
where the rest-frame polarization vectors are
\begin{equation}
 \bm e_{+1}=\frac{1}{\sqrt2}(-1,-i,0),
 \qquad
 \bm e_0=(0,0,1),
 \qquad
 \bm e_{-1}=\frac{1}{\sqrt2}(1,-i,0).
 \label{eq:restpolarization32}
\end{equation}
\subsection{LF-helicity representation}
\label{subsec:LFRSspinors32}
We use the LF coordinates of Eq.~\eqref{eq:LFconvention32} and order
four-vector components as $(+,-,1,2)$. The expressions below follow
the standard LF boost and phase
convention~\cite{Soper:1971wn,Melosh:1974cu,Brodsky:1997de}. With
$p^+>0$, we define
\begin{equation}
 p^R\equiv p^1+ip^2,
 \qquad
 p^L\equiv p^1-ip^2,
 \qquad
 p^-=\frac{m^2+\bm p_\perp^2}{2p^+}.
 \label{eq:LFspinorMomentumNotation32}
\end{equation}
In the Dirac representation, the two spinors of definite LF helicity
are expressed as 
\begin{align}
 u_{\mathrm{LF}}\left(p,+\frac12\right)
 &=
 \frac{1}{\sqrt{2\sqrt2\,p^+}}
 \begin{pmatrix}
  \sqrt2\,p^++m\\
  p^R\\
  \sqrt2\,p^+-m\\
  p^R
 \end{pmatrix},
 \nonumber\\[2pt]
 u_{\mathrm{LF}}\left(p,-\frac12\right)
 &=
 \frac{1}{\sqrt{2\sqrt2\,p^+}}
 \begin{pmatrix}
  -p^L\\
  \sqrt2\,p^++m\\
  p^L\\
  m-\sqrt2\,p^+
 \end{pmatrix}.
 \label{eq:LFDiracSpinors32}
\end{align}
With the transverse phases fixed by
Eq.~\eqref{eq:restpolarization32}, the massive spin-1 LF polarization
vectors are written as 
\begin{align}
 \epsilon_{\mathrm{LF}}^\mu(p,+1)
 &=
 \left(
 0,
 -\frac{p^R}{\sqrt2\,p^+},
 -\frac{1}{\sqrt2},
 -\frac{i}{\sqrt2}
 \right),
 \nonumber\\
 \epsilon_{\mathrm{LF}}^\mu(p,0)
 &=
 \left(
 \frac{p^+}{m},
 \frac{\bm p_\perp^2-m^2}{2mp^+},
 \frac{p^1}{m},
 \frac{p^2}{m}
 \right),
 \nonumber\\
 \epsilon_{\mathrm{LF}}^\mu(p,-1)
 &=
 \left(
 0,
 \frac{p^L}{\sqrt2\,p^+},
 \frac{1}{\sqrt2},
 -\frac{i}{\sqrt2}
 \right).
 \label{eq:LFVectorPolarizations32}
\end{align}
The LF Rarita-Schwinger spinor is obtained by coupling the spin-1 and
spin-$1/2$ LF helicities with the same Clebsch-Gordan convention as
in Eq.~\eqref{eq:RSspinor32}:
\begin{equation}
 u_{\mathrm{LF}}^\mu(p,\lambda)
 =
 \sum_{\lambda_V,s}
 C_{1\lambda_V\,\frac12s}^{\frac32\lambda}
 \epsilon_{\mathrm{LF}}^\mu(p,\lambda_V)
 u_{\mathrm{LF}}(p,s).
 \label{eq:LFRSSpinor32}
\end{equation}

\section{Spin-space matrices for the multipole structures}
\label{app:spinMultipoleMatrices32}
In the ordered $z$-quantized basis of Eq.~\eqref{eq:LSpinStates32},
the six spin structures in Eq.~\eqref{eq:coordinateC32} take the
matrix representations given below. With the operator definitions in
Eq.~\eqref{eq:Qspin32}, the two monopole matrices are given by 
\begin{equation}
 \bm1=
 \begin{pmatrix}
 1&0&0&0\\
 0&1&0&0\\
 0&0&1&0\\
 0&0&0&1
 \end{pmatrix},
 \qquad
 Q^{33}=
 \begin{pmatrix}
 1&0&0&0\\
 0&-1&0&0\\
 0&0&-1&0\\
 0&0&0&1
 \end{pmatrix}.
 \label{eq:spinMonopoleMatrices32}
\end{equation}
The dipole matrices are
\begin{align}
 \epsilon^{ij3}S^iX_1^j(\theta_x)
 &=\frac12
 \begin{pmatrix}
 0&i\sqrt3e^{-i\theta_x}&0&0\\
 -i\sqrt3e^{i\theta_x}&0&2ie^{-i\theta_x}&0\\
 0&-2ie^{i\theta_x}&0&i\sqrt3e^{-i\theta_x}\\
 0&0&-i\sqrt3e^{i\theta_x}&0
 \end{pmatrix},
 \label{eq:spinDipoleMatrix32}\\[1ex]
 \epsilon^{ij3}O^{i33}X_1^j(\theta_x)
 &=\frac15
 \begin{pmatrix}
 0&i\sqrt3e^{-i\theta_x}&0&0\\
 -i\sqrt3e^{i\theta_x}&0&-3ie^{-i\theta_x}&0\\
 0&3ie^{i\theta_x}&0&i\sqrt3e^{-i\theta_x}\\
 0&0&-i\sqrt3e^{i\theta_x}&0
 \end{pmatrix}.
 \label{eq:octupoleDipoleMatrix32}
\end{align}
Finally, the quadrupole and octupole matrices are
\begin{align}
 Q^{ij}X_2^{ij}(\theta_x)
 &=\frac{\sqrt3}{2}
 \begin{pmatrix}
 0&0&e^{-2i\theta_x}&0\\
 0&0&0&e^{-2i\theta_x}\\
 e^{2i\theta_x}&0&0&0\\
 0&e^{2i\theta_x}&0&0
 \end{pmatrix},
 \label{eq:spinQuadrupoleMatrix32}\\[1ex]
 \epsilon^{3ik}O^{ij\ell}X_3^{kj\ell}(\theta_x)
 &=\frac{3i}{4}
 \begin{pmatrix}
 0&0&0&e^{-3i\theta_x}\\
 0&0&0&0\\
 0&0&0&0\\
 -e^{3i\theta_x}&0&0&0
 \end{pmatrix}.
 \label{eq:spinOctupoleMatrix32}
\end{align}
\section{Two-dimensional reduction of the Breit-frame multipole form factors}
\label{app:intrinsicFF32}
In the transverse Breit frame, $\Delta^3=0$, and the matrix elements
in Eqs.~\eqref{eq:BFenergy32}--\eqref{eq:BFstress32} are written in
terms of the two-dimensional tensors $X_n$. The three-dimensional
Breit-frame expansion of Ref.~\cite{Kim:2020lrs} instead uses the
irreducible tensors
\begin{subequations}
\begin{align}
 Y_2^{ij}(\hat{\bm\Delta})
 &=\hat\Delta^i\hat\Delta^j-\frac13\delta^{ij},
 \\
 Y_3^{ijk}(\hat{\bm\Delta})
 &=\hat\Delta^i\hat\Delta^j\hat\Delta^k
 -\frac15\left(
 \delta^{ij}\hat\Delta^k
 +\delta^{ik}\hat\Delta^j
 +\delta^{jk}\hat\Delta^i
 \right),
 \qquad i,j,k=1,2,3,
\end{align}
\end{subequations}
where $\hat\Delta^i=\Delta^i/|\bm\Delta|$. The quadrupole and
octupole terms in Ref.~\cite{Kim:2020lrs} contain $Q^{kl}Y_2^{kl}$
and $i\epsilon^{3ik}O^{ij\ell}Y_3^{kj\ell}$, respectively. At
$\Delta^3=0$, the tracelessness of $Q^{ij}$ and $O^{ijk}$ reduces
them to the two-dimensional forms:
\begin{subequations}
\begin{align}
 Q^{kl}Y_2^{kl}(\hat{\bm\Delta})
 &=-\frac12\left[
 Q^{33}-2Q^{ij}X_2^{ij}(\theta_\Delta)
 \right],
 \label{eq:3Dto2Dquadrupole32}\\
 i\epsilon^{3ik}O^{ij\ell}Y_3^{kj\ell}(\hat{\bm\Delta})
 &=-\frac14\left[
 i\epsilon^{ij3}O^{i33}X_1^j(\theta_\Delta)
 -4i\epsilon^{3ik}O^{ij\ell}X_3^{kj\ell}(\theta_\Delta)
 \right].
 \label{eq:3Dto2Doctupole32}
\end{align}
\end{subequations}
Substituting these forms into the three-dimensional expansion and
comparing the result with
Eqs.~\eqref{eq:BFenergy32}--\eqref{eq:BFangular32}, we obtain
\begin{align}
 \mathcal E_{0,a}
 &=\left.\mathcal E^a_0\right|_{\text{Ref.~\cite{Kim:2020lrs}}},
 &
 \mathcal E_{2,a}
 &=-\frac12
 \left.\mathcal E^a_2\right|_{\text{Ref.~\cite{Kim:2020lrs}}},
 \nonumber\\
 \mathcal J_{1,a}
 &=\left.\mathcal J^a_1\right|_{\text{Ref.~\cite{Kim:2020lrs}}},
 &
 \mathcal J_{3,a}
 &=-\frac14
 \left.\mathcal J^a_3\right|_{\text{Ref.~\cite{Kim:2020lrs}}}.
 \label{eq:standardGFFmapping32}
\end{align}
The factors $-1/2$ and $-1/4$ arise solely from the reduction of the
three-dimensional multipole basis to the two-dimensional transverse
basis.

We define $\mathcal P_X\equiv\sum_a\mathcal P_{X,a}$ for the
conserved total EMT. The same reduction of the $T^{33}$ matrix
element leads to 
\begin{align}
 \mathcal P_0
 &=-\tau
 \left.D_0\right|_{\text{Ref.~\cite{Kim:2020lrs}}},
 &
 \mathcal P_{0Q}
 &=-\tau
 \left.D_2\right|_{\text{Ref.~\cite{Kim:2020lrs}}}
 +2\tau^2
 \left.D_3\right|_{\text{Ref.~\cite{Kim:2020lrs}}},
 \nonumber\\
 \mathcal P_2
 &=-\frac12\left(
 \left.D_2\right|_{\text{Ref.~\cite{Kim:2020lrs}}}
 +2\tau
 \left.D_3\right|_{\text{Ref.~\cite{Kim:2020lrs}}}
 \right).
 \label{eq:standardDmapping32}
\end{align}
Together with Eqs.~\eqref{eq:E0spin32}-\eqref{eq:P2spin32}, these
relations connect the three-dimensional Breit-frame form factors of
Ref.~\cite{Kim:2020lrs} to the ten covariant EMT form factors used in
this work.
\section{\texorpdfstring{Finite-$P_z$ multipole form factors in the EF}
{Finite-Pz multipole form factors in the EF}}
\label{app:DMD32}
This appendix lists the eighteen EF multipole form factors for
$T^{00}$, $T^{03}$, and $T^{33}$ at arbitrary $P_z$. They are
obtained by applying Eq.~\eqref{eq:EFamplitudes32factorized} to the
transverse Breit-frame matrix elements and matching the six
multipole structures in Eq.~\eqref{eq:EFmultipoledecomposition32}.
We use the abbreviations $c_n$ and $s_n$ defined in
Eq.~\eqref{eq:Wignertrigshorthand32} and write the mixing matrix of
the Lorentz components compactly as
\begin{equation}
 \mathsf L(\beta)
 =\begin{pmatrix}
 1&2\beta&\beta^2\\
 \beta&1+\beta^2&\beta\\
 \beta^2&2\beta&1
 \end{pmatrix}.
 \label{eq:appLorentzmixingmatrix32}
\end{equation}
The $0$ monopole form factors read
\begin{widetext}
\begin{align}
\begin{pmatrix}
C_{0,a}^{00}\\[2pt] C_{0,a}^{03}\\[2pt] C_{0,a}^{33}
\end{pmatrix}
={}&\frac{\gamma}{\gamma_P}\mathsf L(\beta)
\begin{pmatrix}
\dfrac12\left[
\big(c_1+c_3\big)\mathcal E_{0,a}
-4\tau\big(c_1-c_3\big)\mathcal E_{2,a}
\right]\\[7pt]
-\dfrac{\sqrt\tau}{2}
\big(s_1+3s_3\big)\mathcal J_{1,a}
+\dfrac{12\tau^{3/2}}{5}
\big(3s_1-s_3\big)\mathcal J_{3,a}\\[7pt]
\dfrac12\big(c_1+c_3\big)\mathcal P_{0,a}
+\dfrac14\big(c_1-c_3\big)\mathcal P_{0Q,a}
+\dfrac{3\tau}{2}
\big(c_1-c_3\big)\mathcal P_{2,a}
\end{pmatrix}.
\label{eq:appC0spin32}
\end{align}

The $0Q$ monopole yields
\begin{align}
\begin{pmatrix}
C_{0Q,a}^{00}\\[2pt] C_{0Q,a}^{03}\\[2pt] C_{0Q,a}^{33}
\end{pmatrix}
={}&\frac{\gamma}{\gamma_P}\mathsf L(\beta)
\begin{pmatrix}
\dfrac14\left[
\big(c_1-c_3\big)\mathcal E_{0,a}
+4\tau\big(5c_1-c_3\big)\mathcal E_{2,a}
\right]\\[7pt]
\dfrac{\sqrt\tau}{4}
\big(-s_1+3s_3\big)\mathcal J_{1,a}
-\dfrac{6\tau^{3/2}}{5}
\big(7s_1-s_3\big)\mathcal J_{3,a}\\[7pt]
\dfrac14\big(c_1-c_3\big)\mathcal P_{0,a}
+\dfrac18\big(7c_1+c_3\big)\mathcal P_{0Q,a}
-\dfrac{3\tau}{4}
\big(c_1-c_3\big)\mathcal P_{2,a}
\end{pmatrix}.
\label{eq:appC0Qspin32}
\end{align}

The $1S$ dipole takes the form
\begin{align}
\begin{pmatrix}
C_{1S,a}^{00}\\[2pt] C_{1S,a}^{03}\\[2pt] C_{1S,a}^{33}
\end{pmatrix}
={}&\frac{\gamma}{\gamma_P}\mathsf L(\beta)
\begin{pmatrix}
\dfrac{1}{10\sqrt\tau}\left[
\big(s_1+3s_3\big)\mathcal E_{0,a}
+4\tau\big(-s_1+3s_3\big)\mathcal E_{2,a}
\right]\\[7pt]
\dfrac1{10}\big(c_1+9c_3\big)\mathcal J_{1,a}
-\dfrac{36\tau}{25}
\big(c_1-c_3\big)\mathcal J_{3,a}\\[7pt]
\dfrac{1}{20\sqrt\tau}\left[
2\big(s_1+3s_3\big)\mathcal P_{0,a}
+\big(s_1-3s_3\big)\mathcal P_{0Q,a}
+6\tau\big(s_1-3s_3\big)\mathcal P_{2,a}
\right]
\end{pmatrix}.
\label{eq:appC1Sspin32}
\end{align}

The $1O$ dipole leads to
\begin{align}
\begin{pmatrix}
C_{1O,a}^{00}\\[2pt] C_{1O,a}^{03}\\[2pt] C_{1O,a}^{33}
\end{pmatrix}
={}&\frac{\gamma}{\gamma_P}\mathsf L(\beta)
\begin{pmatrix}
\dfrac{1}{8\sqrt\tau}\left[
\big(3s_1-s_3\big)\mathcal E_{0,a}
+4\tau\big(7s_1-s_3\big)\mathcal E_{2,a}
\right]\\[7pt]
\dfrac38\big(c_1-c_3\big)\mathcal J_{1,a}
+\dfrac{\tau}{5}
\big(23c_1-3c_3\big)\mathcal J_{3,a}\\[7pt]
\begin{aligned}
\dfrac{1}{16\sqrt\tau}\Big[&
2\big(3s_1-s_3\big)\mathcal P_{0,a}
+\big(13s_1+s_3\big)\mathcal P_{0Q,a}
\\[-2pt]
&+2\tau\big(-s_1+3s_3\big)\mathcal P_{2,a}
\Big]
\end{aligned}
\end{pmatrix}.
\label{eq:appC1Ospin32}
\end{align}

The quadrupole form factors are
\begin{align}
\begin{pmatrix}
C_{2Q,a}^{00}\\[2pt] C_{2Q,a}^{03}\\[2pt] C_{2Q,a}^{33}
\end{pmatrix}
={}&\frac{\gamma}{\gamma_P}\mathsf L(\beta)
\begin{pmatrix}
\dfrac{1}{8\tau}\left[
\big(c_1-c_3\big)\mathcal E_{0,a}
-4\tau\big(3c_1+c_3\big)\mathcal E_{2,a}
\right]\\[7pt]
\dfrac{1}{8\sqrt\tau}
\big(-s_1+3s_3\big)\mathcal J_{1,a}
+\dfrac{\sqrt\tau}{5}
\big(19s_1+3s_3\big)\mathcal J_{3,a}\\[7pt]
\dfrac{1}{16\tau}\left[
2\big(c_1-c_3\big)\mathcal P_{0,a}
-\big(c_1-c_3\big)\mathcal P_{0Q,a}
+2\tau\big(5c_1+3c_3\big)\mathcal P_{2,a}
\right]
\end{pmatrix}.
\label{eq:appC2Qspin32}
\end{align}

Finally, the octupole form factors are given by
\begin{align}
\begin{pmatrix}
C_{3O,a}^{00}\\[2pt] C_{3O,a}^{03}\\[2pt] C_{3O,a}^{33}
\end{pmatrix}
={}&\frac{\gamma}{\gamma_P}\mathsf L(\beta)
\begin{pmatrix}
\dfrac{1}{24\tau^{3/2}}\left[
\big(3s_1-s_3\big)\mathcal E_{0,a}
-4\tau\big(9s_1+s_3\big)\mathcal E_{2,a}
\right]\\[7pt]
\dfrac{1}{8\tau}
\big(c_1-c_3\big)\mathcal J_{1,a}
-\dfrac15\big(19c_1+c_3\big)\mathcal J_{3,a}\\[7pt]
\dfrac{1}{48\tau^{3/2}}\left[
2\big(3s_1-s_3\big)\mathcal P_{0,a}
-\big(3s_1-s_3\big)\mathcal P_{0Q,a}
+6\tau\big(5s_1+s_3\big)\mathcal P_{2,a}
\right]
\end{pmatrix}.
\label{eq:appC3Ospin32}
\end{align}
\end{widetext}
Equations~\eqref{eq:appC0spin32}-\eqref{eq:appC3Ospin32} give all
eighteen multipole form factors in
Eq.~\eqref{eq:EFmultipoledecomposition32}. Although some expressions
contain inverse powers of $\tau$, their $\tau\to0$ limits are finite
because $\theta(P_z,t)=\mathcal O(\sqrt\tau)$.
\section{\texorpdfstring{Direct LF multipole form factors for
$T^{+-}$ and $T^{--}$}{Direct LF multipole form factors for T+- and T--}}
\label{app:directLF32}
Here we list the multipole form factors that enter the $T^{+-}$ and
$T^{--}$ matrix elements in Eq.~\eqref{eq:LFmultipoledecomposition32}.
\subsection{\texorpdfstring{$T^{+-}$}{T+-}}
The six $T^{+-}$ multipole form factors share the factor
$m/[2P^+(1+\tau)]$.
\begin{widetext}
\begin{subequations}
\label{eq:LFplusminusMultipoles32}
\begin{align}
 C_{0,a,\mathrm{LF}}^{+-}
={}&\frac{m}{2P^+(1+\tau)}\Big[
 (1-\tau)(\mathcal E_{0,a}-\mathcal P_{0,a})
 -\tau\mathcal P_{0Q,a}
 -8\tau^2\mathcal E_{2,a}
 -6\tau^2\mathcal P_{2,a}
 \Big],
 \\[2pt]
 C_{0Q,a,\mathrm{LF}}^{+-}
={}&\frac{m}{2P^+(1+\tau)}\Big[
 \tau(\mathcal E_{0,a}-\mathcal P_{0,a})
 -\frac{2+\tau}{2}\mathcal P_{0Q,a}
 +4\tau(1+2\tau)\mathcal E_{2,a}
 +3\tau^2\mathcal P_{2,a}
 \Big],
 \\[2pt]
 C_{1S,a,\mathrm{LF}}^{+-}
={}&\frac{m}{2P^+(1+\tau)}\Big[
 \frac{\tau-5}{5}(\mathcal E_{0,a}-\mathcal P_{0,a})
 +\frac{\tau-2}{5}\mathcal P_{0Q,a}
 +\frac{8}{5}\tau(\tau-2)\mathcal E_{2,a}
 +\frac{6}{5}\tau(\tau-2)\mathcal P_{2,a}
 \Big],
 \\[2pt]
 C_{1O,a,\mathrm{LF}}^{+-}
={}&\frac{m}{2P^+(1+\tau)}\Big[
 -\frac{\tau}{2}(\mathcal E_{0,a}-\mathcal P_{0,a})
 +\frac{4+3\tau}{4}\mathcal P_{0Q,a}
 -2\tau(1+2\tau)\mathcal E_{2,a}
 -\frac{\tau(\tau-2)}{2}\mathcal P_{2,a}
 \Big],
 \\[2pt]
 C_{2Q,a,\mathrm{LF}}^{+-}
={}&\frac{m}{2P^+(1+\tau)}\Big[
 \frac12(\mathcal E_{0,a}-\mathcal P_{0,a})
 +\frac14\mathcal P_{0Q,a}
 -2\mathcal E_{2,a}
 +\frac{\tau-2}{2}\mathcal P_{2,a}
 \Big],
 \\[2pt]
 C_{3O,a,\mathrm{LF}}^{+-}
={}&\frac{m}{2P^+(1+\tau)}\Big[
 -\frac16(\mathcal E_{0,a}-\mathcal P_{0,a})
 -\frac1{12}\mathcal P_{0Q,a}
 +\frac{2(2\tau+3)}{3}\mathcal E_{2,a}
 +\frac{\tau+2}{2}\mathcal P_{2,a}
 \Big].
\end{align}
\end{subequations}
\end{widetext}
\subsection{\texorpdfstring{$T^{--}$}{T--}}
The six $T^{--}$ multipole form factors share the factor
$(P^-)^2/[mP^+(1+\tau)^2]$, with $P^-$ fixed by
Eq.~\eqref{eq:LFDYonshell32}.
\begin{widetext}
\begin{subequations}
\label{eq:LFminusminusMultipoles32}
\begin{align}
 C_{0,a,\mathrm{LF}}^{--}
={}&\frac{(P^-)^2}{mP^+(1+\tau)^2}\Big[
 (1-\tau)(\mathcal E_{0,a}+\mathcal P_{0,a})
 +\tau\mathcal P_{0Q,a}
 +2\tau(\tau-5)\mathcal J_{1,a}
 \nonumber\\[-2pt]
 &\hspace{37mm}
 -8\tau^2\mathcal E_{2,a}
 +6\tau^2\mathcal P_{2,a}
 +\frac{96}{5}\tau^3\mathcal J_{3,a}
 \Big],
 \\[2pt]
 C_{0Q,a,\mathrm{LF}}^{--}
={}&\frac{(P^-)^2}{mP^+(1+\tau)^2}\Big[
 \tau(\mathcal E_{0,a}+\mathcal P_{0,a})
 +\frac{2+\tau}{2}\mathcal P_{0Q,a}
 +2\tau(2-\tau)\mathcal J_{1,a}
 \nonumber\\[-2pt]
 &\hspace{37mm}
 +4\tau(1+2\tau)\mathcal E_{2,a}
 -3\tau^2\mathcal P_{2,a}
 -\frac{48}{5}\tau^2(1+2\tau)\mathcal J_{3,a}
 \Big],
 \\[2pt]
 C_{1S,a,\mathrm{LF}}^{--}
={}&\frac{(P^-)^2}{mP^+(1+\tau)^2}\Big[
 \frac{\tau-5}{5}(\mathcal E_{0,a}+\mathcal P_{0,a})
 +\frac{2-\tau}{5}\mathcal P_{0Q,a}
 +\frac{2(13\tau-5)}{5}\mathcal J_{1,a}
 \nonumber\\[-2pt]
 &\hspace{37mm}
 +\frac{8}{5}\tau(\tau-2)\mathcal E_{2,a}
 +\frac{6}{5}\tau(2-\tau)\mathcal P_{2,a}
 +\frac{288}{25}\tau^2\mathcal J_{3,a}
 \Big],
 \\[2pt]
 C_{1O,a,\mathrm{LF}}^{--}
={}&\frac{(P^-)^2}{mP^+(1+\tau)^2}\Big[
 -\frac{\tau}{2}(\mathcal E_{0,a}+\mathcal P_{0,a})
 -\frac{4+3\tau}{4}\mathcal P_{0Q,a}
 -3\tau\mathcal J_{1,a}
 \nonumber\\[-2pt]
 &\hspace{37mm}
 -2\tau(1+2\tau)\mathcal E_{2,a}
 +\frac{\tau(\tau-2)}{2}\mathcal P_{2,a}
 -\frac{8}{5}\tau(5+8\tau)\mathcal J_{3,a}
 \Big],
 \\[2pt]
 C_{2Q,a,\mathrm{LF}}^{--}
={}&\frac{(P^-)^2}{mP^+(1+\tau)^2}\Big[
 \frac12(\mathcal E_{0,a}+\mathcal P_{0,a})
 -\frac14\mathcal P_{0Q,a}
 +(2-\tau)\mathcal J_{1,a}
 \nonumber\\[-2pt]
 &\hspace{37mm}
 -2\mathcal E_{2,a}
 +\left(1-\frac{\tau}{2}\right)\mathcal P_{2,a}
 +\frac{8}{5}\tau(7+4\tau)\mathcal J_{3,a}
 \Big],
 \\[2pt]
 C_{3O,a,\mathrm{LF}}^{--}
={}&\frac{(P^-)^2}{mP^+(1+\tau)^2}\Big[
 -\frac16(\mathcal E_{0,a}+\mathcal P_{0,a})
 +\frac1{12}\mathcal P_{0Q,a}
 -\mathcal J_{1,a}
 \nonumber\\[-2pt]
 &\hspace{37mm}
 +\frac{2(2\tau+3)}{3}\mathcal E_{2,a}
 -\frac{\tau+2}{2}\mathcal P_{2,a}
 +\frac{8}{5}(5+4\tau)\mathcal J_{3,a}
 \Big].
\end{align}
\end{subequations}
\end{widetext}
Together with the $T^{++}$ results in
Eqs.~\eqref{eq:directLFformfactors32},
Eqs.~\eqref{eq:LFplusminusMultipoles32} and
\eqref{eq:LFminusminusMultipoles32} complete the set of direct LF
multipole form factors for the three independent components
$T^{++}$, $T^{+-}$, and $T^{--}$. Their relation to the leading EF
matrix elements in the IMF is stated in
Eq.~\eqref{eq:IMFLFmatrixrelation32}.
\bibliographystyle{apsrev4-2}
\bibliography{Refs}

@article{Won:2026ljg,
    author = "Won, Ho-Yeon and Lorc{\'e}, C{\'e}dric",
    title = "{Transverse energy-momentum tensor distributions in polarized nucleons}",
    eprint = "2604.07616",
    archivePrefix = "arXiv",
    primaryClass = "hep-ph",
    doi = "10.1103/l7qb-49nz",
    journal = "Phys. Rev. D",
    volume = "114",
    number = "1",
    pages = "014012",
    year = "2026"
}

@article{Alexandrou:2009hs,
    author = "Alexandrou, Constantia and Korzec, Tomasz and Koutsou, Giannis and Lorc{\'e}, C{\'e}dric and Negele, John W. and Pascalutsa, Vladimir and Tsapalis, Antonios and Vanderhaeghen, Marc",
    title = "{Quark transverse charge densities in the Delta(1232) from lattice QCD}",
    eprint = "0901.3457",
    archivePrefix = "arXiv",
    primaryClass = "hep-ph",
    doi = "10.1016/j.nuclphysa.2009.04.005",
    journal = "Nucl. Phys. A",
    volume = "825",
    pages = "115--144",
    year = "2009"
}

@article{Ji:1995sv,
    author = "Ji, Xiang-Dong",
    title = "{Breakup of hadron masses and energy - momentum tensor of QCD}",
    eprint = "hep-ph/9502213",
    archivePrefix = "arXiv",
    reportNumber = "MIT-CTP-2407",
    doi = "10.1103/PhysRevD.52.271",
    journal = "Phys. Rev. D",
    volume = "52",
    pages = "271--281",
    year = "1995"
}

@article{Panteleeva:2021iip,
    author = "Panteleeva, Julia Yu. and Polyakov, Maxim V.",
    title = "{Forces inside the nucleon on the light front from 3D Breit frame force distributions: Abel tomography case}",
    eprint = "2102.10902",
    archivePrefix = "arXiv",
    primaryClass = "hep-ph",
    doi = "10.1103/PhysRevD.104.014008",
    journal = "Phys. Rev. D",
    volume = "104",
    number = "1",
    pages = "014008",
    year = "2021"
}

@article{Kim:2021jjf,
    author = "Kim, June-Young and Kim, Hyun-Chul",
    title = "{Energy-momentum tensor of the nucleon on the light front: Abel tomography case}",
    eprint = "2105.10279",
    archivePrefix = "arXiv",
    primaryClass = "hep-ph",
    reportNumber = "INHA-NTG-04/2021",
    doi = "10.1103/PhysRevD.104.074019",
    journal = "Phys. Rev. D",
    volume = "104",
    number = "7",
    pages = "074019",
    year = "2021"
}

@article{Kim:2022bia,
    author = "Kim, June-Young",
    title = "{Electromagnetic multipole structure of a spin-one particle: Abel tomography case}",
    eprint = "2204.08248",
    archivePrefix = "arXiv",
    primaryClass = "hep-ph",
    doi = "10.1103/PhysRevD.106.014022",
    journal = "Phys. Rev. D",
    volume = "106",
    number = "1",
    pages = "014022",
    year = "2022"
}

@article{Carlson:2009ovh,
    author = "Carlson, Carl E. and Vanderhaeghen, Marc",
    title = "{Empirical transverse charge densities in the deuteron}",
    eprint = "0807.4537",
    archivePrefix = "arXiv",
    primaryClass = "hep-ph",
    doi = "10.1140/epja/i2009-10800-0",
    journal = "Eur. Phys. J. A",
    volume = "41",
    pages = "1--5",
    year = "2009"
}

@article{Cosyn:2019aio,
    author = "Cosyn, Wim and Cotogno, Sabrina and Freese, Adam and Lorc{\'e}, C{\'e}dric",
    title = "{The energy-momentum tensor of spin-1 hadrons: formalism}",
    eprint = "1903.00408",
    archivePrefix = "arXiv",
    primaryClass = "hep-ph",
    doi = "10.1140/epjc/s10052-019-6981-3",
    journal = "Eur. Phys. J. C",
    volume = "79",
    number = "6",
    pages = "476",
    year = "2019"
}

@article{Panteleeva:2020ejw,
    author = "Panteleeva, Julia Yu. and Polyakov, Maxim V.",
    title = "{Quadrupole pressure and shear forces inside baryons in the large $N_c$ limit}",
    eprint = "2004.02912",
    archivePrefix = "arXiv",
    primaryClass = "hep-ph",
    doi = "10.1016/j.physletb.2020.135707",
    journal = "Phys. Lett. B",
    volume = "809",
    pages = "135707",
    year = "2020"
}

@article{Cotogno:2019vjb,
    author = "Cotogno, Sabrina and Lorc{\'e}, C{\'e}dric and Lowdon, Peter and Morales, Manuel",
    title = "{Covariant multipole expansion of local currents for massive states of any spin}",
    eprint = "1912.08749",
    archivePrefix = "arXiv",
    primaryClass = "hep-ph",
    doi = "10.1103/PhysRevD.101.056016",
    journal = "Phys. Rev. D",
    volume = "101",
    number = "5",
    pages = "056016",
    year = "2020"
}

@article{Chen:2022smg,
    author = "Chen, Yi and Lorc{\'e}, C{\'e}dric",
    title = "{Pion and nucleon relativistic electromagnetic four-current distributions}",
    eprint = "2210.02908",
    archivePrefix = "arXiv",
    primaryClass = "hep-ph",
    doi = "10.1103/PhysRevD.106.116024",
    journal = "Phys. Rev. D",
    volume = "106",
    number = "11",
    pages = "116024",
    year = "2022"
}

@article{Won:2025dgc,
    author = "Won, Ho-Yeon and Lorc{\'e}, C{\'e}dric",
    title = "{Relativistic energy-momentum tensor distributions in a polarized nucleon}",
    eprint = "2503.07382",
    archivePrefix = "arXiv",
    primaryClass = "hep-ph",
    doi = "10.1103/PhysRevD.111.094021",
    journal = "Phys. Rev. D",
    volume = "111",
    number = "9",
    pages = "094021",
    year = "2025"
}

@article{Lorce:2022cle,
    author = "Lorc{\'e}, C{\'e}dric and Schweitzer, Peter and Tezgin, Kemal",
    title = "{2D energy-momentum tensor distributions of nucleon in a large-Nc quark model from ultrarelativistic to nonrelativistic limit}",
    eprint = "2202.01192",
    archivePrefix = "arXiv",
    primaryClass = "hep-ph",
    doi = "10.1103/PhysRevD.106.014012",
    journal = "Phys. Rev. D",
    volume = "106",
    number = "1",
    pages = "014012",
    year = "2022"
}

@article{Lorce:2020onh,
    author = "Lorc{\'e}, C{\'e}dric",
    title = "{Charge Distributions of Moving Nucleons}",
    eprint = "2007.05318",
    archivePrefix = "arXiv",
    primaryClass = "hep-ph",
    doi = "10.1103/PhysRevLett.125.232002",
    journal = "Phys. Rev. Lett.",
    volume = "125",
    number = "23",
    pages = "232002",
    year = "2020"
}

@article{Lorce:2025oot,
    author = "Lorc{\'e}, C{\'e}dric and Schweitzer, Peter",
    title = "{Pressure inside hadrons: criticism, conjectures, and all that}",
    eprint = "2501.04622",
    archivePrefix = "arXiv",
    primaryClass = "hep-ph",
    doi = "10.5506/APhysPolB.56.3-A17",
    journal = "Acta Phys. Polon. B",
    volume = "56",
    pages = "3--A17",
    year = "2025"
}

@article{Lorce:2018egm,
    author = "Lorc{\'e}, C{\'e}dric and Moutarde, Herv{\'e} and Trawi{\'n}ski, Arkadiusz P.",
    title = "{Revisiting the mechanical properties of the nucleon}",
    eprint = "1810.09837",
    archivePrefix = "arXiv",
    primaryClass = "hep-ph",
    doi = "10.1140/epjc/s10052-019-6572-3",
    journal = "Eur. Phys. J. C",
    volume = "79",
    number = "1",
    pages = "89",
    year = "2019"
}

@article{Burkert:2023wzr,
    author = "Burkert, V. D. and Elouadrhiri, L. and Girod, F. X. and Lorc{\'e}, C. and Schweitzer, P. and Shanahan, P. E.",
    title = "{Colloquium: Gravitational form factors of the proton}",
    eprint = "2303.08347",
    archivePrefix = "arXiv",
    primaryClass = "hep-ph",
    reportNumber = "JLAB-PHY-23-3774",
    doi = "10.1103/RevModPhys.95.041002",
    journal = "Rev. Mod. Phys.",
    volume = "95",
    number = "4",
    pages = "041002",
    year = "2023"
}

@article{Freese:2021czn,
    author = "Freese, Adam and Miller, Gerald A.",
    title = "{Forces within hadrons on the light front}",
    eprint = "2102.01683",
    archivePrefix = "arXiv",
    primaryClass = "hep-ph",
    reportNumber = "NT@UW-21-02",
    doi = "10.1103/PhysRevD.103.094023",
    journal = "Phys. Rev. D",
    volume = "103",
    pages = "094023",
    year = "2021"
}

@article{Freese:2022yur,
    author = "Freese, Adam and Cosyn, Wim",
    title = "{Spatial densities of momentum and forces in spin-one hadrons}",
    eprint = "2207.10787",
    archivePrefix = "arXiv",
    primaryClass = "hep-ph",
    reportNumber = "NT@UW-2206",
    doi = "10.1103/PhysRevD.106.114013",
    journal = "Phys. Rev. D",
    volume = "106",
    number = "11",
    pages = "114013",
    year = "2022"
}

@article{Kim:2022wkc,
    author = "Kim, June-Young and Sun, Bao-Dong and Fu, Dongyan and Kim, Hyun-Chul",
    title = "{Mechanical structure of a spin-1 particle}",
    eprint = "2208.01240",
    archivePrefix = "arXiv",
    primaryClass = "hep-ph",
    reportNumber = "INHA-NTG-06/2022",
    doi = "10.1103/PhysRevD.107.054007",
    journal = "Phys. Rev. D",
    volume = "107",
    number = "5",
    pages = "054007",
    year = "2023"
}

@article{Polyakov:2002yz,
    author = "Polyakov, M. V.",
    title = "{Generalized parton distributions and strong forces inside nucleons and nuclei}",
    eprint = "hep-ph/0210165",
    archivePrefix = "arXiv",
    reportNumber = "RUB-TP2-14-02",
    doi = "10.1016/S0370-2693(03)00036-4",
    journal = "Phys. Lett. B",
    volume = "555",
    pages = "57--62",
    year = "2003"
}

@article{Goeke:2007fp,
    author = "Goeke, K. and Grabis, J. and Ossmann, J. and Polyakov, M. V. and Schweitzer, P. and Silva, A. and Urbano, D.",
    title = "{Nucleon form-factors of the energy momentum tensor in the chiral quark-soliton model}",
    eprint = "hep-ph/0702030",
    archivePrefix = "arXiv",
    reportNumber = "RUB-TP2-05-2006",
    doi = "10.1103/PhysRevD.75.094021",
    journal = "Phys. Rev. D",
    volume = "75",
    pages = "094021",
    year = "2007"
}

@article{Polyakov:2018zvc,
    author = "Polyakov, Maxim V. and Schweitzer, Peter",
    title = "{Forces inside hadrons: pressure, surface tension, mechanical radius, and all that}",
    eprint = "1805.06596",
    archivePrefix = "arXiv",
    primaryClass = "hep-ph",
    doi = "10.1142/S0217751X18300259",
    journal = "Int. J. Mod. Phys. A",
    volume = "33",
    number = "26",
    pages = "1830025",
    year = "2018"
}

@article{Pagels:1966zza,
    author = "Pagels, Heinz",
    title = "{Energy-Momentum Structure Form Factors of Particles}",
    doi = "10.1103/PhysRev.144.1250",
    journal = "Phys. Rev.",
    volume = "144",
    pages = "1250--1260",
    year = "1966"
}

@book{Varshalovich:1988ifq,
    author = "Varshalovich, D. A. and Moskalev, A. N. and Khersonskii, V. K.",
    title = "{Quantum Theory of Angular Momentum}: {Irreducible Tensors, Spherical Harmonics, Vector Coupling Coefficients, 3nj Symbols}",
    doi = "10.1142/0270",
    isbn = "978-981-4415-49-1, 978-9971-5-0107-5",
    publisher = "World Scientific Publishing Company",
    year = "1988"
}

@article{Kim:2020lrs,
    author = "Kim, June-Young and Sun, Bao-Dong",
    title = "{Gravitational form factors of a baryon with spin-3/2}",
    eprint = "2011.00292",
    archivePrefix = "arXiv",
    primaryClass = "hep-ph",
    doi = "10.1140/epjc/s10052-021-08852-z",
    journal = "Eur. Phys. J. C",
    volume = "81",
    number = "1",
    pages = "85",
    year = "2021"
}

@article{Polyakov:2019lbq,
    author = "Polyakov, Maxim V. and Sun, Bao-Dong",
    title = "{Gravitational form factors of a spin one particle}",
    eprint = "1903.02738",
    archivePrefix = "arXiv",
    primaryClass = "hep-ph",
    doi = "10.1103/PhysRevD.100.036003",
    journal = "Phys. Rev. D",
    volume = "100",
    number = "3",
    pages = "036003",
    year = "2019"
}

@article{Belitsky:2005qn,
    author = "Belitsky, A. V. and Radyushkin, A. V.",
    title = "{Unraveling hadron structure with generalized parton distributions}",
    eprint = "hep-ph/0504030",
    archivePrefix = "arXiv",
    reportNumber = "JLAB-THY-04-34",
    doi = "10.1016/j.physrep.2005.06.002",
    journal = "Phys. Rept.",
    volume = "418",
    pages = "1--387",
    year = "2005"
}

@article{Diehl:2003ny,
    author = "Diehl, M.",
    title = "{Generalized parton distributions}",
    eprint = "hep-ph/0307382",
    archivePrefix = "arXiv",
    reportNumber = "DESY-THESIS-2003-018",
    doi = "10.1016/j.physrep.2003.08.002",
    journal = "Phys. Rept.",
    volume = "388",
    pages = "41--277",
    year = "2003"
}

@article{Goeke:2001tz,
    author = "Goeke, K. and Polyakov, Maxim V. and Vanderhaeghen, M.",
    title = "{Hard exclusive reactions and the structure of hadrons}",
    eprint = "hep-ph/0106012",
    archivePrefix = "arXiv",
    doi = "10.1016/S0146-6410(01)00158-2",
    journal = "Prog. Part. Nucl. Phys.",
    volume = "47",
    pages = "401--515",
    year = "2001"
}

@article{Diehl:2005jf,
    author = "Diehl, M. and Hagler, Ph.",
    title = "{Spin densities in the transverse plane and generalized transversity distributions}",
    eprint = "hep-ph/0504175",
    archivePrefix = "arXiv",
    reportNumber = "DESY-05-061",
    doi = "10.1140/epjc/s2005-02342-6",
    journal = "Eur. Phys. J. C",
    volume = "44",
    pages = "87--101",
    year = "2005"
}

@article{Ji:1996ek,
    author = "Ji, Xiang-Dong",
    title = "{Gauge-Invariant Decomposition of Nucleon Spin}",
    eprint = "hep-ph/9603249",
    archivePrefix = "arXiv",
    reportNumber = "MIT-CTP-2517",
    doi = "10.1103/PhysRevLett.78.610",
    journal = "Phys. Rev. Lett.",
    volume = "78",
    pages = "610--613",
    year = "1997"
}

@article{Leader:2013jra,
    author = "Leader, E. and Lorc\'e, C.",
    title = "{The angular momentum controversy: What\textquoteright{}s it all about and does it matter?}",
    eprint = "1309.4235",
    archivePrefix = "arXiv",
    primaryClass = "hep-ph",
    doi = "10.1016/j.physrep.2014.02.010",
    journal = "Phys. Rept.",
    volume = "541",
    number = "3",
    pages = "163--248",
    year = "2014"
}

@article{Wigner:1939cj,
    author = "Wigner, Eugene P.",
    title = "{On Unitary Representations of the Inhomogeneous Lorentz Group}",
    doi = "10.2307/1968551",
    journal = "Annals Math.",
    volume = "40",
    number = "1",
    pages = "149--204",
    year = "1939"
}

@article{Keister:1991sb,
    author = "Keister, B. D. and Polyzou, W. N.",
    title = "{Relativistic Hamiltonian Dynamics in Nuclear and Particle Physics}",
    doi = "10.1007/0-306-47120-5_2",
    journal = "Adv. Nucl. Phys.",
    volume = "20",
    pages = "225--479",
    year = "1991"
}

@article{Belinfante:1939,
    author = "Belinfante, F. J.",
    title = "{On the spin angular momentum of mesons}",
    doi = "10.1016/S0031-8914(39)90090-X",
    journal = "Physica",
    volume = "6",
    number = "7--12",
    pages = "887--898",
    year = "1939"
}

@article{Rarita:1941mf,
    author = "Rarita, William and Schwinger, Julian",
    title = "{On a Theory of Particles with Half-Integral Spin}",
    doi = "10.1103/PhysRev.60.61",
    journal = "Phys. Rev.",
    volume = "60",
    pages = "61",
    year = "1941"
}

@article{Dirac:1949cp,
    author = "Dirac, P. A. M.",
    title = "{Forms of Relativistic Dynamics}",
    doi = "10.1103/RevModPhys.21.392",
    journal = "Rev. Mod. Phys.",
    volume = "21",
    pages = "392--399",
    year = "1949"
}

@article{Kogut:1969xa,
    author = "Kogut, John B. and Soper, Davison E.",
    title = "{Quantum Electrodynamics in the Infinite-Momentum Frame}",
    doi = "10.1103/PhysRevD.1.2901",
    journal = "Phys. Rev. D",
    volume = "1",
    pages = "2901--2913",
    year = "1970"
}

@article{Drell:1969km,
    author = "Drell, Sidney D. and Yan, Tung-Mow",
    title = "{Connection of Elastic Electromagnetic Nucleon Form Factors at Large $Q^2$ and Deep Inelastic Structure Functions Near Threshold}",
    doi = "10.1103/PhysRevLett.24.181",
    journal = "Phys. Rev. Lett.",
    volume = "24",
    pages = "181--185",
    year = "1970"
}

@article{West:1970av,
    author = "West, Geoffrey B.",
    title = "{Phenomenological Model for the Electromagnetic Structure of the Proton}",
    doi = "10.1103/PhysRevLett.24.1206",
    journal = "Phys. Rev. Lett.",
    volume = "24",
    pages = "1206--1209",
    year = "1970"
}

@article{Soper:1971wn,
    author = "Soper, Davison E.",
    title = "{Infinite-Momentum Helicity States}",
    doi = "10.1103/PhysRevD.5.1956",
    journal = "Phys. Rev. D",
    volume = "5",
    pages = "1956--1962",
    year = "1972"
}

@article{Melosh:1974cu,
    author = "Melosh, H. J.",
    title = "{Quarks: Currents and Constituents}",
    doi = "10.1103/PhysRevD.9.1095",
    journal = "Phys. Rev. D",
    volume = "9",
    pages = "1095--1112",
    year = "1974"
}

@article{Ahluwalia:1993ff,
    author = "Ahluwalia, D. V. and Sawicki, Mikolaj",
    title = "{Front-form spinors in the Weinberg-Soper formalism and generalized Melosh transformations for any spin}",
    eprint = "nucl-th/9603019",
    archivePrefix = "arXiv",
    doi = "10.1103/PhysRevD.47.5161",
    journal = "Phys. Rev. D",
    volume = "47",
    pages = "5161--5168",
    year = "1993"
}

@article{Brodsky:1997de,
    author = "Brodsky, Stanley J. and Pauli, Hans-Christian and Pinsky, Stephen S.",
    title = "{Quantum Chromodynamics and Other Field Theories on the Light Cone}",
    eprint = "hep-ph/9705477",
    archivePrefix = "arXiv",
    doi = "10.1016/S0370-1573(97)00089-6",
    journal = "Phys. Rept.",
    volume = "301",
    pages = "299--486",
    year = "1998"
}

@article{Burkardt:2000za,
    author = "Burkardt, Matthias",
    title = "{Impact parameter dependent parton distributions and off-forward parton distributions for $\zeta\to0$}",
    eprint = "hep-ph/0005108",
    archivePrefix = "arXiv",
    doi = "10.1103/PhysRevD.62.071503",
    journal = "Phys. Rev. D",
    volume = "62",
    pages = "071503",
    year = "2000",
    note = "[Erratum: Phys. Rev. D 66, 119903 (2002)]"
}

@article{Carlson:2003je,
    author = "Carlson, Carl E. and Ji, Chueng-Ryong",
    title = "{Angular conditions, relations between Breit and light-front frames, and subleading power corrections}",
    eprint = "hep-ph/0301213",
    archivePrefix = "arXiv",
    doi = "10.1103/PhysRevD.67.116002",
    journal = "Phys. Rev. D",
    volume = "67",
    pages = "116002",
    year = "2003"
}

@article{vonLaue:1911,
    author = "von Laue, Max",
    title = "{Zur Dynamik der Relativit{\"a}tstheorie}",
    doi = "10.1002/andp.19113400808",
    journal = "Annalen Phys.",
    volume = "340",
    number = "8",
    pages = "524--542",
    year = "1911"
}

@article{Miller:2007uy,
    author = "Miller, Gerald A.",
    title = "{Charge Densities of the Neutron and Proton}",
    eprint = "0705.2409",
    archivePrefix = "arXiv",
    primaryClass = "nucl-th",
    doi = "10.1103/PhysRevLett.99.112001",
    journal = "Phys. Rev. Lett.",
    volume = "99",
    pages = "112001",
    year = "2007"
}

@article{Burkardt:2002hr,
    author = "Burkardt, Matthias",
    title = "{Impact parameter space interpretation for generalized parton distributions}",
    eprint = "hep-ph/0207047",
    archivePrefix = "arXiv",
    doi = "10.1142/S0217751X03012370",
    journal = "Int. J. Mod. Phys. A",
    volume = "18",
    pages = "173--208",
    year = "2003"
}

@article{Carlson:2007xd,
    author = "Carlson, Carl E. and Vanderhaeghen, Marc",
    title = "{Empirical transverse charge densities in the nucleon and the nucleon-to-$\Delta$ transition}",
    eprint = "0710.0835",
    archivePrefix = "arXiv",
    primaryClass = "hep-ph",
    doi = "10.1103/PhysRevLett.100.032004",
    journal = "Phys. Rev. Lett.",
    volume = "100",
    pages = "032004",
    year = "2008"
}

@article{Miller:2010nz,
    author = "Miller, Gerald A.",
    title = "{Transverse Charge Densities}",
    eprint = "1002.0355",
    archivePrefix = "arXiv",
    primaryClass = "nucl-th",
    doi = "10.1146/annurev.nucl.012809.104508",
    journal = "Ann. Rev. Nucl. Part. Sci.",
    volume = "60",
    pages = "1--25",
    year = "2010"
}

@article{Jaffe:2020ebz,
    author = "Jaffe, R. L.",
    title = "{Ambiguities in the definition of local spatial densities in light hadrons}",
    eprint = "2010.15887",
    archivePrefix = "arXiv",
    primaryClass = "hep-ph",
    reportNumber = "MIT-CTP-5245",
    doi = "10.1103/PhysRevD.103.016017",
    journal = "Phys. Rev. D",
    volume = "103",
    number = "1",
    pages = "016017",
    year = "2021"
}

@article{Epelbaum:2022fjc,
    author = "Epelbaum, E. and Gegelia, J. and Lange, N. and Mei{\ss}ner, U. -G. and Polyakov, M. V.",
    title = "{Definition of Local Spatial Densities in Hadrons}",
    eprint = "2201.02565",
    archivePrefix = "arXiv",
    primaryClass = "hep-ph",
    doi = "10.1103/PhysRevLett.129.012001",
    journal = "Phys. Rev. Lett.",
    volume = "129",
    number = "1",
    pages = "012001",
    year = "2022"
}

@article{Cotogno:2019xcl,
    author = "Cotogno, Sabrina and Lorc{\'e}, C{\'e}dric and Lowdon, Peter",
    title = "{Poincar{\'e} constraints on the gravitational form factors for massive states with arbitrary spin}",
    eprint = "1905.11969",
    archivePrefix = "arXiv",
    primaryClass = "hep-th",
    doi = "10.1103/PhysRevD.100.045003",
    journal = "Phys. Rev. D",
    volume = "100",
    number = "4",
    pages = "045003",
    year = "2019"
}

@article{Lorce:2019sbq,
    author = "Lorc{\'e}, C{\'e}dric and Lowdon, Peter",
    title = "{Universality of the Poincar{\'e} gravitational form factor constraints}",
    eprint = "1908.02567",
    archivePrefix = "arXiv",
    primaryClass = "hep-th",
    doi = "10.1140/epjc/s10052-020-7779-z",
    journal = "Eur. Phys. J. C",
    volume = "80",
    number = "3",
    pages = "207",
    year = "2020"
}

@article{Kim:2025iis,
    author = "Kim, June-Young and Kim, Hyun-Chul",
    title = "{Quadrupole forces between quark and gluon subsystems inside higher-spin particles}",
    eprint = "2508.21319",
    archivePrefix = "arXiv",
    primaryClass = "hep-ph",
    reportNumber = "INHA-NTG-07/2025",
    doi = "10.1103/r2t1-f4mc",
    journal = "Phys. Rev. D",
    volume = "112",
    number = "7",
    pages = "074014",
    year = "2025"
}
\end{document}